\documentclass[acmsmall,nonacm]{acmart} 
\usepackage{xcolor}
\usepackage{xspace}
\usepackage{amsmath}
\usepackage{amsthm}
\usepackage{subcaption}
\usepackage{makecell}
\usepackage{wrapfig}
\usepackage[linesnumbered,boxed,ruled,noend,commentsnumbered,norelsize]{algorithm2e}

\AtBeginDocument{%
  \providecommand\BibTeX{{%
    \normalfont B\kern-0.5em{\scshape i\kern-0.25em b}\kern-0.8em\TeX}}}

\definecolor{wtcolor}{RGB}{98, 172, 189}
\newcommand{\vls}[1]{{\color{blue} \textbf{Vibhaa:} {#1}}}
\newcommand{\wt}[1]{{\color{wtcolor} \textbf{Weizhao:} {#1}}}

\newcommand{\gf}[1]{{\color{red} \textbf{Giulia:} {#1}}}
\newcommand{\ma}[1]{{\color{red} \textbf{Mohammad:} {#1}}}
\newcommand{\blue}[1]{{\color{purple} {#1}}}

\newcommand{\ie}{{\em i.e., }}
\newcommand{\Sec}[1]{\S\ref{#1}}
\newcommand{\App}[1]{App.~\ref{#1}}
\newcommand{\Alg}[1]{Algorithm~\ref{alg:#1}}

\newcommand{\denseness}{density\xspace}

\newcommand{\Fig}[1]{Fig.~\ref{fig:#1}}
\newcommand{\NewPara}[1]{\noindent{\bf #1}}

\newcommand{\eat}[1]{}
\newcommand{\Tab}[1]{Tab.~\ref{tab:#1}\xspace}
\newcommand{\Eqn}[1]{Eq.~\ref{eqn:#1}\xspace}
\newcommand{\Thm}[1]{Theorem.~\ref{thm:#1}}

\newtheorem{thm}{Theorem}

\newtheorem{definition}{Definition}
\newtheorem{lemma}[thm]{Lemma}

\newtheorem{corollary}[thm]{Corollary}
\newtheorem{proposition}[thm]{Proposition}

\def\one{\boldsymbol{1}}
\def\zero{\boldsymbol{0}}
\def\ab{\boldsymbol{a}}
\def\bb{\boldsymbol{b}}
\def\cb{\boldsymbol{c}}
\def\db{\boldsymbol{D}}
\def\fb{\boldsymbol{f}}
\def\gb{\boldsymbol{g}}
\def\paths{\mathcal{P}}
\def\capvec{\boldsymbol{C}}
\def\capscalar{C}
\def\pb{\boldsymbol{p}}
\def\db{\boldsymbol{d}}

\def\qb{\boldsymbol{q}}
\def\xb{\boldsymbol{x}}
\def\yb{\boldsymbol{y}}
\def\gammab{\boldsymbol{\gamma}}
\def\xib{\boldsymbol{\xi}}
\def\T{^{\top}}
\def\brk{^{(\pi)}}
\def\dlk{\mathbb{L}}
\def\rch{\mathbb{X}}
\def\feasflow{\mathbb{F}}
\def\efree{E_{\mathrm{free}}}
\def\ffree{\Pi_{\mathrm{free}}}
\def\allbfree{\widetilde{\mathbb{B}}}
\def\interiorfree{\widetilde{\mathbb{B}}_{+}}
\def\rchfreeorig{\rch^{\mathrm{free}}}
\def\rchfree{\widetilde{\rch}}

\newcommand*{\qeda}{\ifmmode \tag*{\hfill\ensuremath{\blacksquare}} \else \hfill\ensuremath{\blacksquare} \fi}
\newcommand{\annot}[1]{}

\newcommand{\allb}{\mathbb{B}}
\newcommand{\interior}{\mathbb{B_{+}}}
\newcommand{\corners}{\mathbb{B}_{\text{corner}}} 
\newcommand{\boundaries}{\mathbb{B}_{\text{boundary}}}
\newcommand{\allr}{\mathbb{R}}
\newcommand{\alln}{\mathbb{N}}
\newcommand{\tpt}{\phi}
\newcommand{\ktpt}{\tpt^{(k)}}
\newcommand{\ostpt}{one-step throughput\xspace}
\newcommand{\wctpt}{worst-case throughput\xspace}
\newcommand{\dpp}{deadlock peeling process\xspace}
\newcommand{\cn}{credit network\xspace}
\newcommand{\cns}{credit networks\xspace}
\newcommand{\mintpt}{\Phi_{\min}}
\newcommand{\maxtpt}{\Phi_{\max}}
\newcommand{\maxdlk}{\boldsymbol{B}_{\mathrm{worst}}}
\newcommand{\dir}{\mathbb{D}}
\newcommand{\reachb}{\rch_{\bb}}

\newcommand{\routemat}{\Delta R}

\newcommand{\er}{Erd\H{o}s-R\'{e}nyi\xspace}
\newcommand{\sftopo}{scale-free\xspace}
\newcommand{\pltopo}{power-law\xspace}
\newcommand{\sw}{small-world\xspace}
\newcommand{\rr}{random regular\xspace}
\newcommand{\lntopo}{Lightning Network\xspace}

\newcommand{\mvec}[2]{\begin{bmatrix} #1 \\ #2 \end{bmatrix}}

\newcommand{\hveclong}[4]{\begin{bmatrix} #1 & #2 &  #3 &  #4 \end{bmatrix}^T}

\SetCommentSty{mycommfont}

\SetProgSty{normalfont}
\SetKwInOut{Input}{input}
\SetKwInOut{Output}{output}
\SetKw{true}{True}
\SetKw{false}{False}
\SetKw{Break}{break}
\SetKw{nil}{Nil}
\SetKwRepeat{Do}{do}{while}
\SetKwProg{Fn}{Function}{:}{}
\SetKw{Continue}{continue}

\SetKw{push}{\textbf{push}}
\SetKw{pop}{\textbf{pop}}

\setcopyright{acmcopyright}
\copyrightyear{2018}
\acmYear{2018}
\acmDOI{10.1145/1122445.1122456}

\acmJournal{JACM}
\acmVolume{37}
\acmNumber{4}
\acmArticle{}
\acmMonth{8}

\setcopyright{none}

\begin{document}

\title{The Effect of Network Topology on Credit Network Throughput}


\author{Vibhaalakshmi Sivaraman}
\affiliation{Massachusetts Institute of Technology, USA}
\author{Weizhao Tang}
\affiliation{Carnegie Mellon University, USA}
\author{Shaileshh Bojja Venkatakrishnan}
\affiliation{The Ohio State University, USA}
\author{Giulia Fanti}
\affiliation{Carnegie Mellon University, USA}
\author{Mohammad Alizadeh}
\affiliation{Massachusetts Institute of Technology, USA}


\begin{abstract}
Credit networks rely on decentralized, pairwise trust relationships (channels) to exchange money or goods. 
Credit networks arise naturally in many financial systems, including the recent construct of \emph{payment channel networks} in blockchain systems. 
An important performance metric for these networks is their transaction throughput.
However, predicting the throughput of a credit network is nontrivial.
Unlike traditional communication channels, credit channels can become imbalanced; they are unable to support more transactions in a given direction once the credit limit has been reached. 
This potential for imbalance creates a complex dependency between a network's throughput and its topology, path choices, and the credit balances (state) on every channel. 
Even worse, certain combinations of these factors
can lead the credit network to \emph{deadlocked} states where no transactions
can make progress. 
In this paper, we study the relationship between the throughput of a credit network and its topology and credit state. 
We show that the presence of deadlocks completely characterizes a network's throughput sensitivity to different credit states.
Although we show that identifying deadlocks in an arbitrary topology is NP-hard,
we propose a peeling algorithm inspired by decoding algorithms for erasure codes that  
upper bounds the severity of the deadlock. We use the peeling algorithm as a tool
to compare the performance of different topologies as well as to aid in the synthesis
of topologies robust to deadlocks.

\end{abstract}




\maketitle

\section{Introduction}
\label{sec:intro}

The global economy relies on digital transactions between entities who do not  trust one another. 
Today, such transactions are handled by intermediaries who 
extract fees (e.g., credit card providers). 
A natural question is how to build financial systems that limit the need for such middlemen.

\emph{Credit networks} (similarly, \emph{debit networks}) are systems in which parties can bootstrap pairwise, distributed trust relations to enable transactions between parties who do not trust each other. 
The core idea is that even if Alice does not trust Charlie directly, if they both share a pairwise trust relationship with Bob, then Alice and Charlie can execute a credit- (or debit-) based transaction through Bob.
The trust relationships that comprise such a network can be based on prior experience or observations\footnote{One of the earliest examples is the {\em hawala} system~\cite{hawala}, a  credit network in existence since the 8th century, which relies on the past performance of a large network of money brokers.}  (e.g., credit scores in a credit network), or they can be based on escrowed funds that are managed either by a third party or an algorithm (debit networks).  Recent debit/credit networks from the blockchain community establish pairwise trust relationships through cryptographically secured data structures stored on a blockchain. Prominent examples include {\em payment channel networks (PCNs)}~\cite{lightning,raiden,deckerPCN} such as the Lightning Network~\cite{lightning} and the Ripple credit network~\cite{moreno2018mind}.

\begin{figure}
\vspace{-3mm}
\centering 
\begin{subfigure}{0.45\textwidth}
  \includegraphics[width=\linewidth]{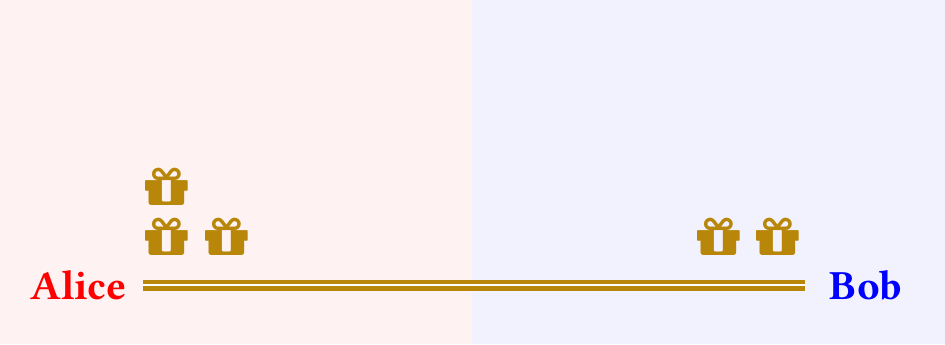}
  \caption{Alice and Bob commit 3 and 2 tokens respectively to a channel for a week.} 
  \label{fig:channel1}
\end{subfigure}\hfil 
\begin{subfigure}{0.45\textwidth}
  \includegraphics[width=\linewidth]{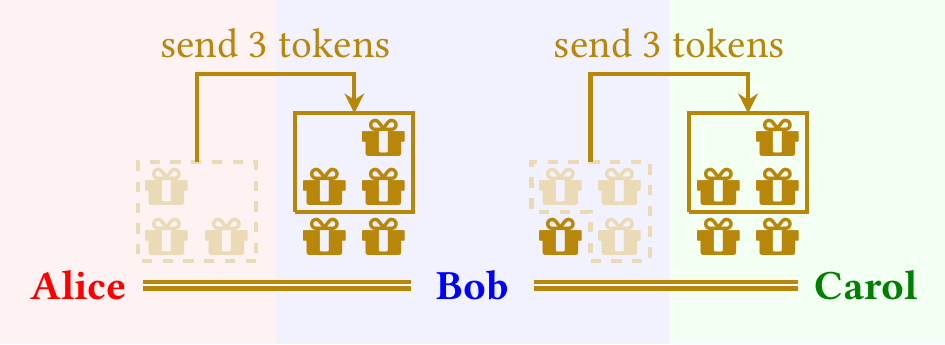}
  \caption{Alice sends 3 tokens to Carol via Bob securely and privately.}
  \label{fig:channel2}
\end{subfigure}\hfil 

\medskip
\begin{subfigure}{0.45\textwidth}
  \includegraphics[width=\linewidth]{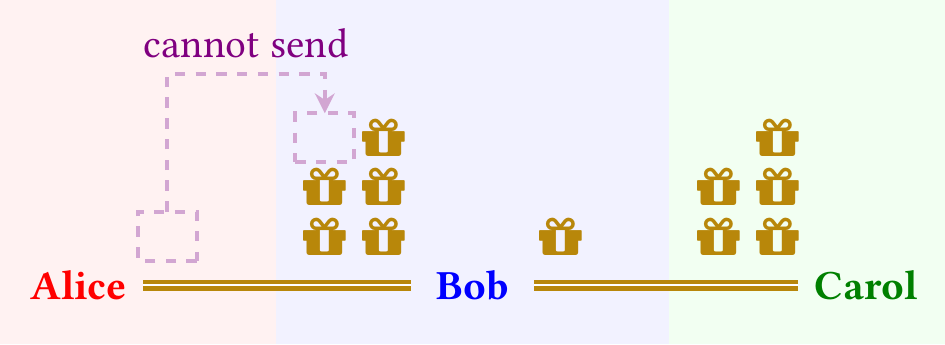}
    \caption{Alice's channel with Bob is \emph{imbalanced} preventing
    Alice from making any more payments.}
  \label{fig:channel3}
\end{subfigure}\hfil 
\begin{subfigure}{0.45\textwidth}
  \includegraphics[width=\linewidth]{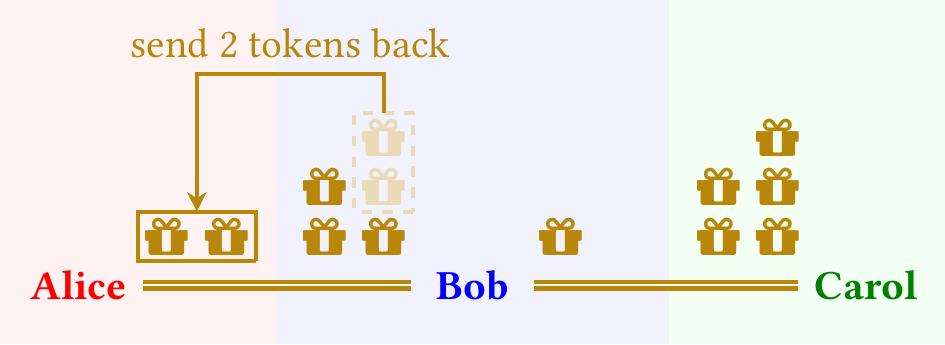}
    \caption{Once Bob sends some tokens back to Alice, some of the balance is restored on the channel.}
  \label{fig:channel4}
\end{subfigure}\hfil 
\caption{\small PCN-style credit network between three parties consisting of two channels through a common intermediary. The network allows each party to transact
with all other parties securely and privately without incurring consensus overhead.}
\label{fig:channel example}
\vspace{-7mm}
\end{figure}

\Fig{channel example} depicts the operation of a pairwise trust channel (or a \emph{payment channel}) in PCNs. If Alice and Bob want to establish a payment channel, they each cryptographically escrow some number of tokens into a contract that is stored on the blockchain and ensures the money can only be used to transact between them for a predefined time period. 
In \Fig{channel1}, Alice commits 3 tokens and Bob commits 2 tokens to a payment channel for a week. 
While the channel is active, Alice and Bob can exchange funds between themselves without committing to the public ledger. 
However, once the active period expires (or once either of the participants choose to close the channel), Alice and Bob commit the final state of the channel back to the blockchain. 
The cryptographic construction of these channels ensures that neither Alice nor Bob can default on their obligation.

PCNs are a network of these pairwise payment channels;
if Alice wants to transact with Carol, she can use Bob as a relay (\Fig{channel2}). We call (Alice, Carol) a demand pair, and the flow of money over route (Alice, Bob, Carol) a \emph{flow} (defined more precisely in Section \ref{sec:model}).  
No party incurs risk of nonpayment because PCNs are implemented as debit networks, with settlement processed on the public blockchain.
PCNs are viewed as an important class of techniques for improving the scalability of legacy blockchains with slow consensus mechanisms \cite{maxwell}. 


Like traditional communication networks, a central performance metric in credit and debit networks is \textbf{throughput}: the total number of transactions a \cn can process per unit time.
However, reasoning about \cn throughput is more difficult than in traditional communication networks because  of  \emph{imbalanced channels}.
That is, because channels impose upper limits on credit (respectively debit) in either direction, transactions cannot flow indefinitely in one direction over a channel.
For instance, in \Fig{channel3}, once Alice sends 3 tokens to Bob, she cannot send any more tokens to Bob (or Carol). If Bob sends some money back to Alice, she would then again have credit to send new payments to Bob or route payments via Bob (\Fig{channel4}). 



Imbalanced channels can affect the throughput of credit networks in unusual ways that depend on topology, user transaction patterns, and transaction routes. An imbalanced channel can 
harm throughput in other parts of the network due to dependencies between flows (e.g., an imbalanced channel blocks a flow, which prevents that flow from balancing other channels, blocking more flows and so on). Certain configurations of imbalanced channels  can even lead to {\em deadlocks} where no transactions can flow over certain edges or even the whole network. Recovering from degraded throughput caused by imbalanced channels requires settlement mechanisms outside of the credit network, such as performing ``on-chain'' transactions on the blockchain to add funds to a channel. 
These mechanisms incur higher cost and overhead compared to transactions within the credit network and should  be avoided. 



In this work, we study the role of network topology and channel imbalance on the throughput of credit networks. While system designers cannot directly control the topology of a decentralized network, existing PCNs (e.g., Lightning Network) indirectly influence network topology, for example, with ``autopilot'' systems that recommend new channels to participants based on a peer's channel degree, size, and longevity~\cite{ln-scoring-blog, ln-routing-guide}.
These systems currently lack an understanding of how network topology and channel imbalance impacts the throughput in credit networks. Our goal is to bridge this gap, paving the way for future autopilot systems that encourage high-throughput topologies with minimal deadlocks. We make the following contributions:




\begin{enumerate}
    \item We present a synchronous round-by-round model to study the throughput of a credit network for a given demand pattern, routing, and channel balance state. We use this model to formulate the best- and worst-case throughput of a credit network as a function of its starting channel balance state as optimization problems. The best-case throughput is easy to compute using a linear program, but determining the worst-case throughput is non-trivial and requires identifying the worst configuration of channel balances. Since transaction patterns and the likelihood of different balance states in \cns are unknown, we instead study the \wctpt in an effort to bound the throughput achievable by a topology.
    \item We introduce the notion of deadlocks\,---\,configurations of channel balances that irrevocably prevent transactions over a subset or all channels in the network. We show that deadlocks precisely capture scenarios in which a credit network's throughput is sensitive to the channel balances, i.e., if a topology is deadlock-free, then its throughput does not depend on the initial state of the channels. 
    
    
    \item We show that the problem of determining whether a given network topology can be deadlocked given a fixed set of demand flows over the topology is NP-hard. Nevertheless, we propose a ``peeling algorithm'' inspired by decoding algorithms for erasure codes~\cite{lubylt} that can be used to bound the number of deadlocked channels in the network. We show empirically that these bounds are very accurate for a variety of topologies, providing a computationally-efficient algorithm to estimate the worst-case throughput. 

    \item We compare standard random graph topologies and a subset of the actual Lightning Network topology in terms of best- and worst-case throughput for randomly-sampled transaction demands. We find that different topologies have different benefits. For example, scale-free graphs have fewer deadlocks and achieve better worst-case throughput than random regular and Erdos-Renyi graphs when the network contains fewer flows, but achieve lower throughput when the network is heavily utilized.
    
    \item We take initial steps towards synthesizing deadlock-resilient topologies by exploiting the surprising connection with erasure codes. In particular, we build on prior approaches for designing efficient LT codes~\cite{lubylt, luby2001efficient} to synthesize topologies with better robustness to deadlocks than existing topologies for a randomized transaction demand model.
    
    
\end{enumerate}

\section{Motivation}
\label{sec:motivation}

\begin{figure*}[t]
    \def\figset{_tikz_tpt}
    \centering
    \begin{tabular}{l c c c}
        \begin{subfigure}{0.23\columnwidth}
            \centering
            \includegraphics[width=\columnwidth]{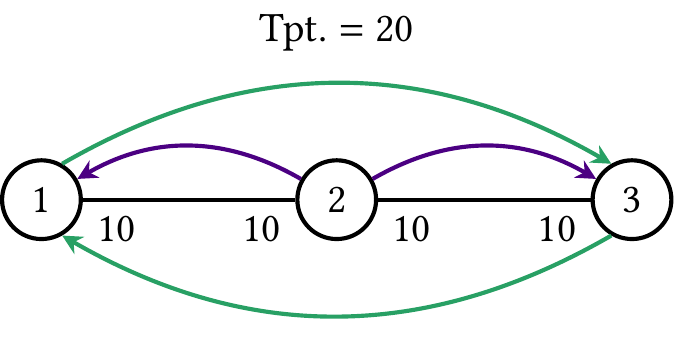}
            \subcaption{Perfect Balance $(\bb_1)$}
\label{fig:perfect balance}
        \end{subfigure} &
        \begin{subfigure}{0.23\columnwidth}
            \centering
            \includegraphics[width=\columnwidth]{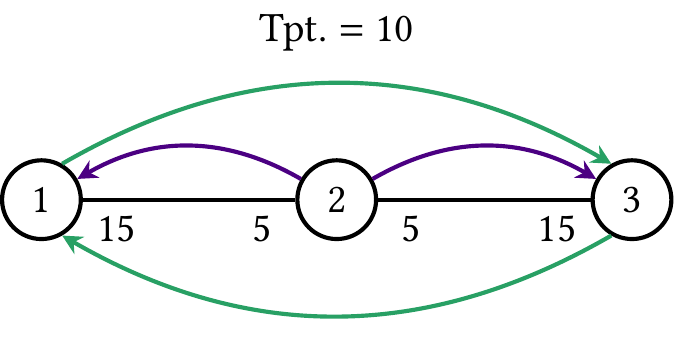}
            \subcaption{Slight Imbalance $(\bb_2)$}
        \label{fig:medtpt}
        \end{subfigure} &
                \begin{subfigure}{0.23\columnwidth}
            \centering
            \includegraphics[width=\columnwidth]{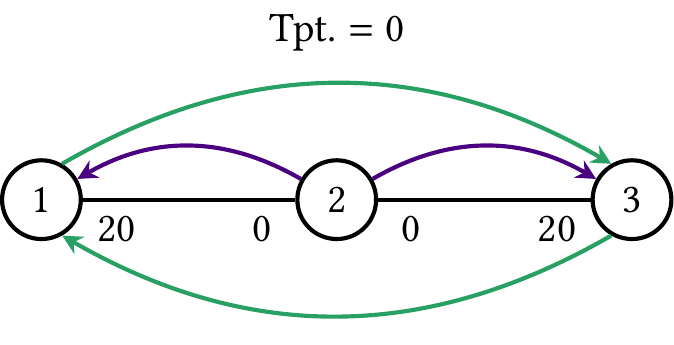}
            \subcaption{Deadlock $(\bb_3)$}
            \label{fig:deadlock}
                \end{subfigure} &
        \begin{subfigure}{0.23\columnwidth}
            \centering
            \includegraphics[width=\columnwidth]{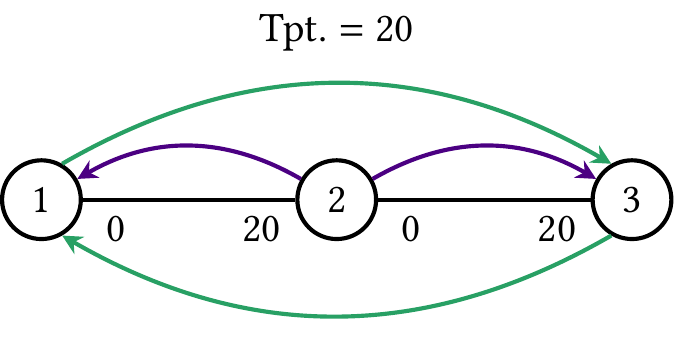}
            \subcaption{Good Boundary $(\bb_4)$}
            \label{fig:good corner}
        \end{subfigure} 
    \end{tabular}
    \vspace{-3mm}
\caption{\small Example illustrating how throughput varies with balance state for the same total collateral}
\vspace{-5mm}
\label{fig:tpt sensitivity}
\end{figure*}

As a motivating example, consider the line topology in \Fig{tpt sensitivity} where nodes $1$ and $3$ have balancing demands to each other, forming a circulation
and node $2$ has one-directional (DAG) demands going out to $1$ and $3$. In steady state, regardless of the balances of the two channels, we only expect
the green flows to contribute to throughput because tokens moved by one green flow is restored by the other, permitting more transactions. 
In \Fig{perfect balance}, both channels are perfectly balanced, and nodes $1$ and $3$ can send at most 10 tokens to each other without changing
the balances on either channel.
If we defined a \emph{throughput} metric for the maximum transaction amount possible in a \emph{round} when send as many tokens as available without changing channel balances at the end of the round, the throughput in the perfect balance state would be $20$ (each green flow moves $10$ tokens).


However, if the purple DAG flows 
were active for a short amount of time in \Fig{perfect balance}, causing 5 tokens
on each channel to move to nodes 1 and 3, we would end up in \Fig{medtpt}. 
Once in \Fig{medtpt}, no amount of future
flows between the shown sender-receiver pairs can ever restore the \cn{} to state \Fig{perfect balance}. 
In other words, \Fig{perfect balance} is \emph{not reachable} from \Fig{medtpt}.
In particular, node $1$ cannot send more than 5 tokens in a given round to node $3$ since it needs funds on the 
$2\rightarrow3$ channel and vice-versa. Sending exactly 5 tokens in a round from each end (node 1 to node 3 and vice versa)
leaves us back at \Fig{medtpt} without any ability to get to \Fig{perfect balance}.
Consequently, the steady-state throughput in \Fig{medtpt} is $10$ instead of $20$ in \Fig{perfect balance}.


Worse still, once in \Fig{medtpt}, if the DAG flows are active for longer, causing the remaining 5 tokens on node $2$ 
on each channel to move outwards, we end up in \Fig{deadlock}.
Notice that \Fig{deadlock} is a \emph{deadlock}; none of the four flows
can make any progress. Consequently, there is no way to get out of \Fig{deadlock} to any of the other
\emph{balance states} for the \cn{}. In other words, the steady-state throughput
for \Fig{deadlock} is $0$, and the only state reachable from \Fig{deadlock} is itself.


The implications of \Fig{tpt sensitivity} are concerning: 
for the same amount of escrowed funds in every channel in a \cn{}, the throughput can vary depending on the starting state. 
In other words, the throughput of the \cn{} in \Fig{tpt sensitivity} is
{\emph{sensitive}} to the balance state.
When every channel is perfectly balanced, the transaction throughput is highest.
If the network stayed in in such a state and only routed circulation flows, the throughput 
would remain high \cite{spider}.
However, transient DAG flows can alter the \cn{} state in ways that it cannot recover from, permanently harming its throughput.
Better routing or rate control cannot eliminate this problem. While a scheme like Spider~\cite{spider} may delay the onset of deadlocks by throttling flows experiencing congestion or more effective load-balancing across multiple paths, DAG demands like the purple flows in \Fig{tpt sensitivity} would eventually move the topology to states with reduced or even zero throughput.
We formalize the notions of throughput, balance states, and deadlocks in \Sec{sec:model} and \Sec{sec:worst tpt}.
\section{Model and Metrics}
\label{sec:model}
\label{sec:model pcn state}
We model the credit network as a graph $G(E,V)$. 
The vertices $V=\{v_1, v_2, \ldots, v_n\}$ of the graph denote the nodes, or users, of the credit
network while the edges $E$ denote pairwise channels. 
Each channel is an undirected edge and is denoted by the pair of nodes constituting the edge.  
Physically, these channels represent a credit relation between the endpoints. 
Each channel $\{u,v\}$ is associated with two {\emph{balances}} $b_{(u,v)}$ and $b_{(v,u)}$ that
denote the maximum number of tokens that $u$ is willing to credit $v$ ($b_{(u,v)}$) and vice versa ($b_{(v,u)}$) in the channel.
These can be thought of as escrowed funds that are allocated to each channel when it is established. 
Each channel $\{u,v\} \in E$ has a {\em capacity} $c_{\{u,v\}}$ that denotes the
total credit limit of the channel;\footnote{We use the notation $\{u ,v\}$ to denote undirected edges, and $(u,v)$ to denote directed edges for nodes $u, v\in V$.} 
that is, $ c_{\{u,v\}} = b_{(u,v)} + b_{(v,u)}$. We assume a static network, i.e., all channel capacities are fixed, and the channels remain open and funded with the same amount of tokens for the entire
duration under consideration. 

We capture the current {\em state} of the credit network using a balance vector $ \bb  \in \allr_{\ge 0}^{|E|}$, containing the balance information of all the $|E|$ channels.
For every channel $\{u, v\} \in E$, the vector $\bb$ contains only one entry corresponding to the balance at either $u$ or $v$'s end in the channel as follows.
Let $\overline{E} = ((v_{i_1}, v_{j_1}), (v_{i_2}, v_{j_2}),\ldots,(v_{i_{|E|}}, v_{j_{|E|}}))$ be an ordering of the channels in $E$, with the nodes within each channel also ordered such that $i_k < j_k$ for any channel $1\leq k \leq |E|$.
Similarly let $\underline{E} = ((v_{j_1}, v_{i_1}), (v_{j_2}, v_{i_2}),\ldots,(v_{j_{|E|}}, v_{i_{|E|}}))$ be an ordering containing edges in a direction reverse to that in $\overline{E}$. 
Now, the $k$-th entry of $\bb$ for $1\leq k \leq |E|$ contains the balance $b_{(v_{i_k}, v_{j_k})}$ of the $k$-th channel in $\overline{E}$.
This allows us to visualize the entire network state space $\allb$ as an $|E|$-dimensional polytope with each perpendicular
axis representing balance on a distinct channel. 
Similar to the balance vector, let the vector $\capvec \in \allr_{> 0}^{|E|}$ contain the capacities of all the channels, with the $k$-th entry of $\capvec$ containing the total capacity $c_{\{v_{i_k}, v_{j_k}\}}$ of the $k$-th channel in $\overline{E}$. 
These definitions allow us to represent the balances on the reverse edges or those channels in $\underline{E}$ using the vector $\capvec - \bb$.
     
\begin{definition}{(Boundary, Corner, and Interior Balance States.)}  
A channel $\{u,v\} \in E$ is \textit{imbalanced} when all of its escrowed funds are on one end of the channel, \ie either $b_{(u,v)} = 0$ or $b_{(v,u)} = 0$.
A balance state $\bb$ is located on the boundary of the $\allb$ polytope when one or more of the channels is imbalanced. A corner of the $\allb$ polytope
is a point on the boundary where all $|E|$ channels are imbalanced in one of the two
directions.
In contrast, if none of the balances on either end of any of the channels is 0, such a balance state $\bb$ is 
an {\em interior} balance state. The set of all interior balance states is denoted by $\allb_+$. Similarly, $\boundaries$ denotes the set of all boundary balance states,
while $\corners \subset \boundaries$ denotes the set of all corner states.
\end{definition}



Transactions attempted in the credit network follow a  demand pattern 
captured through a binary matrix $D$ where $d_{(i,j)} = 1$ if sender $i$ wants to transact
with receiver $j$. 
Such a sender-receiver pair can use one or more paths to transact. 
Let $\paths = (p_1, p_2, \ldots, p_l)$ denote an ordering of all paths (across all sender-receiver pairs with demand)  in the network through which tokens can be routed. 

We assume that transactions in the credit network occur 
synchronously over {\emph{rounds}} or epochs; every transaction starts and completes within the same round. 
We further assume that a transacting sender-receiver pair has unbounded demand; 
tokens can always flow from a sender to a receiver if feasible. We also assume that tokens are
infinitely divisible. 
Suppose $\bb$ is the balance state of the credit network at the beginning of a round. 
Let $\fb \in \allr_{\ge 0}^{|\paths|}$ denote a flow vector, with the $k$-th entry of $\fb$ being the number of tokens sent along path $p_k$ during the round.
For the flow vector $\fb$ to be feasible, we require the total number of tokens sent on any channel $\{u,v\}$ from $u$ towards $v$ to not exceed $b_{(u, v)}$, since only $b_{(u, v)}$ tokens are available for use along channel $\{u, v\}$ (in $u$ to $v$ direction) during the round. 
To define feasibility of a flow precisely, let routing matrix $\overline{R} \in \{0, 1\}^{|E| \times |\paths|}$ be such that $R_{((u, v), p)} = 1$ if the directed edge $(u,v)$ is part of path $p$ and $R_{((u, v), p)} = 0$ otherwise. 
The $k$-th column ($1 \leq k \leq |\paths|$) of $R$ corresponds to the $k$-th path $p_k$ while the $k$-th row ($1 \leq k \leq |E|$) corresponds to the $k$-th directed edge in $\overline{E}$. 
We similarly define routing matrix $\underline{R} \in \{0, 1\}^{|E| \times |\paths|}$ over the directed edges in $\underline{E}$. 
These routing matrices essentially capture which edges are involved in which paths. 
With this notation, for a flow $\fb$ to be feasible we must have 
\begin{align}
\overline{R}\fb \leq \bb \text{ and } \underline{R} \fb \leq \capvec - \bb. 
\label{eqn:feasible flow}
\end{align} 

When such a feasible flow is sent, at the end of the round a node $u$'s balance $b_{(u,v)}$ in channel $\{u,v\}$ is decremented by the total amount of tokens sent from $u$ towards $v$ and incremented by the tokens sent from $v$ towards $u$ during the round.
To compute this balance change, we define $\routemat \in \{1, -1, 0\}^{|E| \times |\paths|}$ as $\routemat := \overline{R} - \underline{R}$.
For an edge $(u, v) \in \overline{E}$ and path $p$, we have
\begin{equation}
    \routemat_{(u,v), p} =
        \begin{cases}
            1, & \text{if } (u,v) \in p \\
            -1,   & \text{if }(v, u) \in p\\
            0,              & \text{otherwise}.
        \end{cases}
\end{equation}
\label{def:routing matrix}
Letting $\bb'$ be the balance state of the network at the end of the round, we then have 
\begin{align}
\bb' = \bb - \routemat \fb. 
\end{align}
\Tab{symbols} summarizes the notation and provides an example for the \cn state described in \Fig{medtpt}. The flow $\fb$ sends $3$ tokens from node $1$ to $3$, and $2$ tokens from node $2$ to $3$, transferring all of the $5$ available tokens on the $(2, 3)$ channel. Sending such a feasible flow results in the new balance state $\bb'$ with no remaining tokens at $2$'s end of the $(2,3)$ channel.

This process of token transfer allows us to define a set of {\emph{reachable states}} from a starting state $\bb$.
We say a state $\bb'$ is reachable from $\bb$ in $k + 1$ rounds if there exists a sequence of flows $\fb^{(0)}, \fb^{(1)}, \ldots, \fb^{(k)}$ for $k \in \mathbb{N}$ such that
\begin{align}
\fb^{(i)} &\succeq 0 \quad \forall i \in [k], \label{eqn:nonnegflow} \\
\overline{R} \fb^{(i)} &\leq \bb^{(i)} \quad  \forall i \in [k], \\
\underline{R} \fb^{(i)} &\leq \capvec - \bb^{(i)} \quad  \forall i \in [k], \\
\bb^{(i+1)} &= \bb^{(i)} - \routemat \fb^{(i)} \quad \forall i \in [k], \label{eqn:statetransition}
\end{align}
with $\bb^{(0)} = \bb$ and $\bb^{(k+1)} = \bb'$. 
A sequence of flows $\fb^{(0)}, \fb^{(1)}, \ldots, \fb^{(k)}$, that satisfies equations~\eqref{eqn:nonnegflow}--\eqref{eqn:statetransition} for a given initial state $\bb^{(0)} = \bb$ is called a feasible flow sequence for initial state $\bb$. 
\begin{definition}{(Set of reachable states.)}
For a credit network with capacity~$\capvec$ and initial balance state $\bb$, the set of states $\rch_{\bb}$ reachable from $\bb$ is given by
    %
    \begin{equation}
        \rch_{\bb} = \left\{ \bb' | \bb' \text{ is reachable from } \bb
        \right\}. \label{eqn:states}
    \end{equation}
\end{definition}

We say a flow $\fb$ is {\em achievable} from state $\bb$ if there exists a feasible flow sequence $\fb^{(0)}, \fb^{(1)}, \ldots, \fb^{(k)}$ from $\bb$ such that $\fb = \sum_{i=0}^k \fb^{(i)}$.  
\begin{definition}{(Set of achievable flows.)}    
The set of flows $\feasflow_{\bb}$ achievable from state $\bb$ is given by
    \begin{equation}
        \feasflow_{\bb} = \left\{ 
        \fb 
        \middle| 
        \fb \text{ is achievable from } \bb
        \right\}. \label{eqn:feasible_flows}
    \end{equation}
    \label{def:reachable_states}

\end{definition}


\subsection{Metrics: Best- and Worst-Case Throughput}
\label{sec:metrics}
We measure the performance of a \cn by its throughput. 
However, the throughput of a \cn depends on many factors, including the demand imposed by users. 
Prior work~\cite{spider} has shown that a demand matrix can be separated into circulation and directed acyclic graph (DAG) components.\footnote{The demand matrix considered in prior work~\cite{spider} is real-valued with the entries denoting how much demand sender-receiver pairs have. In contrast, the demand matrix we consider is binary-valued (\S\ref{sec:model pcn state}) denoting only the presence or absence of demand between sender-receiver pairs. Nevertheless, the concept of decomposing a demand matrix into circulation and DAG components is general and applicable to our model also.} 
The circulation represents the portion of the demand that can be routed in a balanced manner and sustained in the long term. The circulation can be extracted
from the demand matrix by computing the largest union of all cyclic demand pairs 
($a$ has demand to $b$, $b$ has demand to $c$, who has demand back to $a$). 
The remaining demand pairs form the DAG component.
With a circulation demand, transactions can be routed such that every token sent on channel $\{u, v\}$ from $u$ to $v$ is compensated by another token from $v$ to $u$, thus maintaining channel balance. 
This allows a circulation demand to be sustained indefinitely.
In contrast, a DAG demand sends one-way traffic on some channels, eventually leaving them imbalanced and unable to sustain more transactions in the same direction. 
Thus, the DAG portion of the demand
cannot be sustained long-term without rebalancing. Rebalancing is a technique to arbitrarily modify the balance of a channel using a mechanism external to the credit network itself (e.g., ``on-chain'' transactions on a blockchain). Rebalancing usually comes at a substantial cost (high transaction fees, confirmation delay) and should be avoided to the extent possible.
\begin{wraptable}{l}{68mm}
\vspace{-3mm}
\setlength\tabcolsep{4pt}
    \centering
    \footnotesize
            \begin{subfigure}{0.21\columnwidth}
            \centering
            \includegraphics[width=\columnwidth]{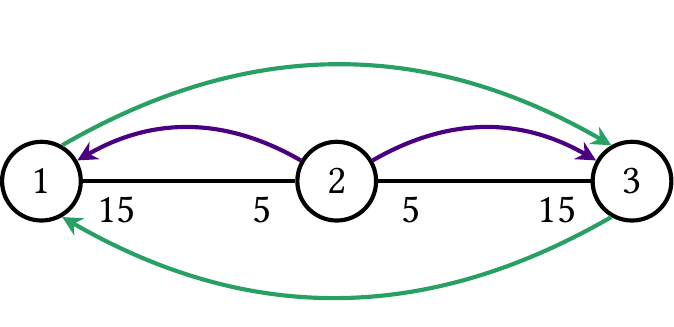}
        \end{subfigure}
    \begin{tabular}{l c c}
    \toprule
         \textbf{Symbol} & \textbf{Meaning} & \textbf{Example} \\
         \midrule
        $\bb$ & \makecell{Initial balance \\ state} & $\mvec{15}{5}$ \\
        $\capvec$ & Capacities & $\mvec{20}{20}$ \\
        $\mathcal{P}$ & Set of paths & \makecell{$(1 \rightarrow 3, 3 \rightarrow 1,$ \\ \ \  $2 \rightarrow 3, 2 \rightarrow 1)$}\\
        $\overline{R}$ & \makecell{Fwd. routing \\ matrix} & $\mvec{1&0&0 &0}{1 &0&1 &0}$ \\
        $\underline{R}$ & \makecell{Bwd. routing \\ matrix} & $\mvec{0&1 &0 &1}{0 &1 &0& 0}$ \\
        $\routemat$ & $\overline{R} - \underline{R}$ & $\mvec{1&-1 &0 &-1}{1 &-1 &1& 0}$ \\
        $\fb$ & Feasible flow & $\ \  \hveclong{3}{0}{2}{0}$ \\
        $\bb'$ & \makecell{Balance state \\ after sending $\fb$} & $\mvec{12}{0}$ \\
        $\psi(\bb)$ & One-step tpt. & 10 \\ 
        $\tpt(\bb)$ & Steady-state tpt. & 10 \\
        $\maxtpt$ & Max. tpt. & 20 \\
        $\mintpt$ & Min. tpt. & 0 \\
        \bottomrule
    \end{tabular}
    \caption{Summary of notations along with examples for above \cn.}
    \label{tab:symbols}
\end{wraptable}

Our goal is to study the throughput of a credit network in the absence of external rebalancing. Consequently our throughput metrics automatically attribute throughput only to the circulations (the long-term throughput of DAG demands is zero without rebalancing). However, we emphasize that our model is general and supports DAG demands.  In particular, transient DAG demands transition the \cn to different states, possibly changing its long-term (circulation) throughput. In fact, even circulation demands can appear ``DAG-like'' over short time scales. We study the impact of such state transitions on the long-term throughput of the \cn.

We define the \ostpt $\psi(\bb)$ of a state $\bb$ to be the maximum flow that is feasible to send within an epoch starting from $\bb$, while ensuring that the balances of all channels remain unchanged after the epoch. Since the balance state does not change, $\psi(\bb)$ can be achieved in every epoch  by repeating the same set of flows.  
\begin{definition}{(One-step throughput.)}
    The \ostpt at state $\bb$ is defined as
    \begin{equation}
        \psi(\bb) = \max \big\{ \one\T \fb \big| \fb \succeq \zero{},~ \overline{R}\fb \leq \bb,~\underline{R} \fb \leq \capvec - \bb,~ \routemat \fb = \zero{} \big\}.
        \label{eqn:psi}
    \end{equation}
    \label{def:psi}
\end{definition}
The one-step throughput is maximized when the channel capacity is equally divided between constituent nodes for all channels, \ie $ \psi\left(\capvec / 2\right) = \max_{\bb \in \allb} \psi(\bb)$. 

We are interested in the {\em maximum steady-state} throughput of the network starting from a state $\bb$. This depends not only on $\bb$ but also on the states reachable from $\bb$. In particular, a credit network starting from state $\bb$ can undergo a series of state transitions facilitated by flows present in the demand to reach a state $\bb'$ with higher $\psi(\bb')$. It can then achieve throughput $\psi(\bb')$ in every subsequent epoch. This motivates the following definition.


\begin{definition}{(Steady-state throughput.)}
    The steady-state throughput of a state $\bb$ is defined as
    \begin{equation}
        \phi(\bb) = \sup_{\ab \in \reachb} \psi(\ab). 
        \label{eqn:phi}
    \end{equation}
    \label{def:phi}
\end{definition}
In \Fig{medtpt}, the balance on the middle node cannot be increased unilaterally because the green flows that contribute to steady-state throughput are tied to each other.  As a result, the maximum throughput is constrained by the $5$ available tokens on each side of node $2$. This means that  $\psi(\bb) = \phi(\bb) = 10$ because no state with higher throughput is reachable from $\bb$ (\Tab{symbols})
\footnote{By Definition \ref{def:psi}, $\psi(\bb)$ is the optimal value of a linear programming problem with a compact feasible set. %
Because the feasible set is related to $\bb$ in a continuous way, the resulting $\psi(\bb)$ is always a continuous function.
In contrast, $\phi$ is typically discontinuous, because the sets of reachable states $\rch_{\bb}$ and $\rch_{\bb'}$ may differ drastically from each other, even if $\bb$ and $\bb'$ are arbitrarily close (particularly in the neighborhood
of the boundaries of $\allb$).
Since we haven't established the closedness of the set of reachable states $\reachb$, we use a supremum definition of steady-state throughput instead of a maximum.}.

Without loss of generality, we use \emph{throughput} of a state to refer to the steady-state throughput $\phi$ of a state 
in the rest of this paper.  
Note that the maximum throughput $\maxtpt$ across all states is also attained at $\capvec /2$: $\maxtpt \triangleq \max_{\bb \in \allb} \phi(\bb) = \psi\left(\capvec / 2 \right)$.

To account for the sensitivity of a \cn{}'s throughput to the balance state, we propose analyzing the lowest throughput it would yield across all possible balance states $\allb$. 
We call this lowest throughput value {\textbf{\wctpt}} and denote it by $\mintpt$. Unlike average-case analysis, considering the worst case lets us compare the resilience of different topologies without making assumptions about the distribution of balance states. Since channel balances are private information, little is known about the balance states observed in credit networks in practice. 


\begin{definition}{(Worst-case Throughput)}
    The \wctpt of a \cn{} with capacity $\capvec$ and a balance space $\allb$ are defined as
    \begin{equation}
        \mintpt = \min_{\ab \in \allb} \phi(\ab)
        \label{eqn:phimin}
    \end{equation}
    \label{def:worst_tpt}
\end{definition}
We denote
its counter-part $\maxtpt$ as the {\textbf{best-case throughput}} achieved at $\bb = \capvec/2$.
The \cn in \Fig{tpt sensitivity} achieves $\maxtpt=20$ at perfect balance (\Fig{perfect balance}) and $\mintpt=0$ at \Fig{deadlock} (\Tab{symbols}). 


\section{Throughput Sensitivity and Deadlocks}
\label{sec:worst tpt}

Best- and worst-case throughput depend on the transaction demand pattern. But since the transaction patterns 
are unknown a priori,
our goal is to design \cn{} topologies such that $\mintpt$ is close to $\maxtpt$ for most or all transaction patterns. 
It is important to ensure that such a design does not compromise on throughput. In other words,
it is not desirable to have $\mintpt = \maxtpt$ in exchange for a very low $\maxtpt$ relative to the escrowed
collateral.

Verifying conditions in which $\mintpt = \maxtpt$ is difficult, because although we know that $\maxtpt$ is attained at $\bb = \capvec/2$, 
it is unclear \emph{a priori} which states achieve $\mintpt$.
In this section, we show that if a topology does not admit  \emph{deadlocks},
then it is always the case that $\mintpt = \maxtpt$.
We call such a topology \emph{insensitive} to the starting balance.

A channel is said to be deadlocked at a particular state $\bb$ if no feasible flow can alter its balance in either direction.\footnote{Notice a subtle difference between our definition of deadlocks and that in \cite{malavolta2017concurrency,werman2018avoiding}. They define a deadlock as a set of flows that starts executing during a round, but cannot complete because (a) different flows utilize the same channels, and (b) a poorly-chosen processing ordering for the path edges for different flows blocks all flows from making progress. Our model does not capture such deadlocks 
because we assume flows complete instantaneously within a round.}
The collection of all such channels forms the deadlocked channel set $\dlk_{\bb}$ for a state $\bb$. A state $\bb$ represents a deadlock if 
$\dlk_{\bb} \ne \varnothing$. 
\begin{definition}{(Deadlocked Channel Set)}
    Let $\dlk_{\bb}$ denote the set of deadlocked channels under balance state $ \bb $. Formally,
    \[
    \dlk_{\bb} =
    \big\{ (u, v) \in \overline{E}~ \big|~ [\routemat \fb]_{(u, v)} = 0, ~ \forall \fb \in \feasflow_{\bb} \big\},
    \]
    where $[\routemat \fb]_{(u, v)}$ denotes the entry corresponding to channel $(u ,v)$ in the vector $\routemat \fb$. 
    \label{def:dlk_channels}
\end{definition} 
Since no feasible flow at $\bb$ can alter the balance values of the deadlocked channels $\dlk_{\bb}$, all such
channels have the same balance values across all reachable states from $\bb$. 
Formally, if $(u, v) \in \dlk_{\bb}$ and $\ab \in \rch_{\bb}$ is a state reachable from $\bb$, then we must have $a_{(u, v)} = b_{(u, v)}$, where $a_{(u, v)}$ and $b_{(u, v)}$ denote the balances of channel $(u, v)$ in vectors $\ab$ and $\bb$ respectively.
This effectively means that the deadlocked channels cannot sustain any flow and thus, make no contributions towards throughput.
In other words, $f_{p} = 0$ for all $\fb \in \feasflow_{\bb}$ if there exists an $(u, v) \in \dlk_{\bb}$ such that $(u,v) \in p$ or $(v, u) \in p$.
We formally state and prove these properties in \App{apdx:subsec:support_thm}.


It is straightforward to verify that for every interior
point $\bb \in \allb_+$, there exists $\epsilon>0$ such that $\epsilon$ tokens can be transmitted over \emph{some} flow in the network.
Hence, for any state $\bb \in \allb_+$, $\dlk_{\bb}=\varnothing$; in other words, deadlocks can \emph{only} happen at the boundary points of $\allb$.
At every boundary point, a nonempty subset of channels is imbalanced in one of the two directions, so flows using those channels in the imbalanced direction(s) are blocked. 

However, merely having imbalanced channels at a boundary point of polytope $\allb$ does not imply the existence of deadlocks at that point.
Consider the two corner balance states $\bb_3$ and $\bb_4$ shown in  \Fig{deadlock} and \Fig{good corner} respectively. $\bb_{3}$ is a deadlock in that none of the four flows can make any progress,  no other state in $\allb$ is reachable from $\bb_{3}$, and $\phi(\bb_3) = 0$. In contrast,  $\bb_{4}$ is a good boundary point that can achieve the best-case throughput $\maxtpt$. For example, 
node $3$ can send $20$ tokens to node $1$ in one round, node $1$ can then send the $20$ tokens back to node 3 in the next round, and so on. The \cn{} returns to $\bb_4$ every two rounds, achieving a throughput of 20 tokens per round. Further, at $\bb_4$, sending $10$ tokens from node $3$ to $1$ would return the \cn{} to the perfectly balanced state $\bb_{1}$; thus $\phi(\bb_{4}) = \psi(\bb_{1}) = 20$.  Effectively, the \cn{} can escape boundary $\bb_{4}$ to reach any  state in $\allb$.



It turns out that to assess a topology's sensitivity to the starting balance state, it is enough to check if its boundary points can be deadlocked. 
Intuitively, if there are no deadlocks then every boundary point, and in particular every corner point of the polytope, can be made balanced by moving into an interior point. 
Since the set of directions along which the corner points can transition to interior points spans the entire state space, this implies one can move in every direction in the $\allb$ polytope.
As a result, 
the \cn must be able to transition from any balance state to any other balance state
without being stuck. 
Such a \cn can always reach its throughput-maximizing state $\capvec /2$ from any other state.
In other words, the throughput of such a topology should be {\emph{insensitive}} to the starting balance state. 
In contrast, if a \cn has deadlocks, the {\em largest} deadlock (i.e., the deadlock involving the most channels) should lead to the maximum unused collateral and such states
would have the lowest throughput.
Therefore, the maximum throughput that can be extracted using only the remaining channels in the largest deadlock state constitutes
the \wctpt of a \cn. The rest of this section formally describes these two results.
\label{sec:dlk_min_tpt}


\begin{thm}{(Insensitive Throughput)}
    If a \cn{} with balance state space $\allb$ is deadlock-free, i.e., for any boundary state $ \cb \in \boundaries$, there exists an interior point $\pb \in \interior$ such that $\pb \in \rch_{\cb}$, then for any state $ \bb $, $\phi(\bb) = \psi(\capvec / 2)$.
    \label{thm:dlk_free}
\end{thm}

\paragraph{Proof Sketch (Full proof in \App{apdx:deadlock free proof}).} We introduce the notion of \textbf{feasible directions} from a balance state.
A unit vector $\db$ is a feasible direction from $\bb$ if there exists $\epsilon > 0$ such that $\bb + \epsilon \db \in \rch_{\bb}$.
%
We list some useful properties about feasible directions below, and provide proofs in \App{apdx:subsec:support_thm}.

\begin{enumerate}

    \item The union of feasible directions of any state is convex. (Prop.~\ref{prop:direction_conv})
    
    \label{propitem:dir_cvx}

    \item Any interior state along a feasible direction of $\bb$ is reachable. (Theorem~\ref{thm:reachable_along_dir})

    \label{propitem:dir_to_state}
    
    
    
    \item Feasible directions of a corner state are also feasible for an \textit{arbitrary} interior state. (Lemma~\ref{lemma:direction_subset}) 
    
    \label{propitem:dir_subset2}
    
\end{enumerate}

Let $\ab \in \interior$ be an arbitrary interior state.
By assumption, every corner state has a corresponding reachable interior state. The transition from
a given corner state to its reachable interior state produces a set of feasible directions which is located in 
an open orthant of the $\allr^{|E|}$ space.
Further, the feasible direction corresponding to a particular corner occupies a distinct
orthant that is not shared by the feasible directions from any other corner state.

By property \ref{propitem:dir_subset2}, the feasible directions of every corner are included in the 
the feasible directions of $\ab$. 
%
%
By property \ref{propitem:dir_cvx}, the set of $\ab$'s feasible directions contains the convex hull of all of 
these $2^{|E|}$ sets of directions, which maps exactly to the full space $\allr^{|E|}$. 
Furthermore, by property \ref{propitem:dir_to_state}, any other interior state $\ab'$ is reachable from $\ab$, including the center state $\capvec / 2$. For simplicity, we skip the extension to boundary states and refer it to \App{apdx:deadlock free proof}. 
Since $\capvec / 2$ globally maximizes the one-step throughput $\psi$, we have for any $\bb \in \allb$, $\phi(\bb) = \sup_{\bb' \in \rch_{\bb}} \psi(\bb') = \psi(\capvec / 2 )$. \qeda \\

While a \cn{} might not be entirely deadlock-free, 
Theorem \ref{thm:dlk_free} can also  be applied to any sub-network of the original \cn{}
that does not have deadlocks.
%
A sub-network is embedded within a subgraph of the original topology and 
only permits paths whose edges are entirely within the subgraph. 
If this sub-network is deadlock-free, then it enjoys throughput insensitivity within the balance states of the sub-network.
In addition, its throughput lower bounds the maximum throughput achievable from 
an arbitrary initial state in the original \cn{}. 

At the same time, 
every \cn{} with deadlocks has a set $\maxdlk$ of balance states that maximizes the number of deadlocked channels.
For any $\bb \in \maxdlk$, the set of channels deadlocked $\dlk_{\bb}$ equals a fixed set $L$ that contains all other sets of deadlocked channels across every balance state $\bb \in \allb$. 
The channels outside $L$ 
form a deadlock-free sub-network. 
The  throughput of the original \cn is at least the throughput achievable on this sub-network, regardless of the initial balance
state of the original network.
 We state this formally in \Thm{worst_corner} and prove it in \App{apdx:subsec:worst_corner}.

\begin{thm}
    A state $\bb$ has the worst throughput $\phi(\bb) = \Phi_{\min}$, if the number of deadlocked channels in $\bb$ is the largest across all states, {\em i.e.,} $|\dlk_{\bb}| = \max_{\bb' \in \allb}|\dlk_{\bb'}|$. 
    Furthermore, the throughput $\Phi_{\min}$ of a worst throughput state $\bb$ can be computed by considering a state $\bb'$ where 
    \[
        b_{(u, v)}' = \left\{ \begin{array}{lr}
            b_{(u, v)}, & \text{ if } (u, v) \in \dlk_{\bb} \\
            C_{(u, v)}/2, & \text{ if } (u, v) \notin \dlk_{\bb}
        \end{array}\right..
    \]
    Then, $\Phi_{\min} = \psi(\bb')$. 
    \label{thm:worst_corner}
\end{thm}


\section{Designing Deadlock-free topologies}
\label{sec:topology guide}

\Sec{sec:worst tpt} suggests that to fully utilize the collateral in a \cn,
the \cn{} needs to be deadlock-free.
In this section, we first show that determining 
whether an arbitrary topology is deadlock-free is NP-hard (\Sec{sec:topo np hard}). 
Next, we propose and analyze
a ``peeling algorithm''  
that bounds the number of deadlock-free edges in a \cn (\Sec{sec:topo peeling}). The peeling algorithm provides a computationally-efficient way to estimate $\mintpt$, and reveals a  surprising connection between designing deadlock-free topologies and LT Codes~\cite{lubylt, decripplelt}, a well-known class of erasure codes. 


\subsection{Deadlock Detection Problem}
\label{sec:topo np hard}
Detecting a deadlock on a \cn{} is equivalent to finding a balance configuration $\bb$ such that one or more channels 
is imbalanced and one or more of the imbalanced channels is deadlocked. 
A full deadlock involves all edges of the \cn  while a partial deadlock can involve any non-empty \emph{proper} subset
of \cn{} edges.
\begin{definition}{(Full and Partial Deadlock)}
    A balance state $\bb$ is a full deadlock on the \cn{} $G(E, V)$ if
    $
    |\dlk_{\bb}| = |E|,
    $
%
    and a partial deadlock if 
    $
    0 < |\dlk_{\bb}| < |E|.
    $
    \label{def:partial_dlk}
\end{definition}

\begin{thm}{(Hardness Result)}
    Given a \cn{} $G(E,V)$ with channel capacity $c_{\{u,v\}}>0$ for each $\{u,v\}\in E$, demand matrix $D$ and set of paths $\paths$ for satisfying the demand, finding a balance state $\bb$ such that $\dlk_{\bb} \ne \varnothing$ is NP-hard. 
    \label{thm:dlk_np_hard}
\end{thm}

\paragraph{Proof Sketch (Full proof in \App{apdx:proof np hard})} When a channel $\{u, v\} \in E $ is imbalanced with
$b_{(u, v)} = 0$, the channel cannot support flow in the $(u, v)$ direction unless the balance state is altered.
As a result, all flows using $\{u, v\}$ from $u$ to $v$ are blocked. If all flows using a channel in both 
directions are blocked, then the channel is deadlocked.
We use this intuition to present a reduction from the Boolean Satisfiability Problem (SAT)
to the full deadlock-detection problem for credit networks. 
We consider a boolean expression in Conjunctive Normal Form (CNF) and construct a credit network such that, each variable is mapped to a unique channel and each clause is mapped to a unique path in the constructed 
network.
The truth value of a literal is associated with whether or not the balance at a particular end of a channel is zero, while the truth value of a clause relates to whether or not a particular path is blocked by some channel.  
The boolean expression is therefore
satisfiable if and only if there exists a full deadlock in the \cn{}, proving the $NP$-hardness claim. 

Note that a straightforward mapping of variables to channels, and clauses to paths results in a disconnected network with invalid paths.
To make our constructing valid, we introduce a polynomial number of auxiliary variables (and accompanying clauses), in a way that preserves the connection between the satisfiability of the initial CNF and the existence of deadlocks in the constructed network. We defer more details about our reduction technique to \App{apdx:proof np hard}.  
Our reduction can be easily extended 
to establish
that detecting a partial deadlock is also $NP$-hard. 

\subsection{Peeling Algorithm}
\label{sec:topo peeling}
Although it is NP-hard to check if a topology is deadlock-free, it is possible to identify subsets of edges that are provably deadlock-free in polynomial time.
We design such an algorithm inspired by the peeling algorithm used to decode LT codes \cite{lubylt}.
The key insight of our peeling algorithm is the following: if an edge has a dedicated flow in either direction, tokens can move freely in the direction
of the flow, as long as there are tokens available at the origin end. We call such flows  ``length 1'' flows, since  they traverse a single edge. 
If an edge has \emph{two} such flows in both directions, it can never be deadlocked. 
The peeling algorithms progressively finds edges with  length 1 flows (which cannot be deadlocked), 
and eliminates them as potential bottlenecks for all  flows that traverse them; this can be viewed as ``peeling''  these edges from these flows. 
Once it is established that none of the edges in a flow can be deadlocked, 
the flow can be removed from consideration.
This process
continues until all edges in
the topology can move in either direction, or none of remaining flows traverses edges that can be peeled, causing the process to terminate unsuccessfully.
We describe the terminology borrowed from LT Codes in \Sec{sec:lt process} and define the deadlock
peeling algorithm in \Sec{sec:dlk peeling alg}.

\subsubsection{LT Process}
\label{sec:lt process}
LT codes 
are a \emph{rateless} code, designed
for channel coding under changing or unknown channel conditions \cite{lubylt}. 
LT codes map a sequence of input bits $\boldsymbol x = (x_1,\ldots, x_n)$ to a sequence of encoded bits $\boldsymbol y = (y_1,\ldots, y_m)$.
Each encoded bit $y_j$ is the XOR of a subset of input bits: $y_j = \oplus_{k\in S_j} x_k$, where set $S_j \subseteq [n]$ indexes a subset of input symbols,  chosen at random. 
The LT encoding procedure can be represented as a bipartite graph, where each input symbol $x_i$, $i\in[n]$ corresponds to a node on the right, and each encoded symbol $y_j$, $j\in [m]$ to a node on the left. 
An edge exists between input node $x_i$ and encoded node $y_j$ iff $x_i\in S_j$.
The degree of an encoded node $y_j$ denotes the number of XORed input symbols, $|S_j|$.

 \begin{figure}
    \centering
        \begin{minipage}{.45\textwidth}
            \centering
            \includegraphics[width=\textwidth]{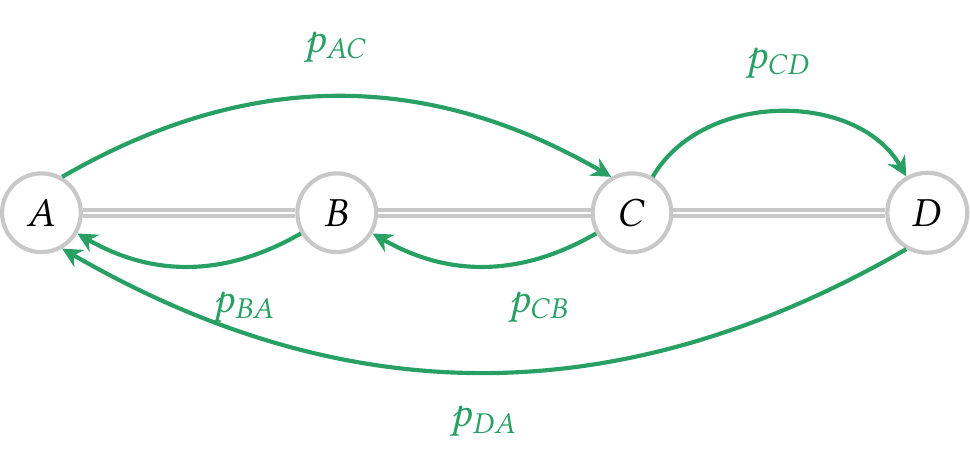}
            \includegraphics[width=\textwidth]{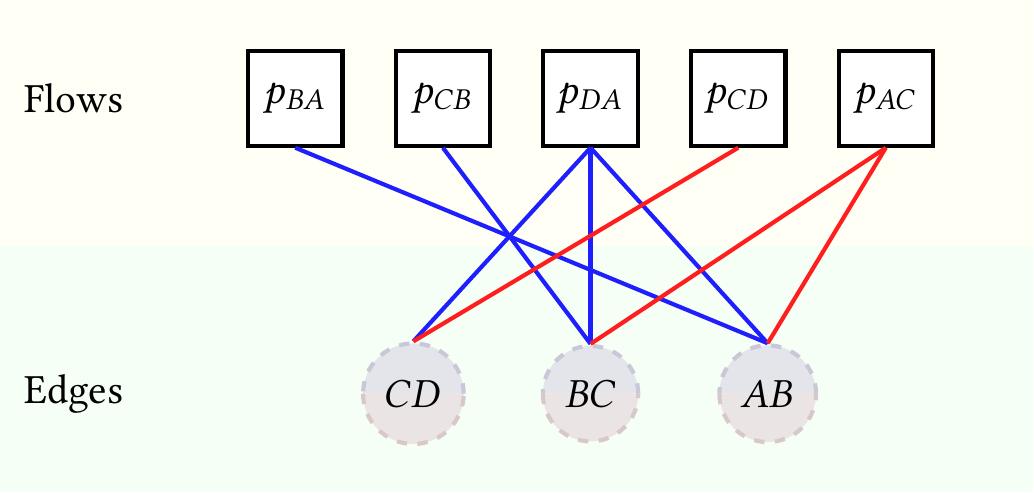}
            \caption{Bipartite graph that is peelable and identifies the lack of deadlock in the topology}
            \label{fig:peelable topo}
        \end{minipage}
        \hfill
        \begin{minipage}{.45\textwidth}
            \centering
            \includegraphics[width=0.8\textwidth,height=3cm]{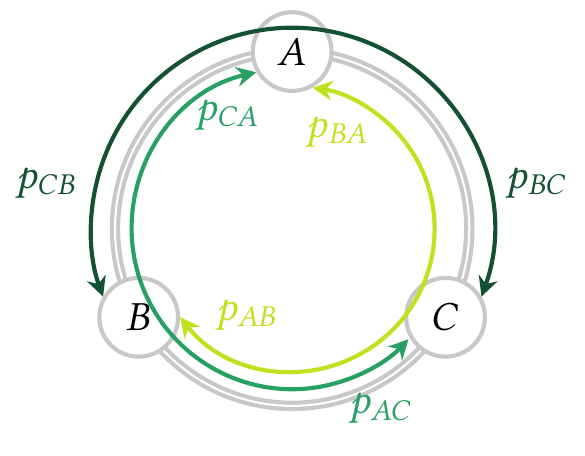}
            \includegraphics[width=\textwidth]{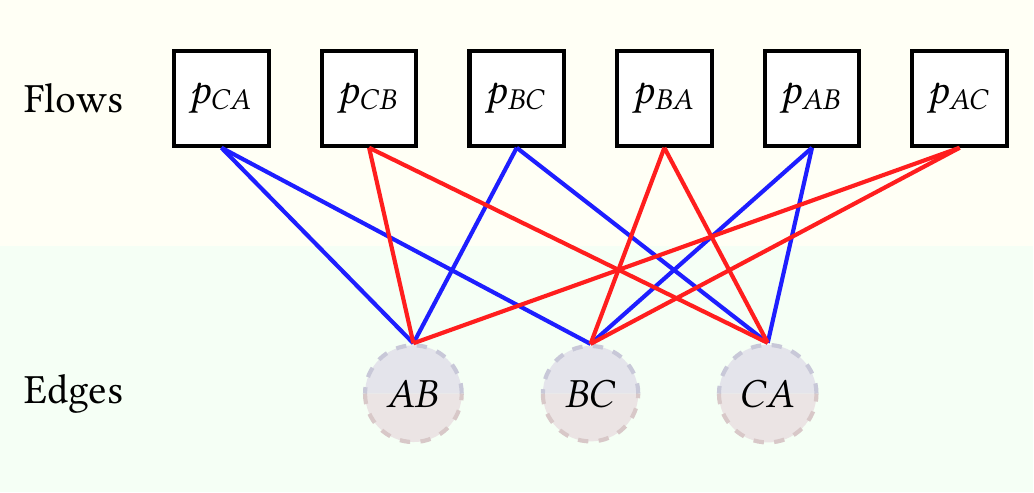}
            \caption{Bipartite graph that cannot be peeled even though there is no deadlock in the topology}
            \label{fig:unpeelable topo}
        \end{minipage}%
        \vspace{-5mm}
    \end{figure}

Decoding the encoded symbols proceeds iteratively.
A degree 1 encoded symbol can be immediately decoded to identify the associated input symbol since
the encoded symbol's value matches the input symbol. 
Such a degree 1 encoded symbol is \emph{released} to \emph{cover} or decode its associated input symbol.
The input symbol is then added (if not already present)
to the \emph{ripple} $\mathcal R$, a set of covered input symbols that are yet to be \emph{processed}.
At every time step, an (unprocessed) input symbol $x_i$ is chosen at random from the ripple $\mathcal R$
and processed: its value $x_i$ is XORed with every encoded symbol $y_j$ that it is a neighbor of (i.e., for which $i \in S_j$). This
reduces the degree of every such encoded symbol, subsequently releasing more encoded symbols that become degree 1. However, the newly covered neighbors of degree 1 flows increase the size of the ripple
only if they were previously \emph{uncovered}. This routine (\emph{LT process}) of releasing encoded symbols, covering
input symbols and adding them to the ripple, and processing input symbols from the ripple,
continues until either the ripple runs empty or
all input symbols are successfully decoded. 

Since encoded symbols incur communication overhead, it is desirable to minimize their quantity. 
A key factor affecting the number of encoded symbols 
is the choice of the degree distribution for the encoded symbols. However, if minimizing the number of encoded symbols leads to a ripple that vanishes before the algorithm terminates, the decoding process fails altogether. Prior work~\cite{lubylt, decripplelt} 
has focused extensively on the design 
of degree distributions that achieve a ripple size that is neither redundant nor
so small that the ripple disappears before the LT process completes. 
For example, under a \emph{robust soliton distribution} \cite{lubylt}, LT codes recover $n$ input symbols with probability at least $1-\delta$ from $n + O(\sqrt{n}\log (n/\delta))$ encoded symbols.

\subsubsection{Deadlock Peeling}
\label{sec:dlk peeling alg}
The idea of peeling in LT codes can be extended to detecting deadlocks with a few modifications. 
Here, channels correspond to input symbols.
A channel can be used (and covered) in one of two directions.
We will consider the two
directions \emph{blue} and \emph{red} with the \emph{red} direction corresponding to usage of the edges in the same
direction as $\overline{E}$ and blue denoting the directions specified by $\underline{E}$.
A flow (an encoded symbol) uses one or more channels (input symbols), each in one of the two directions; the
degree of the flow is the path length, or the number of channels used by the flow.
The \dpp can be represented as a bipartite graph, where each flow $p_i$, $i\in[|\mathcal{P}|]$ corresponds to a node in the top partition, and each channel $e_j\{u, v\}$, $j\in [|E|]$ to a node in the bottom partition. 
An edge exists between channel $e_j\{u, v\}$ and flow $p_i$ iff either $(u, v)$ or $(v, u)$ is part of
$p_i$; the edge is red if $(u, v) \in p_i$ and blue if $(v, u) \in p_i$.
Figures~\ref{fig:peelable topo} and ~\ref{fig:unpeelable topo} show the  the bipartite graph construction for their associated \cn{} topologies and paths.

\begin{figure}[t]
\centering 
\begin{subfigure}{0.3\textwidth}
  \includegraphics[width=\linewidth]{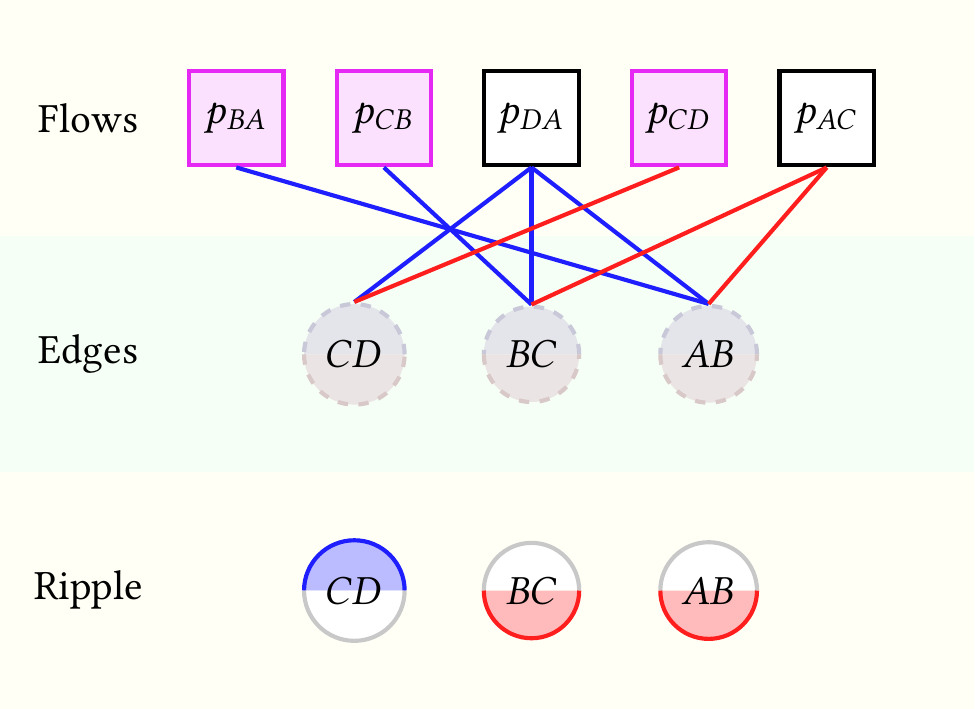}
  \caption{\small At the start, releasing the three degree 1 flows 
    adds three symbols to the ripple.} 
  \label{fig:ripple1}
\end{subfigure}\hfil 
\begin{subfigure}{0.3\textwidth}
  \includegraphics[width=\linewidth]{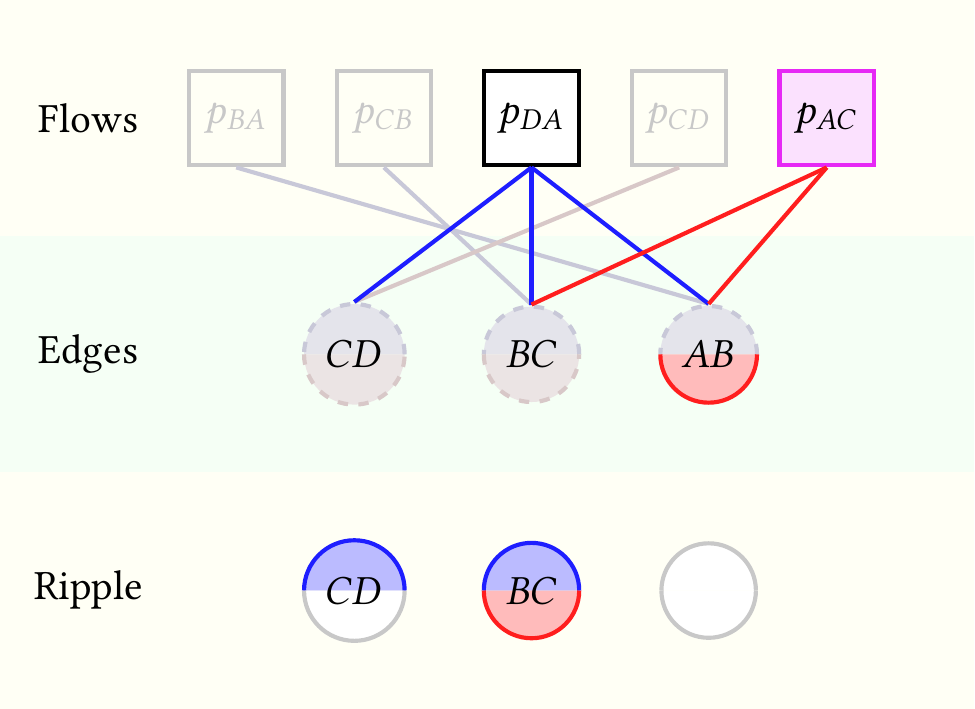}
  \caption{\small After processing red $AB$, flow $p_{AC}$ becomes degree 1 to cover blue $BC$. }
  \label{fig:ripple2}
\end{subfigure}\hfil 
\begin{subfigure}{0.3\textwidth}
  \includegraphics[width=\linewidth]{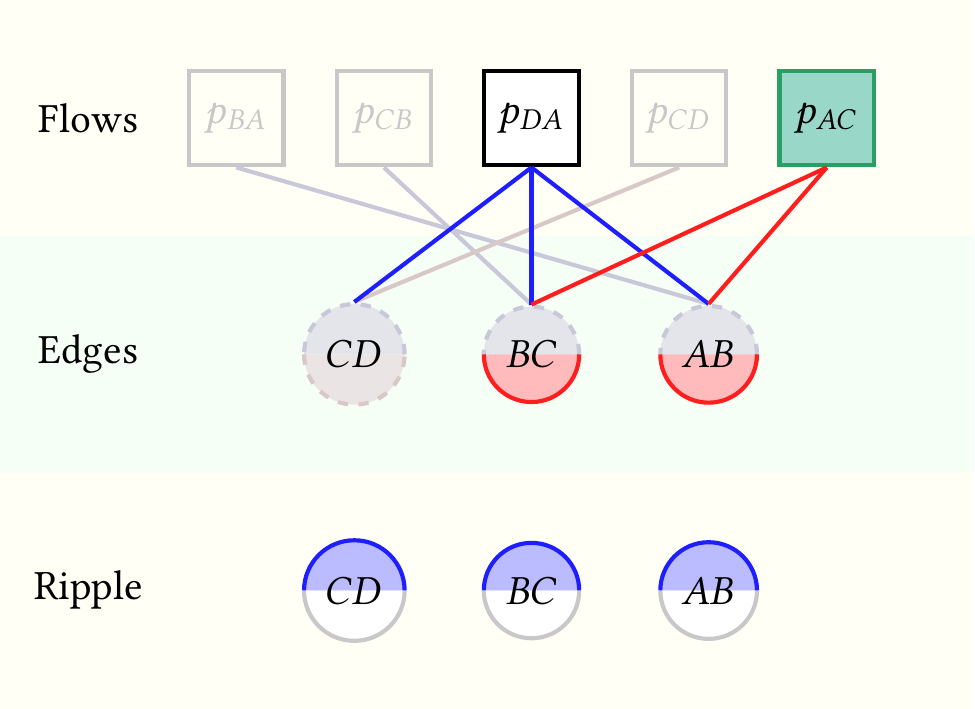}
  \caption{\small Processing red $BC$ makes $p_{AC}$ degree 0, adding blue $AB$ (and blue $BC$) to the ripple.}
  \label{fig:ripple3}
\end{subfigure}

\medskip
\begin{subfigure}{0.3\textwidth}
  \includegraphics[width=\linewidth]{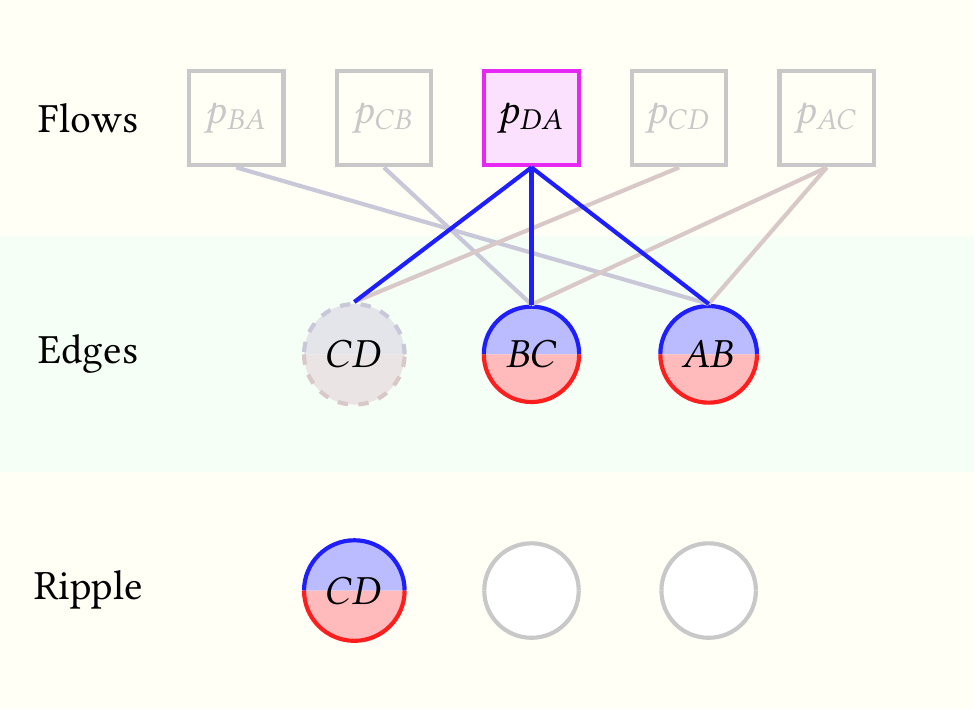}
  \caption{\small Once blue $AB$ and $BC$ are processed, $p_{DA}$ becomes degree 1 and covers red $CD$.}
  \label{fig:ripple4}
\end{subfigure}\hfil 
\begin{subfigure}{0.3\textwidth}
  \includegraphics[width=\linewidth]{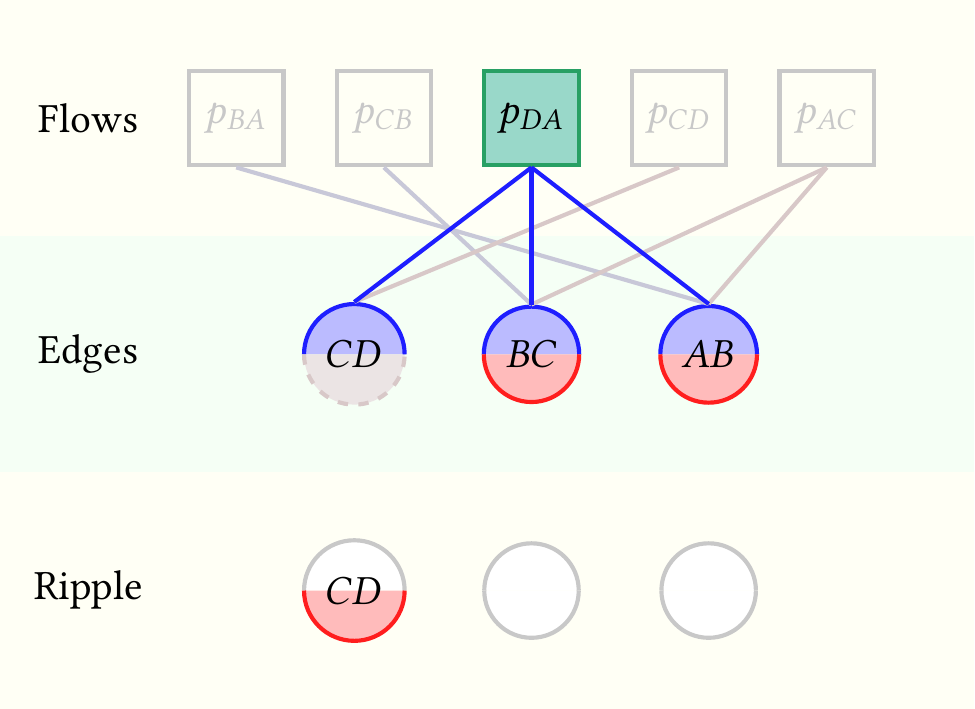}
    \caption{\small Processing blue $CD$ makes $p_{DA}$ degree 0, but its release adds no new symbols to the ripple.}
  \label{fig:ripple5}
\end{subfigure}\hfil 
\begin{subfigure}{0.3\textwidth}
  \includegraphics[width=\linewidth]{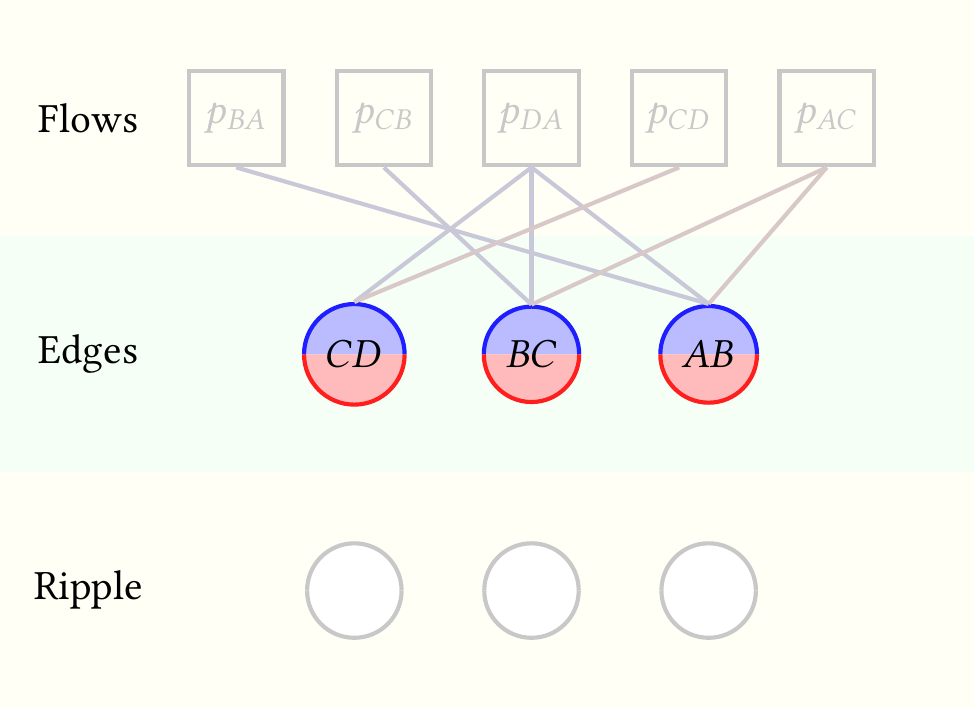}
  \caption{\small Processing red $CD$ ends the peeling algorithm with all channels deadlock-free.}
  \label{fig:ripple6}
\end{subfigure}
\vspace{-3mm}
\caption{\small Evolution of the bipartite graph as channels are processed in the peeling algorithm.}
\vspace{-5mm}
\label{fig:deadlock peeling}
\end{figure}

\Fig{deadlock peeling} shows the \dpp for the example \cn{} in \Fig{peelable topo}.
A flow of degree 1 can be released to cover and add to the ripple 
its last channel in the direction opposite to the direction in use. In other words,
a flow $p_1$ of degree 1 using the edge $e\{u, v\} \ u < v$ in the blue direction $(v, u)$,
can be released to cover $e$ in the red direction $(u, v)$.
Newly covered channels are added to the ripple $\mathcal{R}$.
Covering $e$ in the $(u, v)$ direction signifies that the channel $e$ can 
always support flows in the $(u,v)$ direction because flow $p_1$ would restore tokens back to the $u$ end regardless of $e$'s current balance
state.
\Fig{ripple1} shows the impact of releasing the initial degree $1$ flows.
At every subsequent time step, 
a randomly chosen channel (and its associated color) is processed from the ripple.
Processing a channel is similar to the LT process; the degree of every flow that uses the channel in the processed
color is reduced by 1. 
This may lead to new flows being released to cover more channels (\Fig{ripple2}). However, only the previously uncovered channels
increase the ripple size (e.g. blue $AB$ in \Fig {ripple3}). This peeling process continues as long as the ripple is non-empty and at least one remaining flow is of degree 1.
As is the case with the LT process, to establish that the entire topology is deadlock free,
the ripple should not vanish before the entire peeling process is complete. 
Consequently, prior analyses~\cite{lubylt, decripplelt} that design good degree distributions become relevant for the \dpp also.

\begin{wrapfigure}{r}{0.5\textwidth}
\vspace{-4mm}
    \centering
    \includegraphics[width=0.35\textwidth]{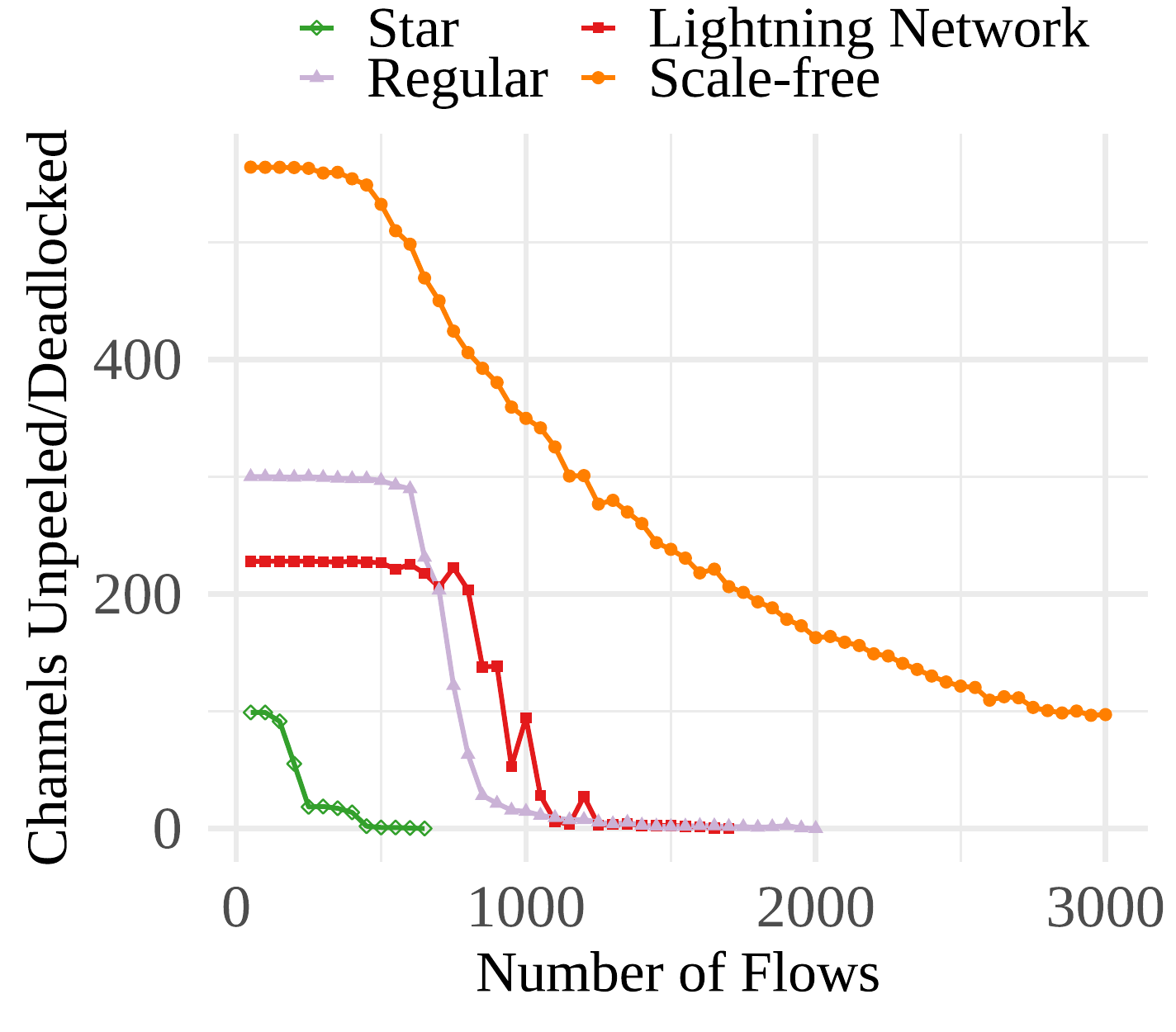}
    \caption{\small Comparison between channels unpeeled at the end of the peeling
    algorithm (line) and channels deadlocked (points on the same lines) based on an ILP.
Across all four topologies, the points fall on the line suggesting that the peeling algorithm's unpeeled channels matches the
number of deadlocked channels exactly.}
    \label{fig:peeling alg ilp comparison}
    \vspace{-4mm}
\end{wrapfigure}

The deadlock peeling process differs  from  the LT process in one key aspect. Unlike the LT process, flows of degree $1$ cannot be removed immediately from the bipartite graph.
A flow is removed only after it reaches degree $0$ when
every channel that it uses is covered in the color opposite to the direction of use.
The intuition behind this is 
that a flow of degree $0$ is unconstrained 
and can move tokens on all the channels it uses (restoring balance
if the opposite direction was imbalanced).
Note the important distinction here that a flow of degree $1$ covers (or frees) only the last remaining channel
in the opposite direction while a flow of degree $0$ covers \emph{all} of its channels
in the opposite direction (Figures~\ref{fig:ripple3},~\ref{fig:ripple4},~\ref{fig:ripple5}).
\Alg{peeling} in \App{apdx:peeling psuedocode} summarizes the \dpp.

The \dpp cannot always detect all deadlock-free edges: the number of edges it successfully peels
is a {\em lower bound} on the actual number of deadlock-free edges in a topology. This is because it identifies patterns of bidirectional dedicated flows on the channels in a topology. Such flows, while guaranteed to render the associated channels deadlock-free, are not required for deadlock-freeness.
For example, the topology in \Fig{unpeelable topo}  is  deadlock-free but it has no length 1 flows. Hence, the peeling algorithm terminates on this topology without peeling any channels.

\NewPara{Empirical evaluation of the \dpp}. While we know that the \dpp is a lower-bound
on the true number of deadlock-free edges in a topology (and consequently its associated $\mintpt$), it is useful to know how good of a lower-bound it is. Since the deadlock detection problem is NP-hard, we formulate an Integer Linear Program (ILP)  to identify the largest deadlock on a given topology. We describe the ILP in \App{apdx:dlk ilp} and compare it to the output of the \dpp. 
Since the ILP is slow, we evaluate it on small samples of four different topologies (described in \Sec{sec:eval-setup}) consisting of only $100$ nodes.

In \Fig{peeling alg ilp comparison}, the number of channels unpeeled (line) matches 
the maximum deadlock reported by the ILP (points on the same lines) in all cases, suggesting that the \dpp provides not just a lower-bound, but rather an 
accurate approximation.
Hence, in our evaluation~(\Sec{sec:eval-metrics}),
we use the \dpp to peel as many channels as it can and compute the \wctpt 
as the \ostpt of the sub-network consisting of all the peeled edges. 

\section{Evaluation}
\label{sec:eval}

We empirically evaluate a number of standard topologies for their
throughput and deadlock characteristics to understand their performance.
We describe our evaluation setup in \Sec{sec:eval-setup},
define metrics in \Sec{sec:eval-metrics}, and compare existing topologies in \Sec{sec:eval-overall}.
We use the peeling algorithm to understand the
behavior of different topologies and propose a preliminary synthesis approach in 
\Sec{sec:eval-synthesis}.

\subsection{Setup}
\label{sec:eval-setup}
\NewPara{Baseline Topologies}. We compare the performance of \cns based on 
random graphs and a graph based on the \lntopo{}
for their throughput and deadlock behaviors. We use five random graph types: a Watts Strogatz \emph{\sw} graph
~\cite{smallworld} with 10\% rewiring probability, a Barab\'{a}si-Albert \emph{\sftopo} graph~\cite{scalefree}, a \emph{\pltopo} graph whose degree distribution follows a power-law, 
an \emph{\er} graph~\cite{er}, and a \emph{\rr} graph. 

We sample 5 different random graphs with 500 nodes and about 2000 edges for each graph type. 
To simulate a topology with properties similar to the \lntopo{}, a PCN currently in use, we retrieved a snapshot of the topology on Oct. 5, 2020 using a c-lightning~\cite{c-lightning} node 
running on the Bitcoin Mainnet. The original full topology has over 5000 nodes and 29000 edges. 
Similar to prior work \cite{spider}, we snowball sample~\cite{snowball-sampling} the
full topology to generate a PCN with 452 nodes and 2051 edges.
We also evaluate stars with 500 nodes as an example of a topology with good 
throughput and deadlock properties at the cost of being highly centralized. 
The throughput of all \cns is normalized to a constant collateral $C$ distributed equally
amongst all their channels. 


\NewPara{Demand Matrix and Path Choice}.
We generate demand matrices by sampling a fixed number of unique source-destination pairs 
from the set of nodes in the graph.
The results in this section use demand pairs sampled uniformly at random, but we include results with 
demand pairs sampled with a skew towards ``heavy-hitting'' nodes in \App{apdx:nonuniform random results}.
Every non-zero entry in the demand matrix adds one sender-receiver pair and allows the shortest path between them to 
move tokens in the \cn across rounds \footnote{Our throughput metrics only depend on the total number of permissible paths. Adding more edge-disjoint paths for a given demand pair and sampling more demand pairs both increase the total paths; we use the latter approach in our evaluations.}. We refer to such a path as a flow in this section since it maps to a single flow node in the bipartite graph associated with the peeling algorithm.
Unlike prior work~\cite{spider}, we do not explicitly control for a circulation demand since we are interested in the effect of DAG demands on circulation throughput.
As the demand matrix becomes more dense, the amount of circulation demand naturally grows.
Unless otherwise mentioned, we sample 4 different demand matrices per random instance of the random graphs, to generate 20 unique points
over which  we average the throughput and deadlock behavior. Since there is only one instance of the star, \lntopo{} and our synthesized
topology, we use 20 different demand matrices instead.
We only present results
for demand \denseness ranges that shows variation between the topologies. If the demand matrix is too sparse, there is not enough
demand for any topology to perform well; if it is too dense, just routing
one-hop demands (between end-points of every edge in the network) uses all the available collateral in the \cn ($\maxtpt = \mintpt = 1$).

\subsection{Metrics}
\label{sec:eval-metrics}
\NewPara{Maximum Throughput}. 
We first compute the maximum \emph{per-epoch} throughput $\maxtpt$ that a topology achieves when none of its channels are imbalanced or constrained. Recall that the throughput  of a \cn is maximized at its perfect balance state $\capvec / 2$.
Further, $\maxtpt = \psi\left(\capvec/2\right)$. We use an LP 
solver to compute $\psi$ based on the constrained optimization problem in \Eqn{psi}.

\NewPara{Worst-case Throughput}.
The \wctpt of a \cn $\mintpt$ is the minimum steady-state throughput achievable in an epoch from any state in its balance
polytope \footnote{Our definitions of $\maxtpt$ and $\mintpt$ assume the ability to send the maximum feasible amount between a sender-receiver pair using an ideal routing algorithm. This allows us to reason about throughput without concerning ourselves with the precise dynamics of any routing algorithm such as transaction sizes or splitting.}. We know from \Thm{worst_corner}
that $\mintpt$ is achieved at the state with the largest deadlock .
Though detecting deadlocked states on an arbitrary topology is NP-hard, we
use our approximate \dpp (\Sec{sec:topo peeling}) to identify the deadlock-free channels and compute 
the \wctpt $\mintpt $ as  $\psi(\capvec'/2)$ where $\capvec'_i=\capvec_i$ if $i$ is reported as a deadlock-free channel, and $\capvec'_i=0$ otherwise. 

\NewPara{Fraction of Channels Unpeeled}. In addition to the above throughput metrics, we also 
report the fraction of the channels in the topology that the \dpp fails to peel. This acts as an upper-bound
for the true number of deadlocked channels in a given topology.


\subsection{Performance of Random Topologies}
\label{sec:eval-overall}
\begin{figure*}[]
\vspace{-5mm}
    \centering
    \includegraphics[width=\textwidth]{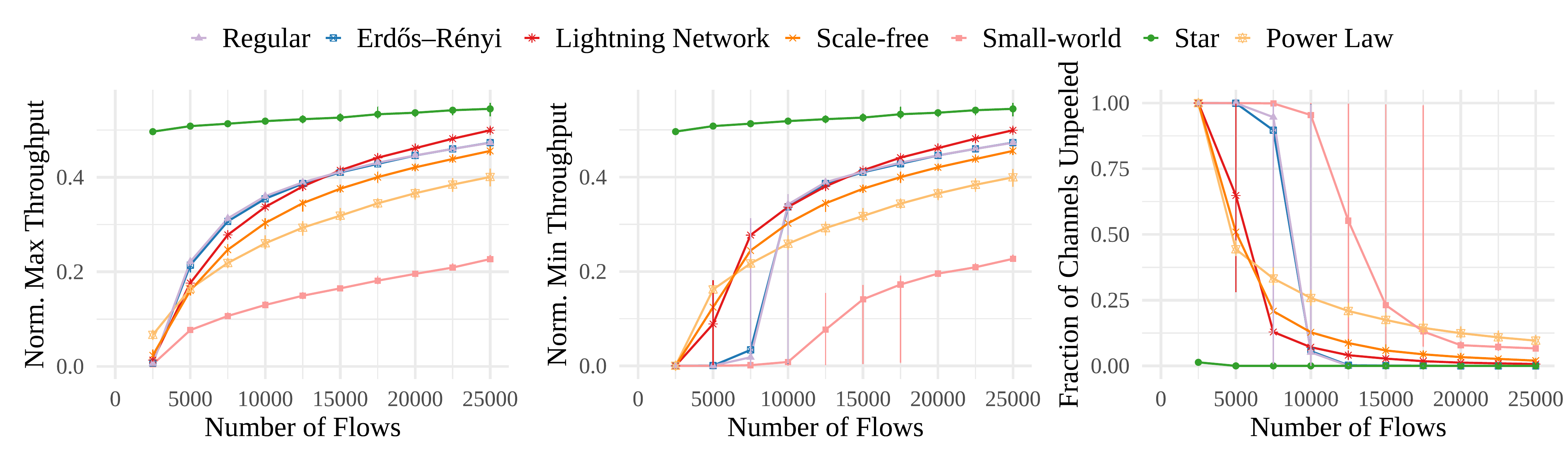}
    \caption{\small Maximum ($\maxtpt$) and minimum throughput ($\mintpt$) achieved by different
    topologies and channels left unpeeled as the number of flows is varied.
    Stars outperform all other topologies but are undesirable due to their centralization.
    The \lntopo, \pltopo and \sftopo graphs are less sensitive than the other topologies: 
    they peel earlier and have higher $\mintpt$.
    The fraction of channels peeled corresponds well with $\mintpt$: topologies that peel well have a higher $\mintpt$. Whiskers denote max and min data point.} 
    \label{fig:overall tpt comparison}
\end{figure*}

\Fig{overall tpt comparison} shows the best-case throughput $\maxtpt$ and the worst-case throughput $\mintpt$
achievable across all starting balance states for a set of random
topologies with 500 nodes and an LN topology with 452 nodes for a fixed collateral budget
for 2500--25000 demand pairs sampled uniformly at random. 
Stars outperform all the other topologies by over 50\% even at the midpoint of the range.
However, stars are highly centralized topologies  (i.e., the hub has degree $n$); consequently, they are undesirable for
decentralized use cases of \cns.
The $\maxtpt$ value is comparable across most topologies; the \sw topology alone stands as an outlier because it has 50\% longer paths than other topologies which results in lower throughput. 

\begin{figure*}
    \vspace{-3mm}
    \centering
    \includegraphics[width=\textwidth]{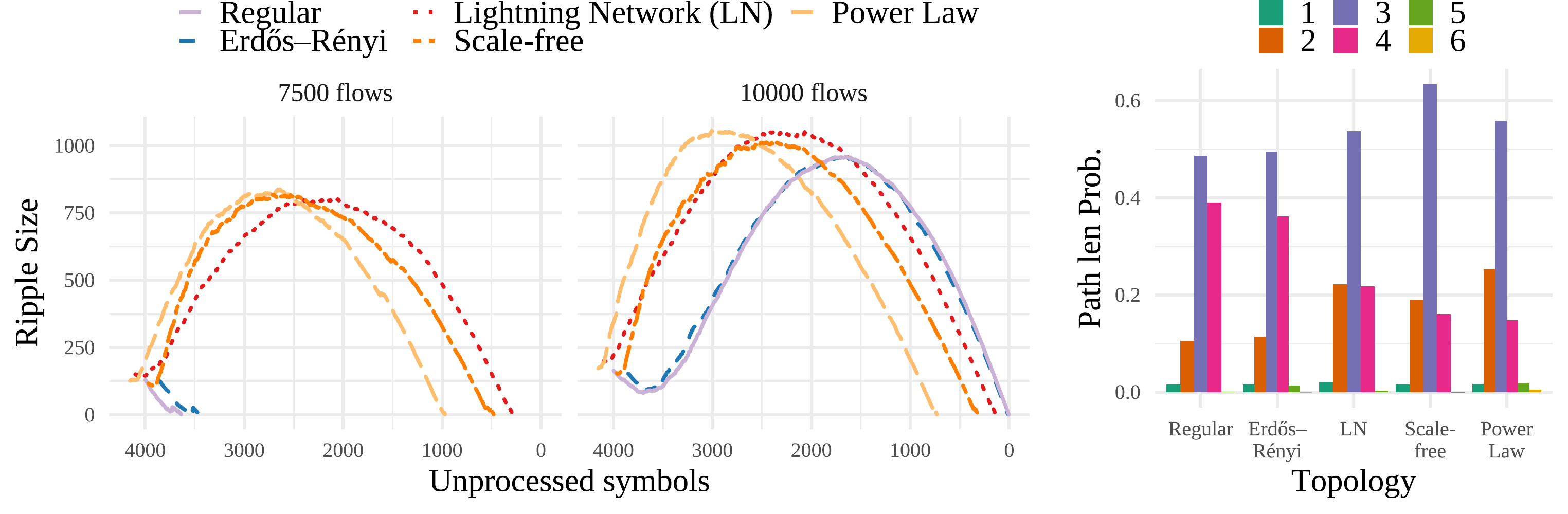}
    \caption{\small Evolution of ripple size during the \dpp on different topologies. 
    At 7500 flows, the ripples of \er and \rr vanish too quickly preventing them from progressing at all
    in contrast to \lntopo, \pltopo and \sftopo topologies. However, while \lntopo, \pltopo and \sftopo
    topologies experience good growth at the beginning, they are unable to sustain a good ripple size resulting
    in the last 6-7\% of channels left unprocessed even at 10000 flows.}
    \label{fig:peeling real evolution}
    \vspace{-5mm}
\end{figure*}


\Fig{overall tpt comparison} also shows the variation in the $\mintpt$ values and the fraction of unpeeled channels
across topologies.
For instance, we notice that the \lntopo, \pltopo and \sftopo topologies have much better $\mintpt$ with fewer flows
when compared to the other random graphs suggesting that they exhibit less throughput 
sensitivity to the channel balance state. At 7500 demand pairs, both \lntopo and \sftopo topologies peel 20-70\%  more channels
than the \sw, \er, and \rr topologies and correspondingly have 7-9x higher $\mintpt$. 
However, the hardship that the \sftopo graph
faces in peeling the last 10\% of its channels manifests as a 12\% hit to its
$\mintpt$ when compared to the \lntopo, particularly with a denser demand matrix. This effect is even more pronounced in the \pltopo graph.
In contrast, the \er and \rr topologies do not peel as well with fewer flows, but quickly
improve to peel all channels, on average, with 12500 demands. 
Beyond this point, their $\mintpt$ is comparable with the \lntopo.
The \sw topology peels only 50\% of the flows even with 12500 demands which when compounded with
its long paths leads to very low $\mintpt$ over the entire range. 
\Fig{nonuniform random overall tpt} in \App{apdx:nonuniform random results} shows similar trends across topologies even when demand matrices are skewed in that some ``heavy-hitter'' nodes are more likely to both send and receive transactions.


\NewPara{Explaining the relative behavior of topologies.}
Given the apparent correlation between $\mintpt$ 
and the fraction of channels peeled, we now consider the effect of topology on the evolution of the \dpp. 
\Fig{peeling real evolution} shows the evolution of the ripple size at 7500 and 10000 demand pairs, along with the path length distributions for the \rr, \er, 
\lntopo, \pltopo and \sftopo topologies. 
We consider the total number of symbols at the start to be twice the number
of channels in the topology, one for each direction of a channel.
Each peeling step involves processing one channel in one of the two
directions. A processed symbol may lead to the release of some flow nodes and consequently, add more 
directed channels to the ripple.

In \Fig{peeling real evolution}, all topologies start with similar initial ripple sizes because
they have the same sparsity. In other words, a randomly sampled demand is equally likely
to be of length 1 (span only one edge) across all topologies. 
However, their ripple evolution
patterns quickly diverge.   
The \lntopo, \pltopo and \sftopo topologies experience fast initial ripple growth, attributed to the
20\%  degree 2 flows (with path length 2) and up to 60\%
flows of degree 3. These short flows are likely to be released early, \
helping the \dpp pick up a robust ripple size. 
In contrast, at 7500 demands, the ripple vanishes quickly for the \rr and \er 
topologies due to 10\% fewer degree $2$ flows. 
Yet, the ripples in \lntopo, \pltopo and \sftopo topologies vanish before the entire topology
is peeled, with 250 and 500 directed channels respectively unpeeled.
This behavior happens with 10000 demands too, albeit to a lesser extent. However,
at 10000 flows, \er and \rr topologies offset the initial dip in 
the ripple size compared to the remaining topologies; they experience
later peaks, but peel the entire topology.
The poor tail behavior of the \sftopo topology
can be attributed to its 6\% less channel coverage for the same number of demand pairs:
the presence of hubs means edges far away from the hub tend to be less used. 
Such channels are difficult to peel without a large increase in the number of flows that ensures the
relevant edges see enough token movement. 
A similar analysis of the ripple evolution for flows sampled in a skewed manner is shown in \Fig{nonuniform random peeling real evolution} in \App{apdx:nonuniform random results}.


\NewPara{Predicting Performance using LT Codes}.
Since the \dpp was inspired by the design on LT Codes, we evaluate
whether the analysis of LT Codes predicts the performance of the \dpp.
We use the same 5 random graphs from \Fig{peeling real evolution} at 7500 flows
and view their predicted ripple evolution in an LT code with the same degree distribution. The predicted ripple evolution computes the probability
that a flow of degree $d$ is released and adds a symbol to the ripple when there are $L$ unprocessed
symbols remaining. Extending this to the expected ripple addition at every step helps build
a ripple evolution curve prediction (Eq. 6 of \cite{decripplelt}) \footnote{The trajectories
use a slightly different expression (\Eqn{expected symbols added}) for the expected symbols added that accounts for overlaps
between symbols covered by the release of different flows at the same step.}.
\begin{wrapfigure}{r}{0.45\textwidth}
    \centering
    \includegraphics[width=0.4\textwidth]{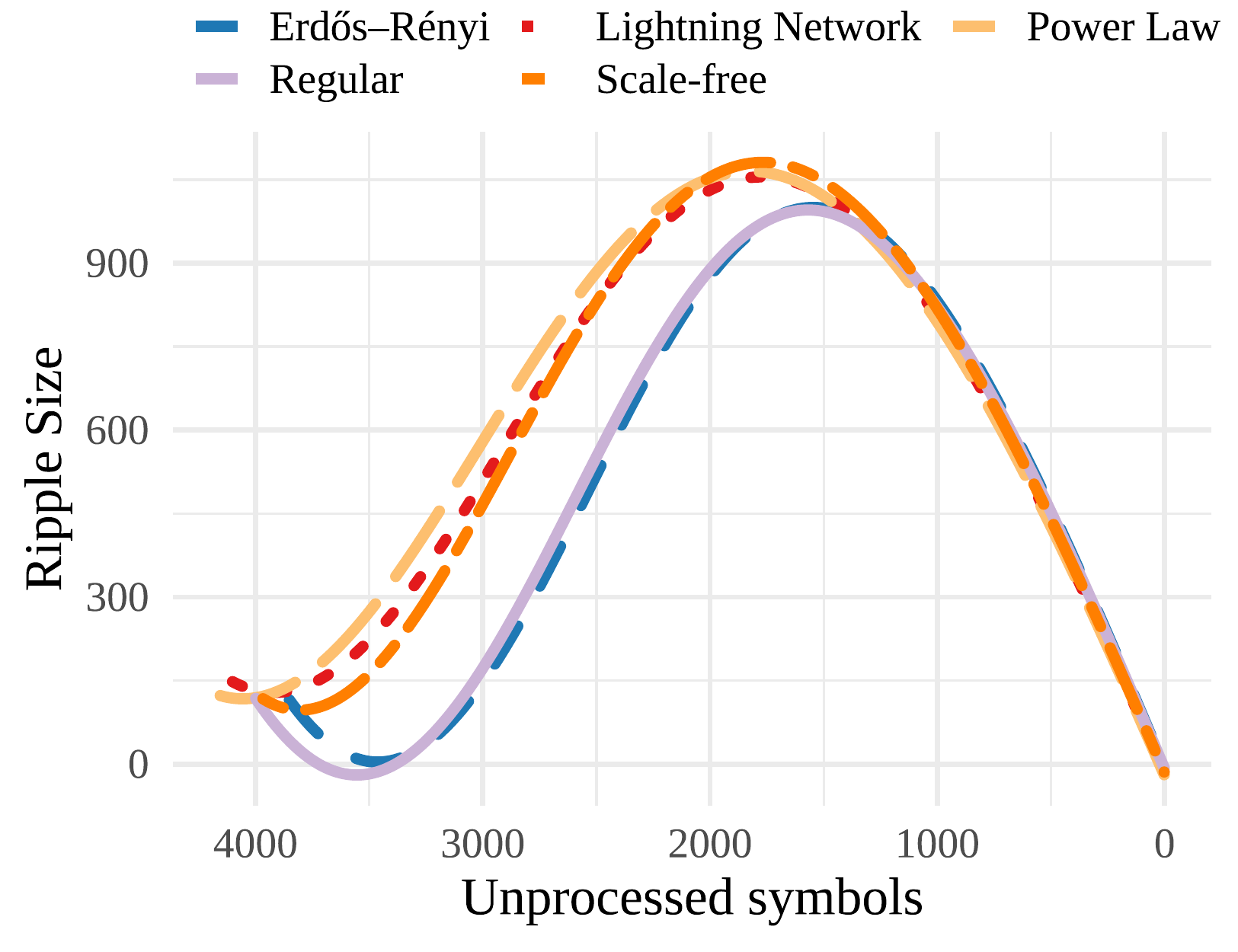}
    \caption{\small Ripple size predicted by the LT codes analysis on different random topologies at 7500 flows. The prediction suggests that \er and \rr ripple sizes dip to zero early on while
    \lntopo, \pltopo and \sftopo experience big growths. However, it departs
    from the real ripple size significantly towards the end of the peeling process.}
    \label{fig:peeling predicted evolution}
    \vspace{-3mm}
\end{wrapfigure}

We notice in \Fig{peeling predicted evolution} that the prediction for the \lntopo, \pltopo and \sftopo
topologies are closer to each other but different from the \er and \rr graphs.
Like the real ripple evolution (\Fig{peeling real evolution}), the initial growth rate for the \lntopo, \pltopo and \sftopo topologies is faster than the other two graphs.
Interestingly, the prediction suggests that \er and \rr topologies will have difficulty peeling at 7500 flows: the predicted ripple sizes approach $0$ with around 3000 unprocessed symbols remaining. Such
a ripple evolution is not robust to the variance encountered
during a typical peeling process as we observe in \Fig{peeling real evolution}
where the ripple decreases drastically early on and vanishes. 
The prediction is more optimistic than the real evolution: the peak ripple
sizes are higher and all topologies peel all their edges.
In reality, the \dpp does not peel all the edges even in the \lntopo, \pltopo and \sftopo topologies.
This difference is anticipated since the LT Codes analyses~\cite{lubylt,decripplelt} rely on
an \emph{i.i.d. bipartite encoding graph} where flows (encoded symbols) choose channels (input symbols) uniformly at random.
On real graphs, the channels traversed by flows are correlated; in fact,  correlation between edges of flows  makes it very hard to peel the last 6\% of channels in the \sftopo topology.
Further, we found that with a skewed demand matrix, the correlations become stronger and the predictions made using the i.i.d. bipartite graph model deviate from reality more significantly (\Fig{nonuniform random peeling real evolution}). Hence, a full analysis of the deadlock peeling process should take into account correlations induced by the topology structure and demand pattern (we leave this to future work). As a first step to understanding the value of such an analysis,  we next investigate whether  the LT code analysis can be used to synthesize good topologies for uniform random demand matrices, where fortuitously LT code predictions are reasonably accurate.

\subsection{Topology Synthesis}
\label{sec:eval-synthesis}
\NewPara{Generating a path length distribution}. 
Our goal is to find topologies that require fewer demand pairs to render a topology deadlock-free.
We start with an approach for LT Codes~\cite{decripplelt}, which fixes a desired ripple evolution
and numerically computes a degree distribution that closely approximates it. 
In our setting, the degree distribution corresponds to a distribution over path lengths in the \cn.
Like \cite{decripplelt}, we choose a ripple evolution of the form $R(L) = 1.7c^{1/2.5}$ where $L$ is the number of unprocessed symbols and $R(L)$ is the ripple size when $L$ symbols remain unprocessed. 
For a target graph with 300 nodes and 1500 edges, this ensures a ripple of $30$ at the start of the peeling algorithm, decaying slowly to $25$ halfway and eventually to $2$ towards the end of the algorithm.
\begin{wrapfigure}{r}{0.4\textwidth}
    \centering
    \includegraphics[width=0.36\textwidth]{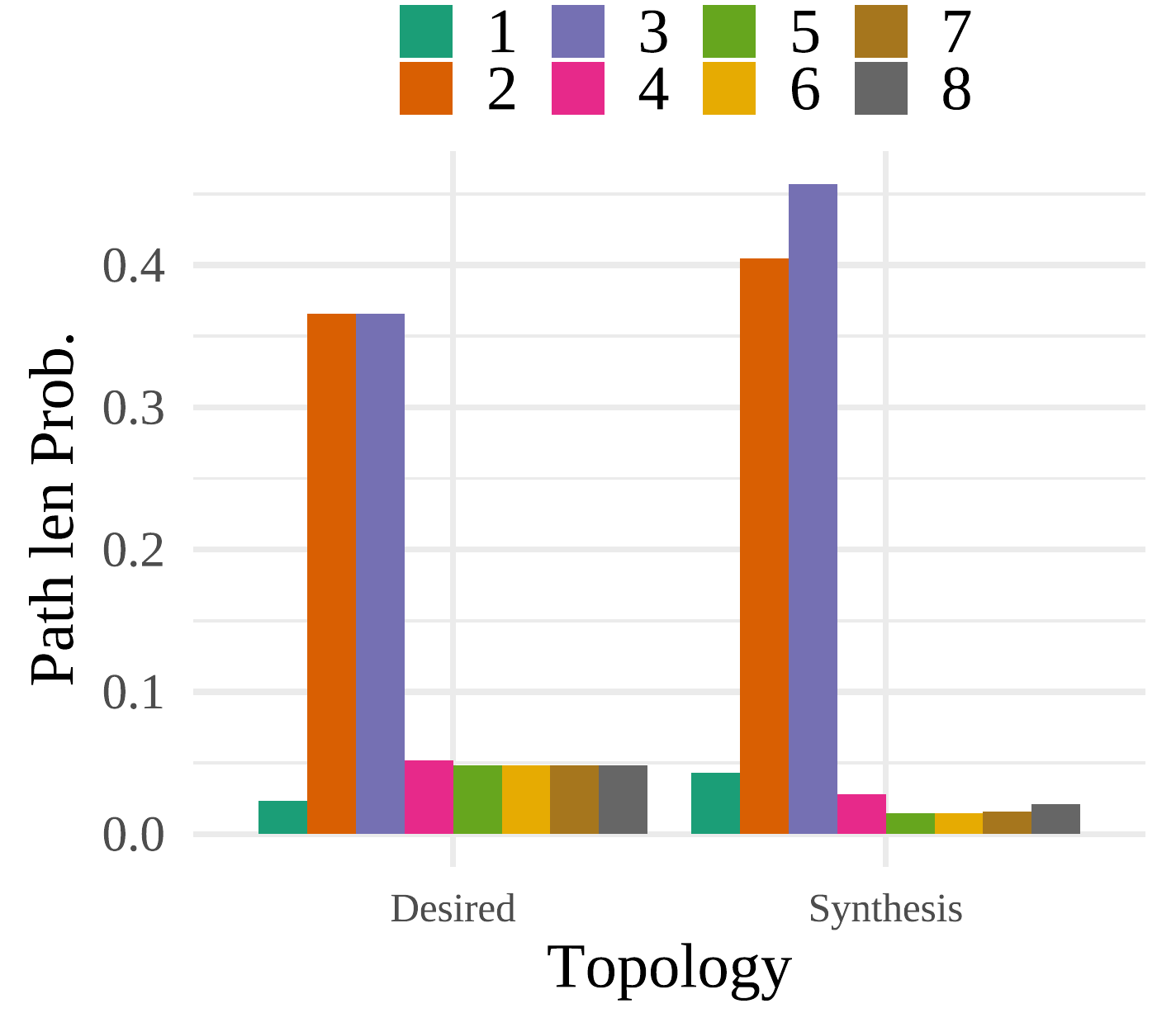}
    \caption{\small Desired path length distribution vs. synthesized topology's distribution.} 
    \label{fig:synthesis path len dist}
    \vspace{-4mm}
\end{wrapfigure}

Next, we choose a path length distribution for the topology that achieves the desired ripple evolution. Prior work shows that
the expected number of symbols added to the ripple at each LT process step is a linear function of the path length distribution~\cite{decripplelt}. The linear transformations and constants correspond to the release probabilities and the desired ripple addition at each step of the algorithm respectively. 
For completeness, we outline the equations in \App{apdx:numerical optimization}.
We minimize the $\ell_2$-norm of the difference between the expected and desired ripple size.
Since our degree distributions map to paths on a graph, 
we constrain the maximum path length and bound path length probabilities 
based on the maximum node degree (to enforce decentralization). We also ensure that the number of length $1$ paths does not exceed the number of edges and that the path length probabilities decrease from length $2$ onwards. Solving this least-squares optimization problem with linear constraints yields a path length distribution.

\NewPara{Synthesizing a matching topology}. 
There are potentially many ways to synthesize a topology with a given shortest-path length distribution.
Our approach exploits a known link between the distribution of shortest path lengths for a random graph and its pairwise joint degree distribution~\cite{analyticalshortestpaths,jntdegconstr}.\footnote{The pairwise joint degree distribution of a graph evaluated at $j$ and $\ell$ specifies the probability that a randomly sampled edge connects nodes of degree $j$ and $\ell$. } 
We first set up an optimization using MATLAB to generate a joint degree distribution
whose path length distribution minimizes the $\ell_2$-norm of the distance from the target path length distribution (output from the previous subsection). 
Given a joint degree distribution, we use
established methods to generate a random graph that matches the joint degree distribution~\cite{nxjointdegree, infocommjointdegree}.
This approach is more expressive than the well-known configuration model \cite{fosdick2018configuring}, 
and can (approximately) recover \er, \sftopo, \sw and \rr graphs.

\begin{figure*}
    \vspace{-3mm}
    \centering
    \includegraphics[width=\textwidth]{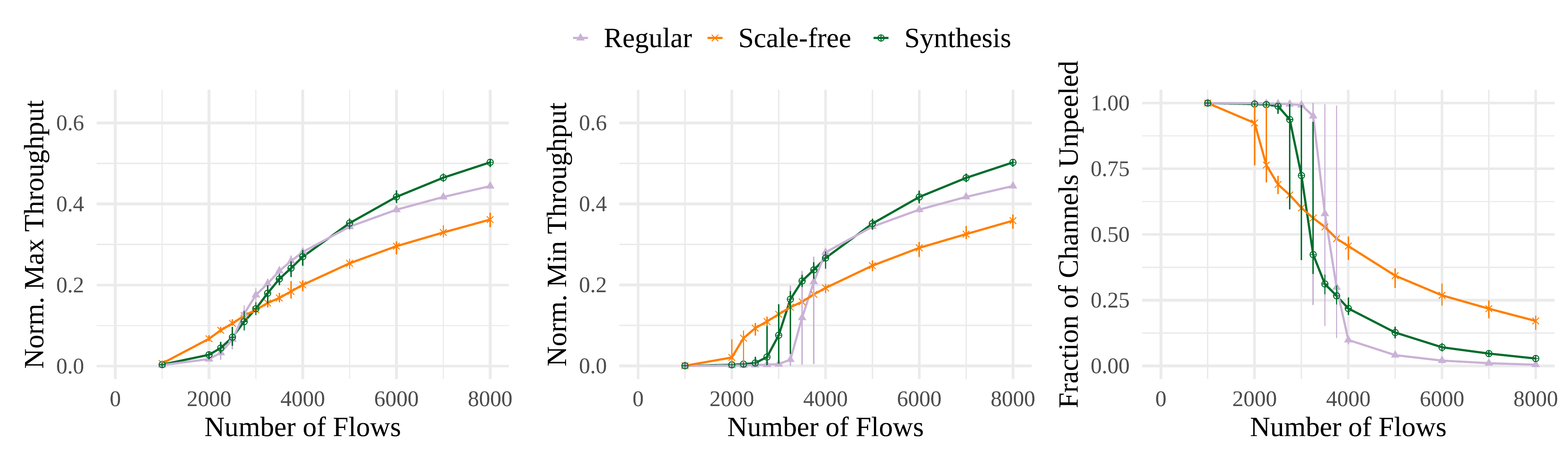}
    \caption{\small Throughput comparison between a synthesized topology with 271 nodes
    and its random counterparts
    at 300 nodes and 1500 edges. The synthesized topology has lower throughput sensitivity than
    the \rr graph and better 
    overall throughput than the \sftopo graph. 
    Whiskers denote max and min points.}
    \vspace{-5mm}
    \label{fig:synthesis comparison}
    
\end{figure*}

\NewPara{Results}. Our target topology has 300 nodes and 1500 edges. We first use the numerical optimizer to find a path length distribution that peels well. We observe that permitting a large maximum node degree generates path length distributions that the downstream MATLAB optimization routine fails to match well, generating graphs with only 60\% of the desired edges. Thus, we ensure that the maximum node degree is 10 and no path is longer than 10 edges. The output distribution from the numerical optimizer is shown in \Fig{synthesis path len dist}.
When using the MATLAB optimization routine to generate a joint degree distribution, we observe that
setting the maximum node degree to 20, 
average degree to 10, and maximum path length to 10 ensures a close match 
to the desired path length distribution while avoiding extremely long paths during the synthesis step. We then synthesize a graph with the desired node distribution and take its largest connected component. The resulting topology has 271 nodes and 1513 channels, and achieves a path length distribution close to the one desired (\Fig{synthesis path len dist}). To evaluate how well the synthesized topology performs, we compare
it to 5 instances of \rr and \sftopo topologies with 300 nodes and about 1500 channels in \Fig{synthesis comparison}. The maximum throughput $\maxtpt$ achieved by the synthesized topology is
comparable to the \rr graph and up to 15\% better than the \sftopo graph, particularly
with more demand pairs. The $\mintpt$ and the fraction
of channels unpeeled of the synthesized topology strikes a balance between
the \rr and \sftopo graphs. It is less sensitive than the \rr topology, notably with fewer demand pairs,
achieving better 10--20\% better $\mintpt$ and peeling 25--50\% more channels. 
While it is more sensitive than the \sftopo graph with sparser demand, it compensates
with a 15\% larger $\mintpt$ at denser demands. 
This approach shows promise in generating topologies
that show good peeling properties and throughput insensitivity. We leave it
to future work to explore generalizing this to different demand models to generate even better topologies.

\NewPara{Remark}. While the above synthesis shows promise, most \cns are formed by individual nodes' connectivity decisions, rather than by a centralized authority. 
However, router nodes in \cns are incentivized to support high throughput via routing fees, and some PCNs already include peer recommendation systems that suggest peers to connect to~\cite{ln-routing-guide, ln-scoring-blog}. While we leave the details to future work, we envision building on these systems to encourage desirable topologies. For example, a \cn software client could ``score", or suggest, peers such that the overall topology obeys a given joint degree distribution.

\section{Related Work}
\label{sec:related}

\NewPara{Network Topology Design}.
The peer-to-peer networking literature~\cite{lua2005survey} has studied the design of structured~\cite{stoica2001chord, maymounkov2002kademlia, ratnasamy2001scalable} and unstructured topologies~\cite{kim2013real, bittorrent}, particularly for efficient content-retrieval.
Meanwhile, fat tree~\cite{al2008scalable}, Clos~\cite{greenberg2009vl2}, small world~\cite{shin2011small}, and random graph~\cite{singla2012jellyfish} topologies have been designed to maximize throughput in datacenters. Adapting these ideas to credit networks is hard due to their centralization---e.g., in a fat tree  \cite{al2008scalable} with $n$ top-of-rack (ToR) switches, the aggregation and core switches require a degree of $O(n^{1/3})$.

\NewPara{Credit Network Performance}.
There has been substantial recent interest in quantifying credit network performance, broadly defined as transaction throughput or success rate; prior work has studied the impact of several categories of influencing factors, including routing and scheduling protocols \cite{spiderhotnets,wang2019flash,spider,malavolta2017concurrency,werman2018avoiding}, privacy constraints \cite{roos2017settling,malavolta2017silentwhispers,tang2020privacy,tikhomirov2020quantitative}, defaulting agents \cite{ramseyer2020liquidity}, and network topology \cite{dandekar2011liquidity,dandekar2015strategic,aumayr2020demand,khamisdemand}.
Prior work exploring the effects of credit network topology on transaction success rate~\cite{dandekar2011liquidity} assumed sequential transactions and a full demand matrix.
In this setting, maximum throughput can be trivially achieved with length-1 flows, so no deadlocks are observed.

\emph{Demand-aware} design of payment channel network topologies~\cite{dandekar2015strategic}
poses the problem as an ILP: given a channel budget, a set of nodes, and a demand matrix, the ILP finds an adjacency matrix that either maximizes the number of connected demand pairs~\cite{aumayr2020demand} or minimizes the number of channels added and the average path length~\cite{khamisdemand}. This approach ignores the effects of channel imbalance, which causes the deadlock-related problems explored in this work.

\NewPara{Erasure Codes}. 
Sparse graph codes have been studied for decades in channel coding  \cite{gallager1962low,byers1998digital,mackay1999good,luby2001efficient,richardson2001design,lubylt,shokrollahi2006raptor}. 
We identify a parallel between such codes and detecting deadlocked edges in a topology. 
However, despite a rich literature on sparse graph codes, our setting cannot fully utilize most existing constructions for two reasons. 
First, we need to be able to change the number of flows (encoding symbols) flexibly, without requiring a totally new encoding.
We therefore use \emph{rateless} codes \cite{luby2001efficient,lubylt,shokrollahi2006raptor}; 
specifically, our peeling algorithm  builds closely on Luby's LT codes \cite{lubylt}. 
Second, our encoding procedure must respect the topology constraints of the underlying graph; 
we cannot assign channels uniformly at random to flows, as in classical sparse graph codes.
To our knowledge, this constraint has not been previously studied. 
This is why our theoretical predictions for the number of flows needed to peel a graph (from \cite{lubylt}) are  lower than the number needed in practice (\Sec{sec:eval}). 
More broadly, this suggests an interesting direction of further study on error-correcting codes that obey encoding constraints imposed by  a graph topology. 
\section{Conclusion}
\label{sec:conclusion}

In this paper, we studied the effects of topology and channel imbalance on the throughput of credit networks. We demonstrated a close relation between worst-case throughput and the presence of deadlocks, and proposed a heuristic peeling algorithm 
for identifying the presence of deadlocks. 
Important directions for future work include synthesizing graphs that are: (a) easy to peel, (b) exhibit high throughput, and (c) limit the maximum degree of any node. 
While we have made progress on that front, we were not able to find topologies that perform substantially better than random graphs. 
Hence, an interesting question is whether this is even possible. 
Another interesting direction is to analyze the peeling algorithm given the correlations induced (in the bipartite graph) by an arbitrary topology and demand pattern.
Further, this work assumes perfect routing; in practice it is unclear whether one can actually achieve $\phi(\bb)$. 
Designing decentralized routing protocols that are tailored to good \cn topologies and understanding their performance with respect to these upper bounds will be important in practice. Finally, designing suitable incentive/recommendation mechanisms to achieve a desired topology in a decentralized manner is an important problem.


\begin{acks}
We thank our anonymous reviewers for their detailed feedback. This work was supported in part by an AFOSR grant FA9550-21-1-0090; NSF grants CNS-1751009, CNS-1910676, and CIF-1705007; a Microsoft Faculty Fellowship; awards from the Cisco Research Center and the Fintech@CSAIL program; and gifts  from Chainlink, the Sloan Foundation, and the Ripple Foundation.
\end{acks}

\bibliographystyle{ACM-Reference-Format}
\bibliography{sigmetrics21}

\newpage
\appendix
\noindent

\section{Theoretical Results}

\label{apdx:sec:theory}

\subsection{Basic Properties}
\label{app:basic_properties}

\begin{proposition}
    For any state $\bb$, $\rch_{\bb}$ is a convex set and $\bb \in \rch_{\bb}$.
    \label{prop:reachable_conv}
\end{proposition}

\begin{proposition}
    For a balance state $ \bb  \in \allr_{\ge 0}^{|E|}$, if $\xb, \yb \in \dir_{\bb}$, then 
    $\alpha \xb + \beta \yb \in \dir_{\bb}$ for any $\alpha, \beta \in \allr_{\ge 0}$.
    \label{prop:direction_conv}
\end{proposition}

\begin{proposition}
    For any $\fb: \fb \geq \zero{}$ and $R \fb \leq \mvec{\bb}{\capvec - \bb}$, we have $\bb - \Delta R \fb \in \rch_{\bb}$ and $\fb \in \feasflow_{\bb}$.
    \label{prop:rchelement}
\end{proposition}

\begin{proposition}
    $\bb \in \rch_{\ab} \Longleftrightarrow \rch_{\bb} \subseteq \rch_{\ab}$. 
    \label{prop:reach_subset}
\end{proposition}

\begin{proposition}
    The center state has maximized $\psi$. 
    Precisely, $\psi\left(\dfrac{\capvec}{2}\right) = \max_{\ab \in \allb} \psi(\ab)$. 
    \label{prop:max_psi}
\end{proposition}


\begin{proposition}
\[ 
    \dlk_{\bb} 
    = \big\{ i \in [|E|]~ \big|~ a_{i} = b_i,~ \forall \ab \in \rch_{\bb}  \big\} 
\] 
\label{prop:dlk_channel_alt_def}
\end{proposition}


\begin{proposition}
    If $i \in \dlk_{\bb}$, then for all $\fb \in \feasflow_{\bb}$ whose decomposition is a sequence of flows $\left( \fb^{(j)} \right)_{j=0}^{k-1}$, we have 
    \begin{enumerate}
        \item For all $j \in [k]$, $\overline{R}_i \fb^{(j)} = \underline{R}_{i} \fb^{(j)} = 0$; 
        \label{prop:zero_flow:subflow}
        
    \item For all $j \in [k]$ and $\pi \in \Pi$ (index of path) where $\overline{R}_{i, \pi} = 1 \vee \underline{R}_{i, \pi} = 1$, $\fb^{(j)}_\pi = 0$.  
        \label{prop:zero_flow:flow_size}
    \end{enumerate}
    \label{prop:zero_flow}
\end{proposition}


\subsection{Supporting Theorems}

\label{apdx:subsec:support_thm}

\begin{lemma}
    If for some $\epsilon \ge 0$, 
    $\epsilon \mvec{\ab}{\capvec - \ab} \leq \mvec{\bb}{\capvec - \bb}$, then for all $\pb \in \rch_{\ab}$, we have $\bb + \epsilon (\pb - \ab) \in \rch_{\bb}$. 
    \label{lemma:reachable_states_mapping}
\end{lemma}

\annot{
    %
    If $\epsilon \mvec{\ab}{\capvec - \ab} \leq \mvec{\bb}{\capvec - \bb}$, a a scaled down version of any ``reachable offset'' from state $\ab$ to can be applied (in the same direction) to $\bb$.}
    
\paragraph{Proof.} Since $\pb \in \rch_{\ab}$,  $\exists k \in \alln$, a sequence of flows $\fb^{(0:k-1)}$ and a sequence of transient states $\ab^{(0:k)}$, where $\ab^{(0)} = \ab$, $\ab^{(k)} = \pb$, and for all $j \in [k]$,
\[ \ab^{(j+1)} = \ab^{(j)} - \Delta R \fb^{(j)},\quad 
\fb^{(j)} \geq \zero{}, \quad
R \fb^{(j)} \leq \mvec{\ab^{(j)}}{\capvec - \ab^{(j)}}.
\]

Consider new flows $\gb^{(j)} = \epsilon \fb^{(j)}$ for $\epsilon \ge 0$ 
and $j \in [k]$. 
Starting from state $\bb^{(0)} = \bb$, we send flows $\left( \gb^{(j)} \right)_{j=0}^{k-1}$ and obtain transient states $\left( \bb^{(j)} \right)_{j=0}^{k}$, 
where for all $j \in [k]$,
\[ \bb^{(j+1)} = \bb^{(j)} - \Delta R \gb^{(j)} = \bb^{(j)} - \epsilon \Delta R \fb^{(j)} = \bb - \epsilon \Delta R \sum_{i=0}^{j} \fb^{(i)} = \bb + \epsilon \left( \ab^{(j+1)} - \ab \right). \]

Note that $\bb^{(0)} = \bb + \epsilon \zero = \bb + \epsilon \left( \ab^{(0)} - \ab \right)$. Staring with $\epsilon \mvec{\ab}{\capvec - \ab} \leq \mvec{\bb}{\capvec - \bb}$, we have
{
\newcommand{\so}{~\Longrightarrow~}
\begin{align*}
    & \epsilon \mvec{\ab}{\capvec - \ab} \leq \mvec{\bb}{\capvec - \bb} \\
     \so~ & \epsilon \mvec{\ab^{(j)}}{\capvec - \ab^{(j)}} \leq \mvec{\bb}{\capvec - \bb} + \epsilon \left( \mvec{\ab^{(j)}}{\capvec - \ab^{(j)}} - \mvec{\ab}{\capvec - \ab} \right) = \mvec{\bb^{(j)}}{\capvec - \bb^{(j)}},\quad \forall j \in [k] \\
     \so~ & R \gb^{(j)} = \epsilon R \fb^{(j)} \leq \epsilon \mvec{\ab^{(j)}}{\capvec - \ab^{(j)}} \leq \mvec{\bb^{(j)}}{\capvec - \bb^{(j)}},\quad \forall j \in [k].
\end{align*}
}

Therefore, combining with $\gb^{(j)} = \epsilon \fb^{(j)} \geq \zero$, we know $\left(\gb^{(j)}\right)_{j=0}^{k-1} = \epsilon\left(\fb^{(j)}\right)_{j=0}^{k-1}$ is a valid sequence of flows that leads to a reachable state from $\bb$. In other words,
\[ 
\rch_{\bb} \ni \bb^{(k)} = 
\bb - \Delta R \sum_{j=0}^{k-1} \gb^{(j)} = \bb - \epsilon \Delta R \sum_{j=0}^{k-1} \fb^{(j)} = \bb + \epsilon (\pb - \ab).
\qeda
\]

\begin{lemma}
   Let $\bb \in \rch_{\ab} - \{\ab\}$ and 
    $\displaystyle \pb \triangleq \arg\max_{\pb' \in \allb} \{ \|\pb' - \ab\|_2 | \exists t \ge 0: \pb' - \ab = t (\bb - \ab) \} $ be the intersection with the boundary of state space with the direction along $\ab$ to $\bb$.
    Let $\displaystyle r = \frac{\| \pb - \bb \|_2}{\|\pb - \ab\|_2} \in [0, 1)$. 
    For all $t \in [0, 1 - r^2]$, $\ab + t (\pb - \ab) \in \rch_{\ab}$ and for all $t \in [1-r, 1-r^2]$, $\ab + t(\pb - \ab) \in \rch_{\bb}$. 
    \label{lemma:ray}
\end{lemma}

\annot{
    If $\bb$ is reachable from $\ab$, then as you extend the direction vector past $\bb$, \emph{some} more states on this vector will be reachable from $\ab$. The proof
    below argues this for points $\pb$ between
    $\ab$ and a hypothetical $\cb$ which 
    satisfies
    \[ \frac{\|\pb - \cb \|_2}{\|\pb - \bb \|_2} = \frac{\|\pb - \bb \|_2}{\|\pb - \ab \|_2} = r. \]
    The proof first discusses points $\pb$
    between $\ab$ and $\bb$, and then
    points between $\bb$ and $\cb$
}

\paragraph{Proof.} By $\bb \in \rch_{\ab}$, there exists a $k \in \alln$, a sequence of flows $\fb^{(0:k-1)}$ and a sequence of transient states $\ab^{(0:k)}$, where $\ab^{(0)} = \ab$, $\ab^{(k)} = \bb$, and for all $j \in [k]$,
\[ \ab^{(j+1)} = \ab^{(j)} - \Delta R \fb^{(j)},\quad 
\fb^{(j)} \geq \zero{}, \quad
R \fb^{(j)} \leq \mvec{\ab^{(j)}}{\capvec - \ab^{(j)}}.
\]

\NewPara{States on segment $[\ab, \bb]$}. These states correspond to $t \in [0, 1-r]$.

For any constant $\epsilon \in [0, 1]$, we have $\epsilon \mvec{\ab}{\capvec - \ab} \leq \mvec{\ab}{\capvec - \ab}$. 
Citing Lemma \ref{lemma:reachable_states_mapping} and using the fact that $\bb \in \rch_{\ab}$, we obtain $\ab + \epsilon(\bb - \ab) \in \rch_{\ab}$, for all $\epsilon \in [0, 1]$.
From $\ab + \epsilon (\bb - \ab) = \ab + \epsilon (1 - r) (\pb - \ab)$, we know the statement is true for all $t \in [0, 1-r]$. \\

\NewPara{States on segment $[\bb, \cb]$}. These states correspond to 
$t \in [1-r, 1-r^2]$, or a part of the segment $[\bb, \pb]$. 

Starting with $0 \le \epsilon \le r < 1$ and $- \mvec{\ab}{\capvec - \ab} \leq \zero \leq \mvec{\pb}{\capvec - \pb}$, we have the following deductions.
{
\newcommand{\so}{~\Longrightarrow~}
\begin{align*}
    & (0 \le \epsilon \le r < 1) \wedge \left(- \mvec{\ab}{\capvec - \ab} \leq \zero \leq \mvec{\pb}{\capvec - \pb} \right) \\
    \so ~& (\epsilon - r) \mvec{\ab}{\capvec - \ab} \leq \zero \leq (1-r)\mvec{\pb}{\capvec - \pb} \\
    \so ~& \epsilon \mvec{\ab}{\capvec - \ab} \leq \mvec{\pb}{\capvec - \pb} - r 
    \left(\mvec{\pb}{\capvec - \pb} - \mvec{\ab}{\capvec - \ab}\right) = \mvec{\bb}{\capvec - \bb}. 
\end{align*}
}

Citing Lemma \ref{lemma:reachable_states_mapping}, we have $\bb +  \epsilon (\bb - \ab) \in \rch_{\bb}$. 
By transitivity of reachable states (Prop. \ref{prop:reach_subset}), we have for all $\epsilon \in [0, r]$, 
\[ 
\rch_{\ab} \supseteq \rch_{\bb} \ni 
\bb + \epsilon (\bb - \ab) = \ab + (1-r)(1+\epsilon) (\pb - \ab). 
\]

In other words, for all $t \in [1-r, 1-r^2]$, $\ab + t (\pb - \ab) \in \rch_{\ab} \cap \rch_{\bb}$. The proof is now complete. \qeda

\begin{thm}
    Assume $\xib \in \dir_{\ab} - \{\zero\}$ is a feasible direction and 
    $\pb \triangleq \arg\max_{\pb' \in \allb} \{ \|\pb' - \ab\|_2 | \exists t \ge 0: \pb' - \ab = t \xib \} $ is the ray's last intersection with the state space. 
    For all $t \in [0, 1)$, $\ab + t (\pb - \ab) \in \rch_{\ab}$. 
    \label{thm:reachable_along_dir}
\end{thm}

\annot{
    If a direction is feasible, consider the segment in that direction between $\ab$ and the boundary of the state space $\allb$. All states on this segment are reachable from $\ab$ except the farthest endpoint. The proof uses the above two lemmas to repeat the creation of point $\cb$ a finite
    number of times until any (interior) point on the direction between $\ab$ and $\allb$ is proved reachable.
}

\paragraph{Proof.} By definition of $\dir_{\ab}$, $\exists \bb \in \rch_{\ab}$, $r \in [0, 1)$, such that $\pb - \bb = r(\pb - \ab)$. 
If $\bb$ itself is the end point on the boundary, then $r = 0$ and $\pb = \bb$. Then, using Lemma \ref{lemma:ray}, we can trivially prove the claim. \\

If $0 < r < 1$, we can take $k = \left\lceil \log_r (1-t) \right\rceil \in \alln$ which satisfies $1 - r^k \ge t$. 
Then, we construct a sequence of transient states $\left(\ab^{(j)}\right)_{j=0}^{k}$, where $\ab^{(j)} = \ab + (1 - r^j) (\pb - \ab)$ for all $j \in [k] \cup \{k\}$. 
By this definition, 
\[ 
\ab^{(0)} = \ab, \quad 
\ab^{(1)} = \bb, \quad
\forall j \in [k]: \pb - \ab^{(j+1)} = r \left( \pb - \ab^{(j)} \right).
\]

We recursively show $\ab^{(j+1)} \in \rch_{\ab^{(j)}}$ for all $j \in [k]$. 
This is true for $j = 0$ given $\bb \in \rch_{\ab}$. 

Suppose $\ab^{(j+1)} \in \rch_{\ab^{(j)}}$ is true for $j = J - 1 < k-1$. Then for $j = J$, we have 
\[ 
\ab^{(J+1)} = \pb - r \left( \pb - \ab^{(J)} \right) = \ab^{(J-1)} + (1-r^2) \left(\pb - \ab^{(J-1)} \right).
\]

By Lemma \ref{lemma:ray}, combined with $\pb - \ab^{(J+1)} = r \left( \pb - \ab^{(J)} \right)$ and $\ab^{(J)} \in \rch_{\ab^{(J-1)}}$, we have $\ab^{(J+1)} \in \rch_{\ab^{(J)}}$. 

Therefore, $\ab^{(j+1)} \in \rch_{\ab^{(j)}}$ for all $j \in [k]$. By Proposition \ref{prop:reach_subset} (transitive reachability), this further implies $\ab^{(k)} \in \rch_{\ab}$. 
Since 
\[ 
\cb \triangleq \ab + t(\pb - \ab) 
= \ab + \dfrac{t}{1 - r^k} \left(\ab^{(k)} - \ab\right), 
\]

we know by $ \dfrac{t}{1 - r^k} \le 1$ that $\cb$ is on segment $\left[ \ab,~ \ab^{(k)} \right]$. By Lemma \ref{lemma:ray}, we have $\ab + t(\pb - \ab) = \cb \in \rch_{\ab}$. \qeda

\begin{corollary}
    If $\xib \in \dir_{\ab}$ is a feasible direction, $\{ \bb~ |~ \bb \in \interior,~ \exists t \ge 0 ( \bb = \ab + t \xib) \} \subseteq \rch_{\ab}$. 
    \label{cor:interior_reachable}
\end{corollary}

\annot{All interior states are reachable along a feasible direction. This, in turn means that if all directions are feasible at an interior point, all states in $\allb$ are reachable. Relevant
in Theorem \ref{thm:dlk_free} to deduce throughput insensitivity from non-existence of deadlocks.}

\begin{lemma}
    For balance states $ \ab, \bb  \in \allb$,
    if $Z_{\bb} \subseteq Z_{\ab}$, then $\dir_{\ab} \subseteq \dir_{\bb}$.
    
        
        
    
    \label{lemma:direction_subset}
\end{lemma}

\annot{
    If balance state $ \bb $ is ``further interior'' than $\ab$, then $ \bb $ has all  feasible directions of $ \ab $. 
    Effectively, directions feasible at any corner point are feasible from any interior point allowing us to show that deadlock-freeness implies throughput insensitivity (Theorem. \ref{thm:dlk_free}).}

\paragraph{Proof.} For $\xib = \zero{} \in \dir_{\ab}$, we trivially obtain that $\xib = \zero{} \in \dir_{\bb}$. 

For any $\xib \in \dir_{\ab} - \{\zero{}\}$, there exists $t > 0$ and $\pb \in \rch_{\ab}$, where $\pb = \ab + t \xib$. 
Take 
\[ 
 \epsilon = \min_{i \notin Z_{\ab}} \dfrac{ \mvec{\bb}{\capvec - \bb}_i }{ \mvec{\ab}{\capvec - \ab}_i }.
\]

By $Z_{\bb} \subseteq Z_{\ab}$, we have $\epsilon > 0$ and $\epsilon \mvec{\ab}{\capvec - \ab} \leq \mvec{\bb}{\capvec - \bb}$. 
By Lemma \ref{lemma:reachable_states_mapping}, we have for all $\pb \in \rch_{\ab}$, $\bb + \epsilon t \xib = \bb + \epsilon (\pb - \ab) \in \rch_{\bb}$. 
Given $\epsilon t > 0$, we have $\xib \in \dir_{\bb}$. Therefore, $\dir_{\ab} \subseteq \dir_{\bb}$. \qeda

\begin{lemma}
    Let $\phi(\bb; \Pi)$ denote the throughput starting from initial state $\bb$ under set of feasible \textbf{paths} $\Pi$. For any state $\bb$, if $\Pi' \subseteq \Pi$,
    \begin{equation}
        \phi(\bb; \Pi) \ge \phi(\bb; \Pi').
    \end{equation}
    
    Let $ \feasflow_{\bb} \subseteq \allr^{|\Pi|},  \feasflow_{\bb}' \subseteq \allr^{|\Pi'|}$ denote the set of feasible flows from $\bb$ under $\Pi, \Pi'$, respectively. 
    Throughput equality is attained when all paths in $\Pi - \Pi'$ are unused or in other words, if $\fb \in \feasflow_{\bb}$, 
    $f_\pi = 0$ holds for any $\pi \notin \Pi'$. 
    \label{lemma:tpt_hierarchy}
\end{lemma}

\annot{
    As more paths and flows become feasible, the optimal throughput can only improve.  The proof uses the fact that
     any flows feasible with fewer path choices are still feasible with more path choices. As a result, with more path choices, the set of reachable states expands. The resulting $\psi$ from maximizing one-step throughput across all feasible flows with more path choices is also higher.
     The two factors combined result in a higher $\phi$.
    If however, some of these additional paths are completely unused, then the throughput is no better. Under this special case, we can show that the set of useful flows are identical resulting in identical reachable state sets and $\psi$ values.
}

\paragraph{Proof.} Let $\rch_{\bb}, \rch_{\bb}'$ denote the set of reachable states from $\bb$, under feasible paths $\Pi, \Pi'$, with routing matrices $R, R'$ respectively. \\
Since, $\Pi' \subseteq \Pi$ implies there exists a matrix $P \in \{0, 1\}^{|\Pi| \times |\Pi'|}$, such that $R' = RP$. 
Note that every row of $P$ can have \emph{utmost} one non-zero entry and every column of $P$ has \emph{exactly} one non-zero entry. In particular, $P_{ij} = 1 $ if $\Pi_i = \Pi_j'$ and $0$ otherwise. 
%
Consequently, for all $\gb \in \allr^{|\Pi'|}$, 
there exists an $\fb = P \gb \in \allr^{|\Pi|}$ satisfying $R \fb = RP \gb = R' \gb$ and $\Delta R \fb = \Delta R' \gb$.  \\

    %
    

\NewPara{Mapping the reachable states}. \\
For any $\pb \in \rch_{\bb}'$, there exists a $k \in \alln$, a sequence of flows $\gb^{(0:k-1)}$ and a sequence of transient states $\bb^{(0:k)}$, where $\bb^{(0)} = \bb$, $\bb^{(k)} = \pb$, and for all $j \in [k]$,
\[ \bb^{(j+1)} = \bb^{(j)} - \Delta R' \gb^{(j)},\quad 
\gb^{(j)} \geq \zero{}, \quad
R' \gb^{(j)} \leq \mvec{\bb^{(j)}}{\capvec - \bb^{(j)}}.
\]

Now let $\fb^{(j)} = P \gb^{(j)}$ for each $j \in [k]$. We may directly obtain
\[ \bb^{(j+1)} = \bb^{(j)} - \Delta R \fb^{(j)},\quad 
\fb^{(j)} \geq \zero{}, \quad
R \fb^{(j)} \leq \mvec{\bb^{(j)}}{\capvec - \bb^{(j)}}.  \]

This implies $\fb = \sum_{j=0}^{k-1} P \gb^{(j)} = P \gb$ is also feasible that leads to a reachable state $\pb = \bb - \Delta R' \gb = \bb - \Delta R \fb$. 
Hence, $\pb \in \rch_{\bb}$, and since $\pb$ is an arbitrary element in $\rch_{\bb}'$, we have $\rch_{\bb'} \subseteq \rch_{\bb}$.

    %


\NewPara{Mapping Throughput Functions}. \\
For any $\gb \in \allr^{|\Pi'|}$, there exists $\fb = P \gb \in \allr^{|\Pi|}$, where $\one{}\T \fb = \one{}\T P \gb = \one{}\T \gb$ and $R \fb = RP \gb = R' \gb$. 
Hence, $\big( \gb \geq \zero{},~ R'\gb \leq \mvec{\bb}{\capvec - \bb},~ \Delta R' \gb = \zero \big) \Longrightarrow \big( \fb \geq \zero{},~ R \fb \leq \mvec{\bb}{\capvec - \bb}, ~ \Delta R \fb = \zero \big )$, which further indicates
\[ 
\Psi' \triangleq 
\left\{\one{}\T \gb \middle| \gb \geq \zero{},~ R'\gb \leq \mvec{\bb}{\capvec - \bb},~ \Delta R' \gb = \zero{} \right\} \subseteq 
\left\{\one{}\T \fb \middle| \fb \geq \zero{},~ R \fb \leq \mvec{\bb}{\capvec - \bb}, ~ \Delta R \fb = \zero{} \right\} 
\triangleq \Psi. \]

Let $\psi(\ab; \widetilde \Pi)$ denote the one-step maximum throughput at state $\ab$ under arbitrary set of feasible flows $\widetilde \Pi$.  By definition of $\psi$, we have 
\[ \psi(\ab; \Pi) = \sup \Psi \ge \sup \Psi' = \psi(\ab; \Pi'). \]

\NewPara{Equality Case}.
        Assume for any feasible flow $\fb \in \feasflow_{\bb}$, $f_\pi = 0$ holds for any $\pi \notin \Pi'$. Based on the construction of $P$, $(PP\T)_{\pi,\pi} = 0$ if $\pi \notin \Pi'$ and $(PP\T)_{\pi,\pi} = 1$ if $\pi \in \Pi'$ . So,
        for this special case, we can argue $PP\T \fb = \fb$.

        Let $\pb$ be an arbitrary state in $\rch_{\bb}$. Then, there exists $\fb \in \feasflow_{\bb}$ such that $\pb = \bb - \Delta R \fb$. 
        Construct $\gb = P\T \fb$, such that $- \Delta R' \gb = - \Delta R P P\T \fb = - \Delta R \fb$. 
        $\gb \in \feasflow_{\bb}'$. Similar to the previous argument, $\pb = \bb  - \Delta R \fb = \bb  - \Delta R' \gb \in \rch_{\bb}'$. Thus, $\rch_{\bb} \subseteq \rch_{\bb}'$. 
        This, with the previously proved $\rch_{\bb}' \subseteq \rch_{\bb}$ implies $\rch_{\bb} = \rch_{\bb}'$.

        %
        Further, $P \gb = PP\T \fb = \fb$. Thus, for any $\fb$ with $\big( \fb \geq \zero{},~ R \fb \leq \mvec{\bb}{\capvec - \bb}, ~ \Delta R \fb = \zero \big )$
        %
        %
        we have $\one{}\T \gb = \one{}\T P\T \fb = \one{}\T \fb$, $R' \gb = R P \gb = R \fb$, and $\big( \gb \geq \zero{},~ R'\gb \leq \mvec{\bb}{\capvec - \bb},~ \Delta R' \gb = \zero \big)$. 
        This indicates 
        \[ 
        \Psi =
        \left\{\one{}\T \fb \middle| \fb \geq \zero{},~ R \fb \leq \mvec{\bb}{\capvec - \bb}, ~ \Delta R \fb = \zero \right\} \subseteq 
        \left\{\one{}\T \gb \middle| \gb \geq \zero{},~ R'\gb \leq \mvec{\bb}{\capvec - \bb},~ \Delta R' \gb = \zero \right\} 
        = \Psi'. 
        \]
        
        Given $\Psi' \subseteq \Psi$, we have $\Psi = \Psi'$ and 
        \[ \psi(\ab; \Pi) = \sup \Psi = \sup \Psi' = \psi(\ab; \Pi'). \]

\NewPara{In summary,}
\[ 
\phi(\bb; \Pi) = 
\sup_{\ab \in \rch_{\bb}} \psi(\ab; \Pi) 
\stackrel{(*)}{\ge} \sup_{\ab \in \rch_{\bb}} \psi(\ab; \Pi') 
\stackrel{(*)}{\ge} \sup_{\ab \in \rchfreeorig_{\bb}} \psi(\ab; \Pi') = 
\phi(\bb; \Pi').  \]

When equality condition holds, both inequalities pointed by $(*)$ are tight. \qeda

\begin{lemma}
    Assume the set of feasible paths $\Pi$ is constant. 
    Let $\dlk_{\bb}$ denote the set of deadlocked channels under balance state $ \bb $. 
    If state $\ab$ satisfies $a_i = b_i, \forall i \in \dlk_{\bb}$, then $\dlk_{\bb} \subseteq \dlk_{\ab}$. \label{lemma:dlk_subset}
\end{lemma}

\annot{
    If a set of channels is deadlocked under some initial state, then they will be still deadlocked if their initial balances are kept the same. The proof below constructs a special state $\bb'$ that reachable from $\bb$ where the only imbalanced channels are those that are imbalanced in $\ab$ and $\bb$. It then uses contradiction to show that if there existed some channels that are deadlocked in $\bb$, but not in $\ab$, the directions at $\ab$ that change the balances of such channels are feasible from $\bb'$. Thus, in moving from $\bb$ to $\bb'$, the deadlocked channels can be "un-deadlocked".
}

\def\dlkdiff{\dlk_{\Delta}}

\paragraph{Proof.} We prove the statement by contradiction. Assume the set $ \dlkdiff{} \triangleq \dlk_{\bb} - \dlk_{\ab} \neq \varnothing$. 
By definition, there exists $\ab' \in \rch_{\ab}$, such that $\forall j \in \dlkdiff{}: a'_j \neq a_j = b_j$. This also means  $\exists \fb \in \feasflow_{\ab}$, such that $- \Delta R_j \fb \neq 0$ for all $j \in \dlkdiff{}$. 

    
    %
    %

Since deadlocked channels' balances are constant across all reachable states (Prop. \ref{prop:dlk_channel_alt_def}) and we can move to any arbitrary interior state along feasible directions (Thm. \ref{thm:reachable_along_dir}), $\exists \bb' \in \rch_{\bb}$, such that $\forall i \in \dlk_{\bb}: b'_i = b_i$ and $\forall j \notin  \dlk_{\bb} : b'_j \neq b_j$,  $0 < b'_j < \capscalar_j$. 

Since $a_i = b_i = b'_i$ for all $i \in \dlk_{\bb}$,
$Z_{\bb'} \subseteq Z_{\ab}$, which leads to $\dir_{\ab} \subseteq \dir_{\bb'}$ by Lemma \ref{lemma:direction_subset}. 

Note that the direction $\ab' - \ab$ corresponds to changing the balances of the channels in $\dlkdiff$. Since $\ab' \in \rch_{\ab}$, $\ab' - \ab \in \dir_{a} \subseteq \dir_{\bb'}$. This means we can construct a new state $\bb'' \in \rch_{\bb'}$ also with changed balances for the channels in $\dlkdiff$. In other words, $t > 0$, $b''_j = b'_j - t \Delta R_j \fb \neq b'_j = b_j$ for all $j \in \dlkdiff{}$. 
Further, transitivity of reachable states (Prop. \ref{prop:reach_subset}) implies $\bb'' \in \reachb$. 

We have constructed $\bb''$, a state reachable from $\bb$ but with changed balances for all
channels in $\dlkdiff \subseteq \dlk_{\bb}$. This contradicts the definition of deadlocked states.
Therefore, $\dlk_{\bb} - \dlk_{\ab} = \varnothing$, which implies $\dlk_{\bb} \subseteq \dlk_{\ab}$. \qeda

\begin{lemma}
    For any state $\ab$ and $\bb$, there exists a state $\cb$, where $\dlk_{\cb} \supseteq \dlk_{\ab} \cup \dlk_{\bb}$. 
    \label{lemma:dlkunion}
\end{lemma}

\annot{
    The deadlocked channels of any two initial states can be combined to generate a new state with a larger deadlock.
    To prove this, we construct a new state $\cb$ by preserving the deadlocked channel states from $\ab$ and porting the balances for all other channels from $\bb$. Since all flows feasible from $\cb$ are also feasible from $\bb$, $\cb$ is just as deadlocked as $\bb$.
    %
}

\paragraph{Proof.} We construct $\cb$ where $\forall i \in \dlk_{\ab}: c_i = a_i$ and $\forall i \notin \dlk_{\ab} : c_i = b_i$. 
By Lemma \ref{lemma:dlk_subset}, we immediately obtain $\dlk_{\ab} \subseteq \dlk_{\cb}$.

Consider any $\fb \in \feasflow_{\cb}$. There exists $k \in \alln$, a sequence of flows $\left( \fb^{(j)} \right)_{j=0}^{k-1}$ and a sequence of transient states $\left(\cb^{(j)}\right)_{j=0}^k$, where $\fb = \sum_{j=0}^{k-1} \fb^{(j)}$, $\cb^{(0)} = \cb$, and for each $j \in [k]$, 
\[ \cb^{(j+1)} = \cb^{(j)} - \Delta R \fb^{(j)},\quad 
\fb^{(j)} \geq \zero{}, \quad
R \fb^{(j)} \leq \mvec{\cb^{(j)}}{\capvec -\cb^{(j)}}.  \]

Let $\bb^{(0)} = \bb$ and for each $j \in [k]$,
if $\fb^{(j)}$ is feasible,
$ \bb^{(j+1)} = \bb^{(j)} - \Delta R \fb^{(j)} $. 
We show that $\fb^{(j)}$ is indeed feasible by considering 
an arbitrary channel $i \in [|E|]$. 

\begin{itemize}
    \item $i \in \dlk_{\ab}$: By construction and Lemma  \ref{lemma:dlk_subset}, $i \in \dlk_{\cb}$.
    %
    %
    Since a deadlocked channel can never sustain any flow in either direction in any feasible flow (Prop. \ref{prop:zero_flow}.\ref{prop:zero_flow:subflow}),  $R_i \fb^{(j)} = R_{i+|E|} \fb^{(j)} = 0$ for each $j \in [k]$. 
    Therefore, $R_i \fb^{(j)} = 0 \le \mvec{\bb^{(j)}}{\capvec - \bb^{(j)}}_i = b^{(j)}_i$ and $R_{i+|E|} \fb^{(j)} = 0 \le \mvec{\bb^{(j)}}{\capvec - \bb^{(j)}}_{i+|E|} = C_i - b^{(j)}_i $ for each $j \in [k]$. 
    
    \item $i \notin \dlk_{\ab}$: Starting with $c^{(0)}_i = c_i = b_i = b^{(0)}_i$, we have for all $j \in [k] \cup \{k\}$,
    \[ 
        c^{(j)}_i = c_i - \Delta R_i \sum_{j'=0}^{j-1} \fb^{(j')} = b_i - \Delta R_i \sum_{j'=0}^{j-1} \fb^{(j')} = b^{(j)}_i. 
    \]
    
    Since $\fb^{(0)}$ is feasible from $\bb$ trivially, by extension, for $j \in [k]$,  $R_i \fb^{(j)} \le \mvec{\cb^{(j)}}{\capvec - \cb^{(j)}}_i = b^{(j)}_i$ 
    and $R_{i+|E|} \fb^{(j)} \leq \mvec{\cb^{(j)}}{\capvec - \cb^{(j)}}_{i+|E|} = C_i - b^{(j)}_i$.
\end{itemize}

To sum up, for all $\fb \in \feasflow_{\cb}$, we have $\fb \in \feasflow_{\bb}$. 
Since for all $\fb \in \feasflow_{\bb}$, channels in $\dlk_{\bb}$ experience no change in balance (Def. \ref{def:dlk_channels}), we have $\dlk_{\bb} \subseteq \dlk_{\cb}$. 
Combining it with $\dlk_{\ab} \subseteq \dlk_{\cb}$, we justify the lemma. \qeda

\subsection{Formal Proof of Theorem \ref{thm:dlk_free}}
\label{apdx:deadlock free proof}

\paragraph{Proof.} For every corner state $ \cb\brk$, let $\pb\brk$ be the corresponding interior point such that $\pb \brk \in \rch_{\cb\brk}$ and $\pi \in \big[2^{|E|}\big]$.  Let $\xib\brk = \pb\brk - \cb\brk$ be the state offset. 
By definition of set of directions $\dir_{\cb\brk}$, we have offset $\xib\brk \in \dir_{\cb\brk}$. For each channel $i$, we have either $i \in Z_{\cb\brk}$ or $i + |E| \in Z_{\cb\brk}$ since one of the two
ends of every channel has 0 balance.


\begin{itemize}
    \item If $i \in Z_{\cb\brk}$, $c\brk_i = 0$. By $0 < p\brk_i < \capscalar_i$, we have $\xi\brk_i > 0$;
    
    \item If $i + |E| \in Z_{\cb\brk}$, $c\brk_i = \capscalar_i$. By $0 < p\brk_i < \capscalar_i$, we have $\xi\brk_i < 0$.
\end{itemize}

State of one channel on the corner determines the $\pm$ sign of its corresponding $\delta$ entry. 
Each open orthant in $\allr^{|E|}$ contains exactly one $\xib\brk$ vector. Across all the corners, all of the $2^{|E|}$ (open) orthants in $\allr^{|E|}$ are covered.


\NewPara{Interior Points}. Consider an arbitrary interior state $\pb \in \interior$.
By Lemma \ref{lemma:direction_subset}, $\xib\brk \in \dir_{\cb\brk} \subseteq \dir_{\pb}$ for all $\pi \in \big[2^{|E|}\big]$, because $Z_{\cb\brk} \supseteq \varnothing = Z_{\pb}$. 
By putting all the $2^{|E|}$ vectors $\big\{ \xib\brk \big\}$ in its columns, we construct a matrix $\Xi$. 


We claim that for any $\xib \in \allr^{|E|}$, there exists $\fb \geq \zero{}$, such that $\Xi \fb = \xib$. 
Let us assume otherwise. By Farkas' Lemma, there must exist a $\gammab \in \allr^{|E|}$, such that $\Xi\T \gammab \geq \zero{}$ and $\xib\T \gammab < 0$.
If $\gammab \neq \zero{}$, because columns of $\Xi$ covers all the orthants, there must exist a column $\xib'$ in $\Xi$, where for all $i$, if $\gamma_i \neq 0$, then $\xi'_i \gamma_i < 0$. 
This gives us a negative entry in $\Xi\T \gammab$, contradicting $\Xi\T \gammab \geq \zero{}$, so $\gammab$ must equal $\zero{}$. However, this contradicts $\xib\T \gammab < 0$. 

Therefore, there exists $\fb \geq \zero{}$ where $\Xi \fb = \xib$ for any $\xib \in \allr^{|E|}$. Since $\xib \in \dir_{\pb} \forall \xib$, by convexity of the feasible direction set (Prop. \ref{prop:direction_conv}), $\dir_{\pb} = \allr^{|E|}$. 

Since all interior states along a feasible direction are reachable (Corollary \ref{cor:interior_reachable}) and $\dir_{\pb} = \allr^{|E|}$, $\interior \subseteq \rch_{\pb}$. Thus, $\capvec / 2$ is reachable from any interior point $\pb$. 



\NewPara{Boundary Points}. We consider an arbitrary boundary state $\qb$ with a corner $\cb\brk$ where $Z_{\cb\brk} \supseteq Z_{\qb}$. 
Let $\underline{q}$ denote the minimum positive entry in vector $\mvec{\qb}{\capvec - \qb}$. We construct
\[ \epsilon = 0.9 \min\left\{1,~ \frac{ \underline{q}}{\|\xib\brk\|_\infty } \right\}. \]

\begin{itemize}
    \item For $i \notin Z_{\qb}$, we have 
    \[ 0 \le q_i - \frac{q_i \big|\xi\brk_i\big|}{\|\xib\brk\|_\infty} < q_i + \epsilon \xi\brk_i < q_i + \frac{(\capscalar_i - q_i) \big|\xi\brk_i\big|}{\|\xib\brk\|_\infty} \le \capscalar_i; \]
    
    \item For $i \in Z_{\qb}$ and $i \le |E|$, we have $q_i = c\brk_i = 0$, $\xi\brk_i > 0$, and
    \[ 0 < \epsilon \xi\brk_i = q_i + \epsilon \xi\brk_i < q_i + \xi\brk_i = c\brk_i + \xi\brk_i = p\brk_i < \capscalar_i; \]
    
    \item For $i \in Z_{\qb}$ and $i > |E|$, we have $0 < q_i + \epsilon \xi\brk_i < \capscalar_i$ analogously.
\end{itemize}

Effectively, $\qb + \epsilon \xib\brk$ is an interior state which lies on the direction  $\xib\brk$ from $\qb$.
Since $\qb$ is less constrained than $\cb \brk$, we know $\xib\brk \in \dir_{\qb}$ (Lemma \ref{lemma:direction_subset}). By Corollary \ref{cor:interior_reachable}, the interior state $\qb + \epsilon \xib\brk \in \rch_{\qb}$ since it is along a feasible direction. 

The previous argument shows that $\capvec / 2$ is reachable from any interior state. Thus, $\capvec / 2 \in \rch_{\qb + \epsilon \xib\brk}$. 
By transitivity of reachable states, (Proposition \ref{prop:reach_subset}), $\rch_{\qb + \epsilon \xib\brk} \subseteq \rch_{\qb}$, 
which implies $\capvec / 2 \in \rch_{\qb}$, i.e., $\capvec / 2$ is also reachable from the boundary state $\qb$. 

\NewPara{Throughput Insensitivity}. Since $\capvec / 2$, the global $\psi$ maximizer by Proposition \ref{prop:max_psi}, is reachable from any state $\pb \in \allb$, 
\[ \phi(\pb) = \sup_{\qb \in \rch_{\pb}} \psi(\qb) = \psi\left(\frac{\capvec}{2} \right). \qeda \]

\subsection{Proof of Theorem \ref{thm:worst_corner}}

\label{apdx:subsec:worst_corner}

\paragraph{Proof.} 

By Lemma \ref{lemma:dlkunion}, we may assert the existence of a grand deadlocking state $\bb$, where $\dlk_{\ab} \subseteq \dlk_{\bb}$ for any other state $\ab$. 
In such cases, $|\dlk_{\bb}|$ is maximized. 

\NewPara{Filtering out unused paths and deadlocked edges}. Let $\efree = E - \dlk_{\bb}$ and $\ffree = \{\pi \in [\Pi]~ |~ \forall j \in \dlk_{\bb}: R_{j, \pi} = R_{j+|E|, \pi} = 0 \}$. 
Based on these subsets, we construct matrices $P \in \{0, 1\}^{|E| \times |\efree|}$ and $Q \in \{0, 1\}^{|\Pi| \times |\ffree|}$ that extracts the rows in $\efree$ and columns in $\ffree$, respectively. 
Every row of $P, Q$ can have \emph{utmost} one non-zero entry and every column of $P, Q$ has \emph{exactly} one non-zero entry. In particular, 
\[
    P_{ij} = \left\{ \begin{array}{lr}
        1, & E_i = (\efree)_j; \\
        0, & \text{otherwise}.
    \end{array} \right., \qquad
    Q_{ij} = \left\{ \begin{array}{lr}
        1, & \Pi_i = (\ffree)_j; \\
        0, & \text{otherwise}.
    \end{array} \right..
\]

In addition, we define $\hat P = \begin{bmatrix} P & P \end{bmatrix}$ and take submatrices $\widetilde R = \hat P\T R Q$ and $\Delta \widetilde R = P\T \Delta R Q$. 
    
    
    
    

We point out that for all $i \in \dlk_{\bb}$ and $\gb \in \allr^{|\ffree|}$, we have the $\pi$-th column $Q_\pi = \zero{}$ for all $\pi : R_{i, \pi} = 1$. Therefore $R_i Q \gb = 0$ and analogously, $R_{i+|E|} Q \gb = 0$. 
An important implication is that any vector $RQ \gb$ has $0$ entries on the positions that the selecting matrix $\hat P$ throws away. 
Similarly, any vector $\Delta R Q \gb$ has $0$ entries on the positions that $P$ throws away. 
Thus, for any $\xb \in \allr^{2|E|}$ and $\yb \in \allr^{|E|}$, we have the following bijection.

\[ 
    \xb = R Q \gb ~\Longleftrightarrow~ \hat P\T \xb = \hat P\T R Q \gb = \widetilde R \gb , \qquad 
    \yb = \Delta R Q \gb ~\Longleftrightarrow~ P\T \yb = P\T \Delta R Q \gb = \Delta \widetilde R \gb .
\]

These equalities support \eqref{eqn:tpt_equivalence}. 


\NewPara{Throughput Mapping}. Since some of the edges used by flows outside of $\ffree$ are deadlocked, by Proposition \ref{prop:zero_flow}.\ref{prop:zero_flow:flow_size}, we have $\forall \pi \notin \ffree: f_\pi = 0$, as long as $\fb \in \feasflow_{\bb}$. 
Since these $0$ flows cannot contribute throughput, by Lemma \ref{lemma:tpt_hierarchy}, 
\begin{equation}
    \phi(\bb; \Pi) = \phi(\bb; \ffree) \label{eqn:tpt_equality}
\end{equation}

    

Let $\rchfreeorig_{\ab}$ denote the set of reachable states in $\allr^{|E|}$ (without removing channels outside $\efree$) using only paths in $\ffree$. 
For any state $\ab$ (not necessarily equal to $\bb$), 
\begin{align}
    \phi(\ab; \Pi) & \ge \phi(\ab; \ffree) \notag \\
    & = \sup_{\ab' \in \rchfreeorig_{\ab}} \psi(\ab; \ffree) \notag \\
    & = \sup_{\ab' \in \rchfreeorig_{\ab}} \left\{ \one{}\T \gb~ \big|~ 
        \gb \geq \zero,~
        R Q \gb \leq \mvec{\ab'}{\capvec - \ab'},~ 
        \Delta R Q \gb = \zero \right\} \notag \\
    & = \sup_{\ab' \in \rchfreeorig_{\ab}} \left\{ \one{}\T \gb~ \big|~ 
        \gb \geq \zero,~ 
        \widetilde R \gb \leq \hat P \mvec{\ab'}{\capvec - \ab'},~ 
        \Delta \widetilde R \gb = \zero \right\} \label{eqn:tpt_equivalence} \\
    & = \widetilde \phi(P\T \ab; \ffree), \label{eqn:tpt_inequality}
\end{align}

where $\widetilde \phi(P\T \ab; \ffree)$ describes the throughput at initial state $P\T \ab$ on a \cn{} characterized by routing matrix $\widetilde R$. 

%
\NewPara{Reachable States Equivalence}. Let $\allbfree = \big\{ \xb \in \allr^{|\efree|} \big| 0 \leq \xb \leq P\T \capvec \big\}$ and $\interiorfree{} = \big\{ \xb \in \allr^{|\efree|} \big| 0 \prec \xb \prec P\T \capvec \big\} $.
We also let $\rchfree{}_{\qb} \subseteq \allr^{|\efree|}$ denote the set of reachable states in the \cn{} (after removing channels outside $\efree$) from state $\qb$.

    

Now we prove that for any state $\ab, \cb \in \allb$, if $\ab \in \rchfreeorig_{\cb}$, then $P\T \ab \in \widetilde \rch_{P\T \cb}$. 

Let $k \in \alln$ and $\left( \fb^{(j)} \right)_{j=0}^{k-1}$ be the sequence of flows where $\cb^{(0)} = \cb,~ \cb^{(k)} = \ab$, and for all $j \in [k]$, 
\[ 
    \cb^{(j+1)} = \cb^{(j)} - \Delta R \fb^{(j)}, \quad
    \fb^{(j)} \geq \zero, \quad 
    R \fb^{(j)} \leq \mvec{\cb^{(j)}}{\capvec -\cb^{(j)}}. 
\]

Since $\ab \in \rch_{\cb}'$, no flow outside $\ffree$ is used, and thus, $\fb^{(j)}_\pi = 0$ for all $\pi \notin \ffree$. 
As a result, $Q Q\T \fb^{(j)} = \fb^{(j)}$, and furthermore, 
\[ 
    P\T \Delta R \fb^{(j)} = P\T \Delta R Q Q\T \fb^{(j)} = \Delta \widetilde R Q\T \fb^{(j)}, \qquad
    \hat P\T R \fb^{(j)} = \hat P\T R Q Q\T \fb^{(j)} = \widetilde R Q\T \fb^{(j)}.
\]

By left-multiplying $P\T $ (or $\hat P\T$), we directly obtain $P\T \cb^{(0)} = P\T \cb$, $P\T \cb^{(k)} = P\T \ab$, and for all $j \in [k]$,
\[ 
    P\T \cb^{(j+1)} = P\T \cb^{(j)} - \Delta \widetilde R Q\T \fb^{(j)}, \quad
    Q\T \fb^{(j)} \geq \zero, \quad 
    \widetilde R Q\T \fb^{(j)} \leq \hat P \mvec{\cb^{(j)}}{\capvec -\cb^{(j)}}. 
\]

Hence, $\sum_{j=0}^{k-1} Q\T \fb^{(j)}$ is a feasible flow in a \cn{} with routing matrix $\widetilde R$ and initial state $P\T \cb$. 
The destination $P\T \ab \in \widetilde \rch_{P\T \cb}$.

    %

\NewPara{Throughput insensitivity of filtered \cn{}}. We consider all the corners $\big\{ \cb\brk \big\}$ where $\forall i \in \dlk_{\bb}$ and $\pi \in 2^{|\efree|} : c\brk_i = b_i$. 
From any $\cb\brk$, there exists $\qb\brk \in \rch_{\cb\brk}$ such that $\forall i \in \efree: 0 < q\brk_i < \capscalar_i$. 
Since $|\dlk_{\bb}|$ is maximized and its deadlocked channels continue to be imbalanced at $\cb\brk$, by Lemma \ref{lemma:dlk_subset}, we have $\dlk_{\cb\brk} = \dlk_{\bb} = E - \efree$. 
Thus, all paths outside $\ffree$ are not used, and we have $\qb\brk \in \rchfreeorig_{\cb}$,
which indicates $P \qb \brk \in \widetilde \rch_{P\T \cb\brk}$.

By definition, the states $\big\{ P\T \cb\brk \big\}$ are corners in $\allbfree$, while $\big\{ P \qb\brk \big\}$ are interior states in $\interiorfree$. 
So this sub-\cn{} is deadlock-free with balance state space $\allbfree$ and routing matrix $\widetilde R$.
By Theorem \ref{thm:dlk_free}, a deadlock-free \cn{} is insensitive to throughput. Thus, we know for all states $\widetilde \qb \in \allbfree$, 
$\phi(\widetilde \qb; \ffree) \equiv C$ for some constant $C \ge 0$. 


\NewPara{Summary}. Combine \eqref{eqn:tpt_equality} with \eqref{eqn:tpt_inequality}, and we get for any state $\ab \in \allb$,
\[ \phi(\ab; \Pi) \ge C = \phi(\bb; \ffree) = \phi(\bb; \Pi). \]

Hence, $\bb$ is the worst corner (more generally, state) with lowest throughput. \qeda

\nopagebreak
\subsection{Formal Proof of Theorem \ref{thm:dlk_np_hard}}
\label{apdx:proof np hard}
\paragraph{Proof.} We first prove that detecting a full deadlock on a \cn{} $G(E, V)$ is NP-hard by reducing the boolean
satisfiability (SAT) problem, a decision
problem known to be NP-complete, to deadlock-detection. We then extend the reduction to establish that
partial deadlock detection is also NP-hard.



We start with an arbitrary instance of the SAT problem, which determines satisfiability of a boolean expression in conjunctive normal form (CNF). The expression consists of $\ell$ literals ($x_1 \cdots x_{\ell}$) and $\mathcal{C}$ clauses 
($c_1 \cdots c_{\mathcal{C}}$).

\NewPara{Preprocessing the CNF expression}. We preprocess the expression to ensure that literals always appear in the same order
in every clause. 
We then map it to an equivalent expression $s'$ by inserting fresh variables ($y_i$'s) between every pair
of adjacent literals in any clause of the original CNF expression. 
For each ordered pair $i, j \ (i < j)$, there are 4 cases in total where 
$x_i$ (or $\neg x_i$) is adjacent to $x_j$ (or $\neg x_j$). 
We create a new literal $y_{i,j,*}$ for each case. 
Notice that the total \# of literals ($\ell'$) in $s'$ is polynomial in $\ell$, the original problem size.
To ensure that the new literals have no effect on the satisfiability of the original expression, 
we add new clauses (exactly $\ell' - \ell$ new clauses).
Let the total number of clauses in $s'$ be $\mathcal{C}'$ respectively.
\Tab{expression conversion} shows an example conversion for a particular CNF expression.
It is fairly straightforward to see that the original CNF $s$ and the altered expression $s'$
are equisatisfiable:  $s'$ is satisfiable if and only if $s$ is satisfiable.
\begin{table}[h!]
\centering
\begin{tabular}{ |c|c|c| }
 \hline
    Original Expr. $s$ & $(x_1 \lor x_2 \lor x_3) \land (x_4 \lor \neg x_2) \land (\neg x_3) $  \\
 Ordered  Expr.& $(x_1 \lor x_2 \lor x_3) \land (\neg x_2 \lor x_4) \land (\neg x_3)$ \\
    Fresh Variables & 
    $(x_1 \lor y_1 \lor x_2 \lor y_2 \lor x_3) \land (\neg x_2 \lor y_3 \lor x_4) \land (\neg x_3)$ \\
    Final Expr. $s'$ & 
    $(x_1 \lor y_1 \lor x_2 \lor y_2 \lor x_3) \land (\neg x_2 \lor y_3 \lor x_4) \land (\neg x_3) 
    \land (\neg y_1) \land (\neg y_2) \land (\neg y_3)$\\
 \hline
\end{tabular}
    \caption{Example showing pre-processing for a CNF expression whose satisfiability is to be determined
    before mapping it to \cn{} deadlock detection.}
\label{tab:expression conversion}
\end{table}



\begin{figure*}
        \centering
        \begin{subfigure}[b]{0.475\textwidth}
            \centering
            \includegraphics[width=\textwidth]{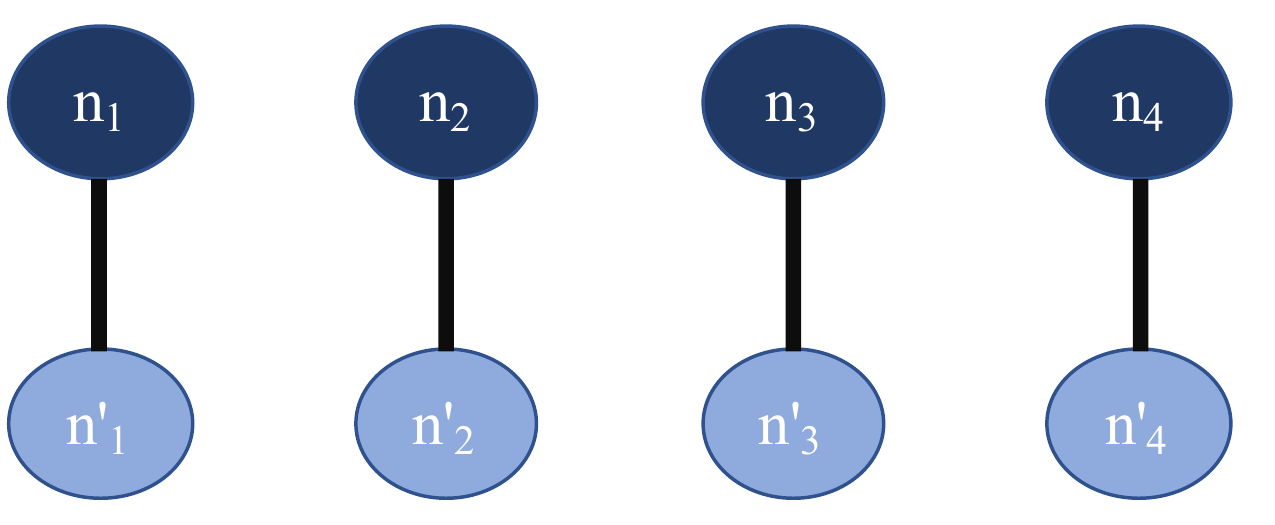}
            \caption{{\small 4 edges with their own endpoints are introduced for each of the 4 literals $x_1 \cdots x_4$.}}%
            \label{fig:reduction original edges}
        \end{subfigure}
        \hfill
        \begin{subfigure}[b]{0.475\textwidth}  
            \centering 
            \includegraphics[width=\textwidth]{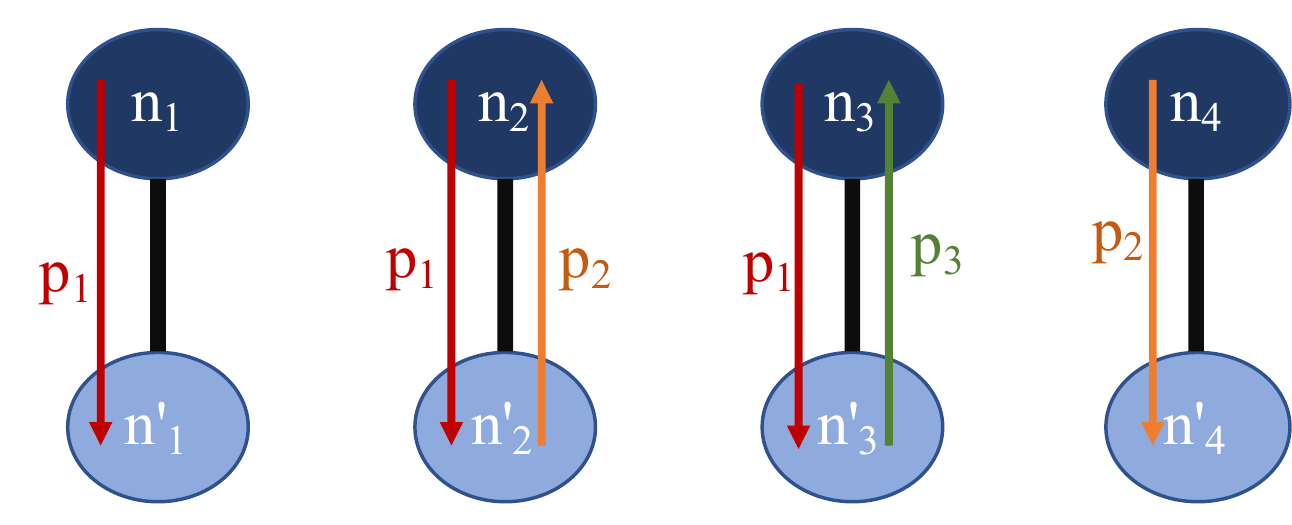}
            \caption{{\small 3 flows added based on the directions given by negation of each literal in each clause.}}
            \label{fig:reduction original flows}
        \end{subfigure}
        \vskip\baselineskip
        \begin{subfigure}[b]{0.475\textwidth}   
            \centering 
            \includegraphics[width=\textwidth]{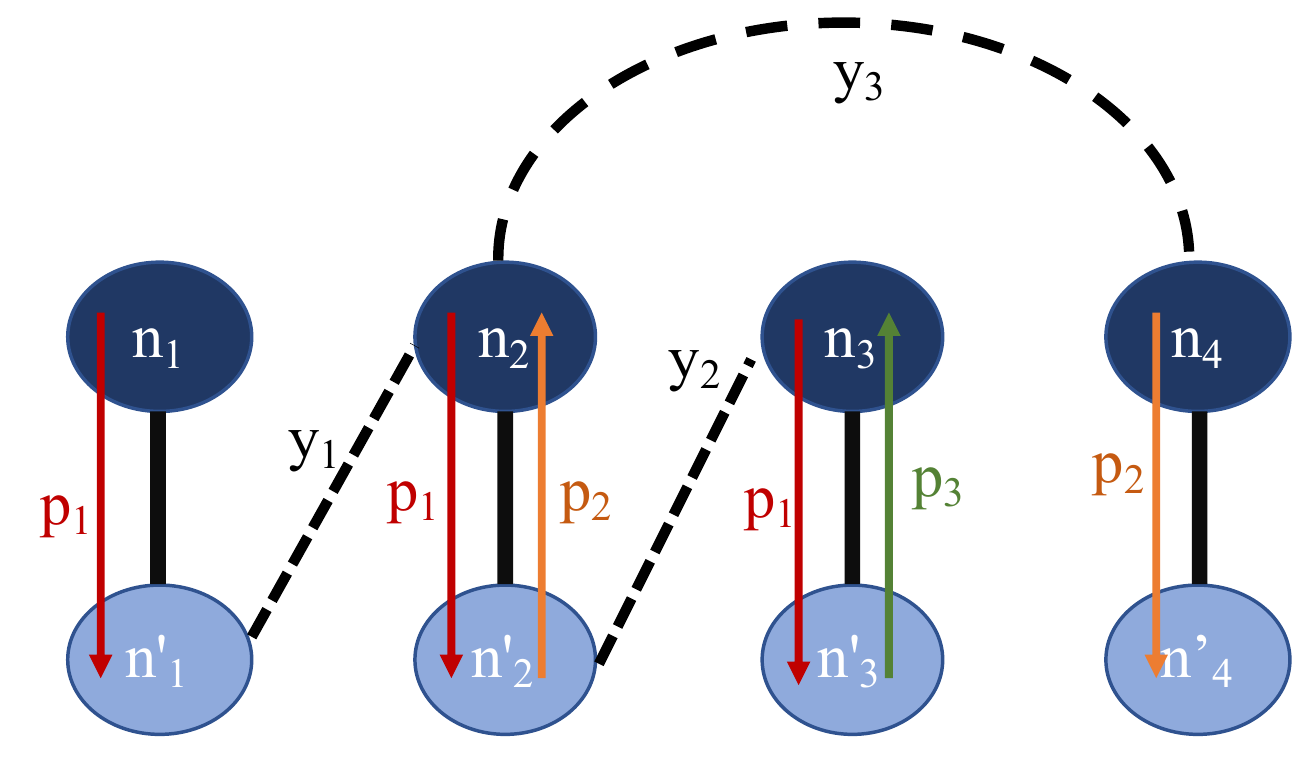}
            \caption{{\small Dashed lines denote edges $y_k$ added to connect the topology.
            Each $y_k$ edge maps to one of $4$ orientations for a pair of literals.}}
            \label{fig:reduction dummy edges}
        \end{subfigure}
        \hfill
        \begin{subfigure}[b]{0.475\textwidth}   
            \centering 
            \includegraphics[width=\textwidth]{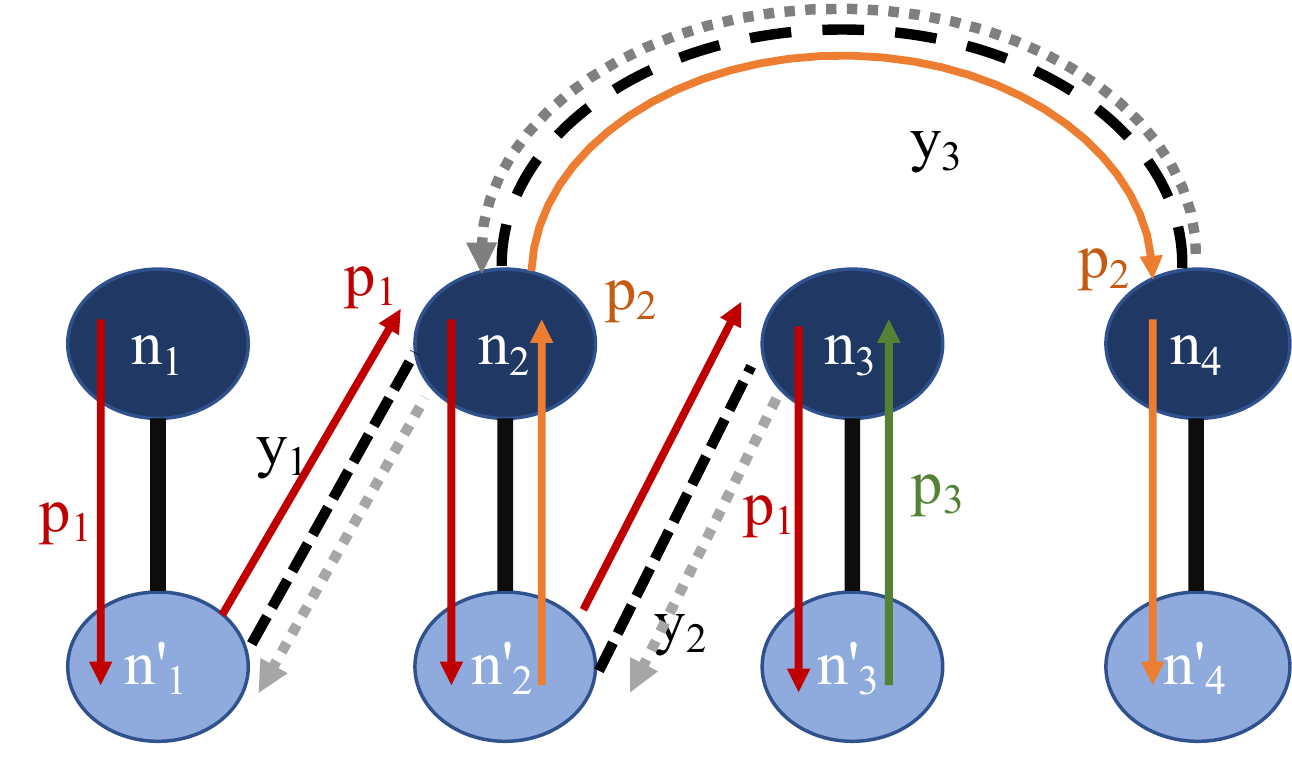}
            \caption{{\small Completed \cn{} by connecting flows corresponding to clauses  
            the left to right and using dedicated (dotted) flows on the $y_k$ edges from right to left.}}
        \label{fig:reduction full}
        \end{subfigure}
        \caption{{\small \cn{} conversion for the sample pre-processed CNF expression in \Tab{expression conversion}. 
        Deadlock detection on this \cn{}
        is equivalent to deciding whether the boolean expression is satisfiable.}}
        \label{fig:mean and std of nets}
    \end{figure*}

\NewPara{Mapping the expression to a \cn{}.}
To construct the \cn{} on which deadlocks are detected, every $x_i$ literal in the original expression
is mapped to an edge $e_i$ that is oriented in the vertical direction with two endpoints $n_i$ and $n_i'$.
\Fig{reduction original edges} shows this for the example expression in \Tab{expression conversion}.
The clauses $c_i$ dictate the paths $p_i$ on the \cn{} and the direction of use on each edge. 
If $x_i$ appears in negated form in any clause $c_i$, the path $p_i$
will send flow on the edge $x_i$ from $n_i'$ towards $n_i$ and vice-versa.
As shown in \Fig{reduction original flows}, based on clauses $c_1$ and $c_2$, $p_1$ in red
sends flow in the downward direction on both edges $e_1$ and $e_2$ 
while $p_2$ (in orange) will send flow 
in the upward direction on $x_2$ and downward on the $x_4$. 

The edges corresponding to the fresh variables $y_k$'s act as intermediary edges that help us complete paths.
If $y_k$ is a fresh variable corresponding to $x_i \lor x_j$, $y_k$ connects $n_i'$ and $n_j$ enabling a path that runs 
downward on both $x_i$ and $x_j$. We also add a \emph{dedicated flow} in the opposite direction from $n_j$ to
$n_i'$ to denote the clause $\neg y_k$. Such dedicated flows ensure that the intermediary edges can never be deadlocked
unless the original edges themselves are deadlocked.
Similarly, if $y_m$ is a fresh variable for $x_i \lor x_j'$, 
$y_m$ connects $n_i'$ to $n_j'$, enabling a path that runs 
downward on $x_i$ and upward on $x_j$. A dedicated flow mapped to the clause $\neg y_m$ is also added
from $n_j'$ to $n_i'$. This can be extended to the other two orientations of $e_i$ and $e_j$
to give four possible ways of drawing $y_k$. \Fig{reduction dummy edges} denotes the \cn{} after adding 
the fresh variables. 
Once we add the dedicated flows and connect the original paths or clauses,
the final \cn{} looks like \Fig{reduction full}.


\NewPara{Reduction.} Effectively, we have a \cn{} instance $G(E,V)$ based on the original expression that we want to
detect a full-deadlock on. In a fully deadlocked configuration for this \cn{},
the vertical edge $e_i(n_i, n_i')$ is unable to support downward flow
if all the tokens are on the $n_i'$ end (on the node that is below).
Since downward flow maps to the non-negated $x_i$, $x_i = 1$ if all the tokens are at $n_i'$ and $x_i = 0$ if the tokens
are at $n_i$.
The $y_k$ edges cannot sustain left-right flow if all the tokens are
all on its right end. Thus, $y_k = 1$ puts all its tokens on its right end.


If there exists a configuration with all edges of the above described \cn{} deadlocked,
there must be a satisfying assignment to the original boolean expression.
If all paths are deadlocked, each of the dedicated
flows (gray dashes) must be bottlenecked because tokens are on the left end of the respective $y_k$ edges 
($y_k = 0$). 
This ensures that all added clauses on the fresh variables are true.
However, if the tokens are on the left end of each $y_k$, the $p_i$ paths associated with 
the original clauses cannot be bottlenecked on any of the $y_k$ edges. Thus, for each 
path, one of the $e_i$'s must be preventing flow in the corresponding direction. 
Such an $e_i$ corresponds to a true literal in every clause
ensuring that the original boolean expression is satisfiable.

If there exists a satisfying assignment to the original boolean expression $s$, then there exists a satisfying assignment to
$s'$ also by just setting all $y_k = 0$. But, this implies a full deadlock exists on
the equivalent \cn{}. This is 
because one or more literals in every original clause must be true. The true literal in every original 
clause maps to a direction of its edge $e_i$ that bottlenecks the associated $p_i$ flow.
The independent gray flows are trivially bottlenecked
because a satisfying assignment sets $y_k = 0$ which puts all tokens on the left end of each of the $y_k$ edges.
Thus, in aggregate all paths in the \cn{} are deadlocked, thereby leaving us with a full deadlock on the edges.

To sum up, a solution to a full-deadlock problem of polynomial size constructed based on a SAT problem instance can be converted in polynomial time to a solution to the SAT problem. This yields the NP-hardness of the full-deadlock problem.

\NewPara{Partial Deadlocks}
While the above construction is for a full deadlock on the \cn{}, but adding extraneous edges with dedicated flows
that cannot be deadlocked allows us to extend the reduction to partial deadlocks.
In particular, after the above \cn{} construction with edges for an arbitrary boolean expression, 
consider adding $k'$ edges with dedicated flows in both directions on each edge that make
them all deadlock-free. The identification of a $k-k'$ deadlock on the $k$-edged final \cn{} is now equivalent to 
finding a satisfying assignment on the original boolean expression. A deadlock involving the $k - k'$ edges
map to the original literals from the expression. 
We can use this to establish finding an $k-k'$ deadlock is NP-hard too for any $k$.
In other words, finding a partial deadlock is also NP-hard. 
\qeda

\noindent

\section{Additional Details and Results}

\label{apdx:dlk ilp}

\subsection{Deadlock Detection Integer Linear Program}
We set up the following Integer Linear Program to define the dependencies
between imbalanced channels, the paths using them and the resulting set of deadlocked channels.
We define $x_{(u,v)} \in \{0, 1\}$ to denote if a channel is imbalanced 
with $x_{(u,v)} = 0 \iff b_{(u,v)} = 0$.
Since an edge can't be imbalanced on both ends,
\begin{equation}
    x_{(u,v)} + x_{(v,u)} \geq 1 \quad \forall \ \  (u,v) \in E
    \label{eqn:positive capacity}
\end{equation}

We next define $y_p \in \{0,1\}$ such that $y_p = 0$ if path $p$ is bottlenecked
on one or more of its edges and hence can't make progress. $y_p = 1$ if no edge is
imbalanced along the path. Consequently, we have
\begin{equation}
    y_p \leq x_{(u,v)} \quad \forall \ \  (u,v) \in p \quad \forall p \in \mathcal{P}
    \label{eqn:path infeasible}
\end{equation}
\begin{equation}
    y_p \geq \sum_{(u,v) \in p} x_{(u,v)} - |p| + 1 \quad \forall p \in \mathcal{P}
    \label{eqn:path feasible}
\end{equation}

Lastly, we capture whether an edge is deadlocked by defining $z_{\{u,v\}} \in \{0, 1\}$. $z_{\{u,v\}} = 0$ if the edge is deadlocked. An edge is deadlocked if and only if it is imbalanced and no path using it in either direction
can make progress in either direction. The former prevents a temporarily imbalanced but deadlock-free channel from blocking a flow, and ensures that a channel that is not at one of its imbalanced states is deadlock-free.  To enforce this, we have
\begin{equation}
    z_{\{u,v\}} = x_{(u,v)} + x_{(v,u)} - 1 \quad \forall \ \  \{u,v\} \in E
    \label{eqn:no deadlock if not imbalanced}
\end{equation}
\begin{equation}
    z_{\{u,v\}} \geq y_p \quad \forall p : (u,v) \in p \text{  or  } (v,u) \in p 
    \label{eqn:deadlocked edge}
\end{equation}

Given the above constraints, detecting largest deadlock is equivalent to 
$$\min \sum_{\{u,v\} \in E } z_{\{u,v\}} $$
We use an ILP solver like Gurobi to solve the above minimization problem to detect the largest deadlock and compare it to the unpeeled channels at the end of the \dpp.


\subsection{Pseudo-code of Peeling Algorithm}
\label{apdx:peeling psuedocode}


Algorithm \ref{alg:peeling} is a description of our graph-embedded peeling algorithm in pseudo-code form. 
It takes as input a graph topology $G = (V, E)$ and a set of demand paths $\mathcal{P}_{\text{init}}$. 
Each demand path is a collection of directed channels, where each directed channel has form $(\text{channel},~ \text{color})$. 
The first entry is the identifier of an undirected channel, and the second entry ``color'' represents the direction, which can only be either red or blue. 
For simplicity, we use $\neg \textrm{color}_0$ to represent the opposite color of $\textrm{color}_0$. 

\begin{algorithm}
    \caption{The Peeling Algorithm}
    \label{alg:peeling}
	\def\Pc{\mathcal{P}}
	\def\blue{\textrm{\color{blue}blue}}
	\def\red{\textrm{\color{red}red}}
	\def\chnl{\textrm{\color{violet}channel}}
	\def\clr{\textrm{\color{gray!50!orange}color}}
	\def\chnls{{\color{violet}\textrm{channel}'}}
	\def\clrs{{\color{gray!50!orange} \textrm{color}'}}
	\def\len{\texttt{length}}
    \def\oppo#1{{\color{green!50!blue} \neg  #1}}
	
	\Input{Graph $G = (V, E)$, set of demand paths $\Pc_{\text{init}}$}
	\Output{\textbf{Success} in peeling every channel in both directions, or \textbf{Failure}}
    
    
    Initialize $E_{\text{processed}} \leftarrow \varnothing$, $R \leftarrow \varnothing$, $\Pc \leftarrow \Pc_{\text{init}}$
    \tcp*{$R$ is the ripple}
    
    
    \For(\tcp*[f]{Add all channels with dedicated flows to ripple}){$p: [p \in \Pc] ~ \wedge ~ [\len(p) = 1] $} {
            $\Pc.\texttt{remove}(p)$\;
            
            $(\chnl,~ \clr) \leftarrow p[0]$\;
            
            $R.\texttt{add}\big( (\chnl,~ \oppo{\clr}) \big) $\;
    }
    
    \While{$R \neq \varnothing$} {
        $(\chnl,~ \clr) \leftarrow R.\texttt{pop}()$\;
        
        $E_{\text{processed}}.\texttt{add}\big( (\chnl,~ \clr) \big)$ \tcp*{Process an arbitrary directed channel in ripple}
        
        \For{$ p: [p \in \Pc]  ~ \wedge ~ [(\chnl,~ \clr) \in p] $} {
            
            $p.\texttt{remove}\big( (\chnl,~ \clr) \big)$\; 
            
            \uIf(\tcp*[f]{Release of a degree 1 flow frees the only remaining channel in the opposite direction}){ $\len(p) = 1$ } {
                $(\chnls,~ \clrs) \leftarrow p[0]$\;

                $R.\texttt{add}\big( (\chnls,~ \oppo{\clrs}) \big)$\;
            }
            \uElseIf(\tcp*[f]{Release of a degree 0 flow frees all channels it originally contained in the opposite direction}){ $\len(p) = 0$} {
                
                $p_{\text{init}} \leftarrow \text{look up for initial } p \text{ in } \Pc_{\text{init}}$\;
                
                $\Pc.\texttt{remove}(p)$\;
                
                \For{ $(\chnls,~ \clrs) \in p_{\text{init}}$ }{
                    \If{ $(\chnls,~ \oppo{\clrs}) \notin E_{\text{processed}} $ }{
                        $R.\texttt{add}\big( (\chnls,~ \oppo{\clrs}) \big)$\;
                    }
                }
            }
        }    
    
    }
    
    \If{ $\left| E_{\text{processed}} \right| = 2 |E|$ }{
        \KwRet{\textbf{Success}}\;
    } 
    \Else{
        \KwRet{\textbf{Failure}}\;
    }
\end{algorithm}

\subsection{Non-uniform Random Demand Matrix Results}
\label{apdx:nonuniform random results}

In addition to the results with demand matrices were senders and receivers are sampled uniformly at random (\Sec{sec:eval}), we investigate the 
throughput and deadlock behavior of different topologies when the senders and receivers are sampled in a skewed manner. Specifically, we designate 10\% of the 500 nodes as ``heavy-hitters'' and sample sources and destinations from them 70\% of the time when sampling a new source-destination pair for the demand matrix. 30\% of the senders and receivers (independently) are sampled from the remaining 450 infrequent nodes. We vary the number of demand pairs and measure the $\maxtpt$ and $\mintpt$ values, repeating the procedure for \Fig{overall tpt comparison}.

\begin{figure*}[]
    \centering
    \includegraphics[width=\textwidth]{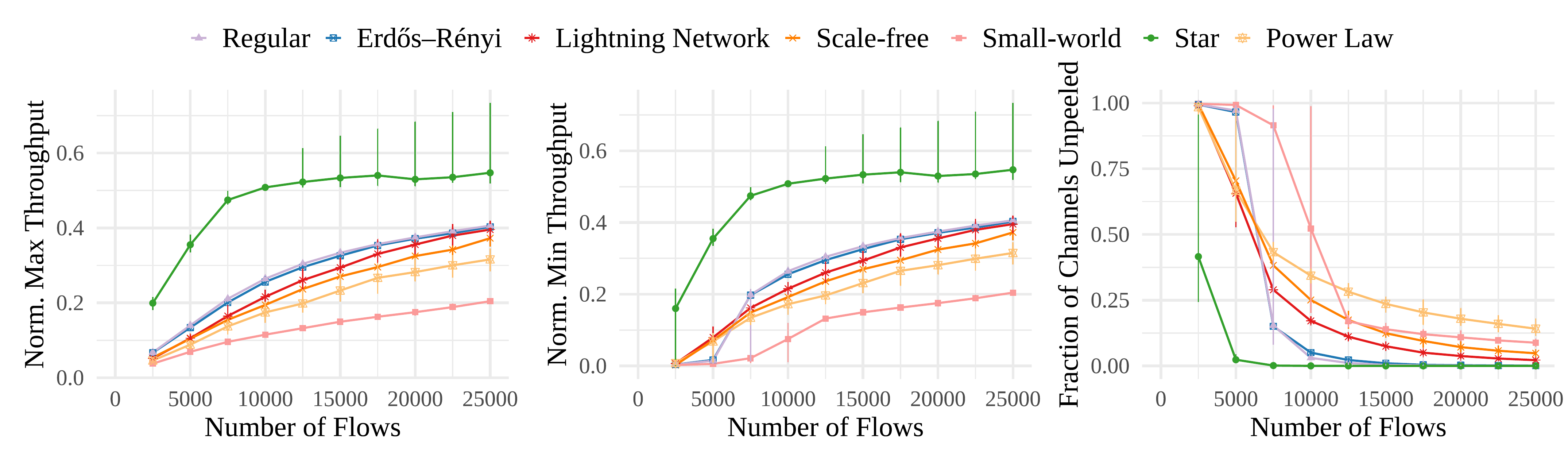}
    \caption{\small Maximum ($\maxtpt$) and minimum throughput ($\mintpt$) achieved by different
    states and the corresponding channels left unpeeled as the number of flows is varied, when flows' senders and receivers are sampled in a skewed manner. While individual throughput values are lower than the uniform random demand matrices, stars still outperform other random topologies. 
    The \lntopo, \pltopo and \sftopo graphs peel earlier and have higher $\mintpt$, but over a smaller range of demand matrix density than with a uniform random demand.
    However, the fraction of channels peeled continues to correlate well with the $\mintpt$ achieved by the topologies. Whiskers denote max and min data point.} 
    \label{fig:nonuniform random overall tpt}
\end{figure*}

\NewPara{Comparing topologies.} \Fig{nonuniform random overall tpt}  shows the variation in the $\mintpt$ values and the fraction of unpeeled channels
across topologies with the skewed demand matrix.
Compared to \Fig{overall tpt comparison}, we notice that the throughput values are lower. This is expected because the skewed sampling results in more flows that are close to each other, resulting in lower throughput due to the shared tokens across channels close to the frequent nodes.
However, we still notice that the \lntopo, \pltopo and \sftopo topologies have much better $\mintpt$ with fewer flows
when compared to the other random graphs. At 5000 demand pairs, both \lntopo and \sftopo topologies peel 25\%  more channels
than the \sw, \er, and \rr topologies and correspondingly have higher $\mintpt$. 
But, the trends start flipping once we have a few more demand pairs. The \sftopo and \pltopo graphs once again struggle to 
peel the last 10-25\% of their channels compared to other random topologies.
In contrast, the \er and \rr topologies do not peel as well with fewer flows, but quickly
improve to peel all channels, on average, with 15000 demands. 
Beyond this point, their $\mintpt$ is comparable with the \lntopo, similar to trends in the uniform random demand matrix case. It is worth noting that the star's throughput shows a slight dip with an increase in flows. This behavior depends on whether the ``heavy-hitter'' nodes includes the hub or not, which in turn affects the average path length across all flows and throughput values.

\begin{figure*}
    \centering
    \includegraphics[width=\textwidth]{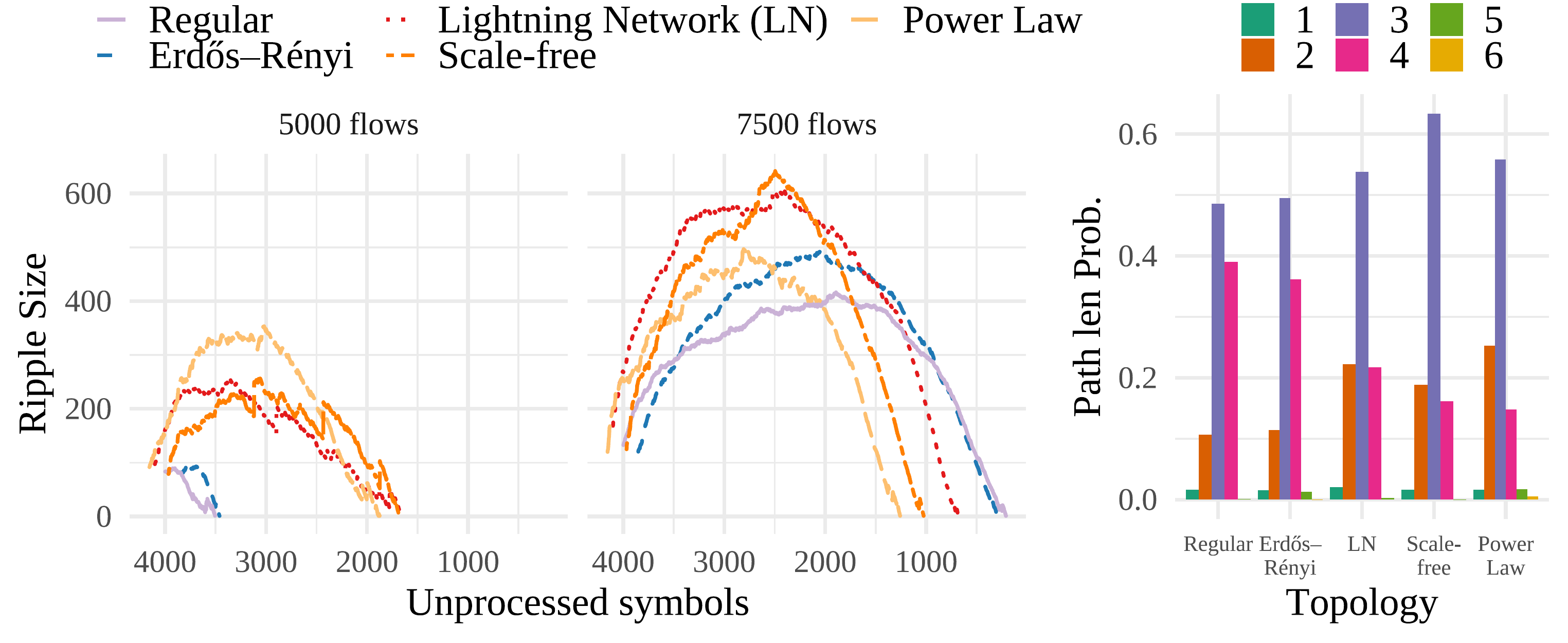}
    \caption{\small Evolution of ripple size during the \dpp on different topologies with a skewed demand matrix. The tendency of the ripple to vanish at 5000 flows explains the poor performance of \er and \rr graphs compared to the other random graphs.
    However, at 7500 flows, despite a lower peak value, the ripples of \er and \rr take longer to vanish than their \lntopo, \pltopo, and \sftopo counterparts. This allows for a larger number of peeled or deadlock-free channels, and consequently slightly better $\mintpt$. This trend is markedly different from  the ripple evolution at the same 7500 flows with a uniform random demand matrix, despite very similar path length distributions.}
    \label{fig:nonuniform random peeling real evolution}
    \vspace{-2mm}
\end{figure*}

\NewPara{Explaining the relative behavior of topologies.}
Like the uniform demand matrix case, we seek to understand if the differences in the $\mintpt$ of different topologies is explained by the evolution of the \dpp. 
\Fig{nonuniform random peeling real evolution} shows the evolution of the ripple size as
the \dpp progresses at 5000 and 7500 demand pairs, along with the path length distributions for the \rr, \er, 
\lntopo, \pltopo and \sftopo topologies for the skewed demand matrix case.
Similar to \Fig{peeling real evolution}, we consider the total number of symbols at the start to be twice the number
of channels in the topology, one for each direction of a channel.
Every step of the \dpp involves processing one channel in one of the two
directions. Each processed symbol may lead to the release of some flow nodes and consequently, add more 
directed channels to the ripple.

Firstly, notice that the trajectory of the ripple explains the relative performance of the topologies at 5000 and 7500 flows. At 5000 flows, the ripples associated with \rr and \er topologies fails to grow after the initial release of degree 1 flows resulting in large number of unpeeled edges, and consequently low $\mintpt$. In contrast, at 7500 flows, the \er and \rr topologies' ripples last longer than its counterparts resulting in higher $\mintpt$ values. However, this is a marked difference from the trajectory with a uniform demand matrix at the same 7500 flows (\Fig{peeling real evolution}) where the \er and \rr topologies' ripples disappear soon into the peeling process. This is despite
nearly identical path length distributions, which suggests an identical prediction based on the LT code analysis (\Fig{peeling predicted evolution}). This difference is due to the presence of correlations between the randomly sampled flows with a skewed 
demand matrix. Particularly, since a small set of nodes are more likely to be senders or receivers, certain channels close to these nodes are more likely to be part of flows than if demand pairs were sampled randomly.
This means two contradicting trends for the ripple evolution: first, a processed symbol is likely to reduce the degree of more nearby flows, which could potentially lead to their release and add more unprocessed channels to the ripple; second, flows are likely to contain many of the same channels, and thus fewer overall symbols are covered through the peeling process. We see both of these trends in \Fig{nonuniform random peeling real evolution}; the maximum ripple size is smaller across all topologies than the uniform random demand case but, the \er and \rr topologies' ripples grow early on because the few initially processed symbols are able to release more flows and add more symbols.

\NewPara{Discussion.} Figs.~\ref{fig:nonuniform random overall tpt} and ~\ref{fig:nonuniform random peeling real evolution} suggests that the the \dpp and ripple trends help compare the best and \wctpt behaviors of different topologies regardless of the nature of the demand matrix. However, the prediction and analysis of the \dpp needs to be revised in the presence of correlated bipartite graphs that deviate significantly from the i.i.d. bipartite encoding graph where channels are sampled randomly to be part of flows. This is not unexpected; in fact, it is rather fortunate that with a uniform random demand matrix, these correlations are minimized to a point that we can assume an i.i.d. bipartite graph and extend the LT codes analysis. However, with a skewed demand matrix, these correlations that affect how likely a channel is to be part of a flow need to be accounted for more carefully. One way to approach this is to consider a channel or input symbol degree distribution in addition to the encoded symbol degree distribution. Such a revised analysis can then be used in the rest of the synthesis pipeline to generate good degree distributions and topologies that are robust to deadlocks across more demand patterns.

\subsection{Topology Synthesis Using LT Codes Analysis}
\label{apdx:topo synthesis details}
\subsubsection{Preliminaries}
\label{apdx:peeling preliminaries}
Just as LT codes' degree distributions are designed to minimize the number of encoded symbols, 
with the \dpp, it is desirable to minimize the number of 
paths that need to sustain active payments in a \cn{} with a fixed number of channels. 
A natural question to ask towards this goal, 
is what path length distribution $\Omega(\cdot)$, akin to the degree distribution
in the LT code design, is optimal.

However, despite the similarities, the \dpp is markedly different from the LT process. 
First, every channel is processed twice: once in the 
blue direction and once in the red direction. This can be considered akin to having twice the number
of input symbols as channels unprocessed at the start of the peeling process.
Second, while the release of flows of degree $1$ covers one unprocessed symbol (as is with the LT process)
the release of flows of degree $0$ does not have an LT counterpart. 
In particular, when all the channels that a flow
of length $d$
uses have been covered in the direction of use, such a flow can freely move tokens on all of the 
associated channels covering up to $d - 1$ new channels in the opposite direction. 
In addition, with LT codes, an encoded symbol of degree $d$ chooses $d$ input symbols \emph{at random}
to XOR. However, a flow of length $d$ can never choose $d$ channels at random: 
the flows belong to an underlying topology with the proximity of two channels dictating
their presence in any given flow.

To account for these differences, we make two assumptions.
First, we analyze the \dpp under an \emph{i.i.d. bipartite construction}.
In other words, we assume that a flow of degree $d$ is allowed to choose $d$ channels at random
regardless of their location in the underlying graph. This is rather optimistic but captures
some important trends in the ripple evolution of a topology,
as has been already demonstrated in \Fig{peeling predicted evolution}.
Next, we assume that when a channel is processed in one direction, it is immediately also processed in the opposite direction. 
We approximate this in practice by ensuring that whenever a channel is processed from the ripple, 
if the channel's opposite direction is also in the ripple, it gets processed immediately thereafter.
Our evaluations suggest that $80-95\%$ of the channels in a random graph can be processed
sequentially. 
This implies that we can treat a channel and its two directions as a single unit,
akin to a single unprocessed symbol in the LT codes terminology.
During the \dpp, an unprocessed channel from the ripple is chosen
and processed (in both directions) reducing the degree of an aggregate set of flows (that use it in
either direction), leading to more channels being added to the ripple. 
An added advantage of this is that we can entirely ignore the effect of flows reaching degree $0$ and
covering $d - 1$ channels since \emph{both directions}
of earlier encountered symbols have already been processed.
Under these assumptions, we can  apply the analyses of LT codes \cite{lubylt, decripplelt}
exactly to the \dpp. 

\subsubsection{Analysis}
\label{apdx:peeling analysis}
Consequently, we are able to reuse the following three results from the LT codes 
analysis \cite{decripplelt, lubylt}, restated here
for clarity in the context of a \cn.
\begin{thm}{(Flow release probability)}
    Given a \cn{} $G(E,V)$, the probability that a flow of degree $d$ is released 
    $L$ out of $|E|$ channels remain unprocessed, 
    is given by
    \[
        q(d, L) = 
        \begin{cases} 
            1, & \text{if } L = |E|, R = 0  \\
            \frac{d(d - 1)L}{(|E|-d + 1) (|E| - d + 2)} \prod_{j=0}^{d-3} \frac{|E| - (L+1) - j}{ |E| - j}, 
            & \text{if } d = 2, \dots, |E|, 1 \le L \le |E| - d + 1 \\ 
            0,              & \text{otherwise}.
        \end{cases}
    \]
    \label{thm:flow release prob}
\end{thm}

\begin{thm}{(Ripple addition probability)}
    Given a \cn{} $G(E,V)$, the probability that a flow of degree $d$ is released 
    and adds a new channel to the ripple when the ripple size 
    is $R$ and $L$ out of $|E|$ channels remain unprocessed, 
    is given by
    \[
        q(d, L, R) = 
        \begin{cases} 
            1, & \text{if } L = |E|, R = 0  \\
            \frac{d(d - 1)(L - R + 1)}{(|E|-d + 1) (|E| - d + 2)} \prod_{j=0}^{d-3} \frac{|E| - (L+1) - j}{ |E| - j}, 
            & \text{if } d = 2, \dots, |E|, 1 \le R \le L \le |E| - d + 1 \\ 
            0,              & \text{otherwise}.
        \end{cases}
    \]
    \label{thm:ripple add prob}
\end{thm}

\begin{thm}{(Expected channels added)}
    Given a \cn{} $G(E,V)$ with $|\mathcal{P}|$ paths whose path lengths $d$ are drawn from a distribution
    $\Omega(d)$,
    the expected number of channels added to the ripple due to the most recently processed symbol,
    when the ripple size is $R(L + 1)$ 
    and $L$ out of $|E|$ channels remain unprocessed, is given by
    \[
        Q(L) = \sum_{d = 1}^{|E|}  |\mathcal{P}|\Omega(d)q(d, L, R(L + 1))
    \]
    \label{thm:expected symbols added}
\end{thm}

\NewPara{Accounting for Overlaps}. 
\Thm{expected symbols added} assumes that there is no overlap across the channels that are covered by
different flows that are released simultaneously after the most recent symbol has been processed 
(when the current ripple size is R and there are $L$ unprocessed symbols).
We find that this isn't the case in practice. To account for this, consider viewing the release 
of encoded symbols as a series of $n$ balls that are
dropped into some bins corresponding to the input symbol. A given ball (or flow) is released with 
probability $r(L) =\sum_{d=1}^{k} \Omega(d)q(d, L)$. 
All the bins they fall into corresponding to the 
covered symbols must be unprocessed; there are $L$ bins in total. 
$L - R$ of these are not in the ripple. 

The expected number of symbols added newly to the ripple, can therefore be computed as the
expected number of non-empty non-ripple bins at the end of the ball drops.
Consider
a random variable $X_i$ per bin where $i = 1, \dots, L$. $X_i = 1$ only if the $i^{th}$ bin
isn't already in the ripple and it is non-empty after all releasable flows are released. 
For the $L - R$ bins that correspond to input symbols not in the ripple, this is determined
by the chance that a bin is non-empty after all $n$ flows are released.
$$E[X_i] = 1\cdot Pr(X_i = 1) + 0 \cdot \mathrm{Pr}(X_i = 0) = \mathrm{Pr}(i^{th} 
\text{bin is not empty}) = \Big(1 - {\Big(1 - \frac{r(L)}{L}\Big)}^{n}\Big)$$
For the remaining $R$ bins, $E[X_i] = 0$ because those input symbols are already in the ripple.
Thus, 
\begin{equation}
    Q(L) =\sum_{i= 0}^{L} E[X_i] = (L - R)\left[1 - {\left(1 - \frac{r(L)}{L}\right)}^{n}\right]
    = (L - R)\left[1 - {\left(1 - \frac{\sum_{d=1}^{k} \Omega(d)q(d, L)}{L}\right)}^{n}\right]
    \label{eqn:expected symbols added}
\end{equation}
We use this expression when evaluating the predicted ripple size as a function of $L$
in \Fig{peeling predicted evolution}.

\subsubsection{Numerical Optimization}
\label{apdx:numerical optimization}
Since we have established that a number of the results translate
from the LT codes analysis to the \dpp, we next try to optimize for a good path length
distribution using a numerical approach~\cite{decripplelt}.
This section describes the steps involved in the numerical optimization 
which ultimately outputs a good path length distribution.

The optimization aims to find a good degree distribution by matching a
desired ripple evolution that is not only efficient, but also robust enough to ensure success of encoding. 
Prior work~\cite{decripplelt} has proposed the use of an exponential
decaying function of the following form where $L$ denotes the number of unprocessed symbols
and $R(L)$ denotes the ripple size when $L$ symbols remain unprocessed:
%


\begin{equation}
    R(L) = \left\{ \begin{array}{lr}
        c_1 L^{1/c_2}, & \text{if } c_1 L^{1/c_2} \le L;  \\
        L, & \text{otherwise.}
    \end{array} \right.
    \label{eqn:dopt:defr}
\end{equation}

We use the same ripple evolution curve with $c_1 = 1.7$ and $c_2=2.5$. Once, we have 
the exact values for the sizes of the ripple at every step of the \dpp,
we can compute the desired number of symbols added to the ripple $Q(L)$ when
$L$ symbols remain unprocessed as follows:
\begin{equation}
    Q(L) = \left\{ \begin{array}{lr}
        R(L), & \text{for } L = k;  \\
        R(L) - R(L+1) + 1, & \text{for } k > L \ge 0.
    \end{array} \right.
    \label{eqn:dopt:defq}
\end{equation}

Let $\Omega(\cdot)$ 
represent the path length 
    distribution of the flows (encoded symbols). For a \cn with $k$ channels,
    the path length can be utmost $k$.
    Ideally, we want to solve for $\Omega(\cdot)$ that matches the desired ripple evolution as closely
    as possible. The key insight that enables searching for such an $\Omega(\cdot)$ is that
    \Thm{expected symbols added} allows us to establish a linear relationship
    between the expected symbols added and the unknown degree distribution.
The desired degree distribution should ideally satisfy Eq.\eqref{eqn:dopt:Ax=b} where $m$ denotes
the total number of flows.
Notice that the variables here are $m_1 \cdots m_k$ rather than $\Omega(1) \cdots \Omega(k)$
directly, so the total number of flows in the satisfying solution can be computed
by summing all the variables.

For simplicity, we denote the matrix below by $A$, vector $m \cdot \begin{bmatrix}\Omega(1) & \cdots & \Omega(k) \end{bmatrix}\T$ by $x$ and the vector on the right hand side by $b$. 

\begin{equation}
    \begin{bmatrix}
        q(1, k, R(k+1)) & 0 & 0 \\
        \vdots & \ddots & 0 \\
        q(1, 1, R(2)) & \cdots & q(k, 1, R(2)) \\
    \end{bmatrix} 
    \begin{bmatrix}
        m \Omega(1) \\
        \vdots \\ 
        m \Omega(k) \\
    \end{bmatrix} =
    \begin{bmatrix}
        Q(k) \\ 
        \vdots \\
        Q(1) \\
    \end{bmatrix} 
    \label{eqn:dopt:Ax=b}
\end{equation}

However, as already mentioned this equation may not be satisfiable, particularly for non-negative
$\begin{bmatrix}\Omega(1) & \cdots & \Omega(k) \end{bmatrix}$. The standard optimization
    procedure used for LT-code degree distribution synthesis seeks to minimize
    the $\ell_2$-norm of $Ax - b$ for $x \succeq \mathbf{0}$. 

While this procedure could be directly useful for generating path length distributions,
it is particularly convenient that this structure lends itself to adding more constraints that
are specific to finding path lengths on a topology.
Assume we want to generate a topology with $k$ channels and $n$ nodes.
We impose constraints on a maximum path length $p_{\max}$ (Eq.\eqref{eqn:dopt:mpl}) since long
paths are undesirable both from a throughput
and deadlock perspective.
We impose a maximum node degree in the graph $d_{\max}$ (Eq.\eqref{eqn:dopt:mdeg}) to
enforce limits of centralization.
We also ensure that the number of paths of length $1$ does not exceed the number of
edges in the graph (Eq.\eqref{eqn:dopt:deg1}) and that
the probability of finding longer paths decreases as path length increases (Eq.\eqref{eqn:dopt:hdeg}).
Further, since $\Omega(\cdot)$ is a probability distribution, 
we enforce constraints Eq.\eqref{eqn:dopt:nz} and Eq.\eqref{eqn:dopt:distr}. 

As a consequence of these constraints, instead of solving $Ax = b$ for $x \succeq 0$, we find a good path length distribution $\hat x$ by solving the following optimization.

\begin{align}
    \mathrm{minimize} \qquad & \|Ax - b \|_2 \label{eqn:dopt:obj} \\
    \textrm{subject to} \qquad &
        x \succeq \mathbf{0} \label{eqn:dopt:nz} \\
        & \mathbf{1}\T x = m \label{eqn:dopt:distr} \\
        & x_1  \le \frac{2km}{n(n-1)} \label{eqn:dopt:deg1} \\ 
        & x_{i + 1} \le x_{i},~~ \forall i \geq 2 \label{eqn:dopt:hdeg} \\ 
        & x_{i} = 0,~~ \forall x > p_{\max} \label{eqn:dopt:mpl} \\ 
        & x_{i} \le \frac{{(d_{\max})}^{i}m}{n} \label{eqn:dopt:mdeg} 
\end{align}

The solution to this optimization problem gives us the path length distribution that satisfies
all the constraints and mimics the desired ripple evolution as closely as possible.


\subsubsection{Generating a matching topology}  
\label{apdx:topsynth}
Once we have a path length distribution from the constrained optimization 
outlined above, we aim to synthesize a graph topology that
achieves the desired path length distribution. We approach this by
searching over the space of random graphs captured using their joint degree distribution.
The joint degree distribution captures the probability that a randomly sampled edge in
the graph connects nodes of degree $j$ and $k$.
We choose to focus on the joint degree distribution in the hope that
its expressiveness will lend itself to a higher likelihood of matching our desired
path length distribution. There are still two steps to the process of using a joint degree distribution:a) finding the right distribution and b) synthesizing a graph to match it. The latter problem has been well-studied, with popular graph packages like networkx implementing generators~\cite{nxjointdegree}.
For the former portion, there exist recurrent analytical and exact expressions
that capture the probability that nodes are located within distance $d$ of each
based on the likelihood that an edge connects nodes of degree $j$ and $k$. 
(Eq.(5-10) in \cite{analyticalshortestpaths}).
We use this relationship to express the path length distribution as a function of the joint degree distribution
and iteratively search for a joint degree distribution whose corresponding path length 
distribution moves closer and closer to the desired path length distribution.
We perform this using MATLAB's black-box optimizer (~\cite{fmincon}).
We then verify that the output joint degree
distribution is valid (Theorem 2 in~\cite{jntdegconstr}) since not all joint degree distributions can be mapped
to a simple graph. Once we produce a valid degree distribution,
we sample from it to generate a joint degree sequence that denotes how many edges
need to be connected between a nodes of degree $j$ and $k$. Such a sequence also needs
to be validated (~\cite{nxjointdegreevalidation}) after which it generates a graph.
In the final output graph, we take the largest connected component and use
it as our synthesized graph. 

In the context of the experiments in \Sec{sec:eval-synthesis}, we sought to generate a topology with 300 nodes and 1500 edges. We first obtained a desirable path length distribution from the numerical optimization with a maximum path length constraint of 10 and a maximum degree constraint of 10. We then feed it into the MATLAB optimizer to generate a valid joint degree distribution.
Direct sampling from the output of the MATLAB optimizer did not give us a valid joint
degree sequence; we improved it by adding some edges until the conditions for a 
valid joint degree sequence were met. This resulted in a sequence for 301 nodes and 1540 edges. The synthesized graph was disconnected with the largest component containing 271 nodes and 1513 edges. We use this largest component as our synthesized topology in \Fig{synthesis comparison}.

\end{document}